\documentclass[aps,11pt,english,tightenlinesletterpaper,superscriptaddress,secnumarabic,nofootinbib]{revtex4}

\usepackage{amsmath,amssymb}
\usepackage[T1]{fontenc}
\usepackage[latin9]{inputenc}
\usepackage[letterpaper]{geometry}
\usepackage{amssymb}
\usepackage{esint}
\usepackage{color}
\usepackage{graphicx}
\usepackage{verbatim}
\usepackage{slashed} %Allows to use \slashed{xxx}

\makeatletter

\@ifundefined{textcolor}{}
{%
 \definecolor{BLACK}{gray}{0}
 \definecolor{WHITE}{gray}{1}
 \definecolor{RED}{rgb}{1,0,0}
 \definecolor{GREEN}{rgb}{0,1,0}
 \definecolor{BLUE}{rgb}{0,0,1}
 \definecolor{CYAN}{cmyk}{1,0,0,0}
 \definecolor{MAGENTA}{cmyk}{0,1,0,0}
 \definecolor{YELLOW}{cmyk}{0,0,1,0}
 }

\makeatother

\usepackage{babel}
\usepackage{hyperref} %Automatically links \label and \ref commands; Always load last
\hypersetup{
    colorlinks=false,       % false: boxed links; true: colored links
    linkcolor=red,          % color of internal links
    citecolor=green,        % color of links to bibliography
    filecolor=magenta,      % color of file links
    urlcolor=cyan           % color of external links
}

\numberwithin{equation}{section}

\newcommand{\GeV}{\ \mathrm{GeV}}
\newcommand{\eref}[1]{\ensuremath{\mathrm{Eq.}\;(\ref{#1})}}
\newcommand{\sv}{\left< \sigma v \right>}
\newcommand{\ord}[1]{O\left( #1 \right)}
\newcommand{\ex}{\mathrm{ex}}

\newcommand{\hs}{\left\{ h,s \right\}}
\newcommand{\vx}{\left\{ v,0 \right\}}
\newcommand{\const}{\mathrm{const.}}

\begin{document}

\title{Cosmological Constant, Dark Matter, and Electroweak Phase Transition}

\author{Daniel Chung and Andrew Long \\ 
Department of Physics, University of Wisconsin, Madison, WI 53706 \\ 
\ E-mail:\ danielchung@wisc.edu, ajlong@wisc.edu
}

\begin{abstract}

%We consider an effective cosmological constant obtained by summing the
%energy density of the vacuum condensate and unknown UV physics.  
Accepting the fine tuned cosmological constant hypothesis, we  
%If the cosmological constant is small today because of a fine tuning
%between vacuum condensate and quantum fluctuation contributions, one
%can calculate it during an electroweak scale phase transition as the
%vacuum state changes.  In principle the tuning hypothesis can be
have recently proposed that this hypothesis can be tested if the dark
matter freeze out occurs at the electroweak scale and if one were to
measure an anomalous shift in the dark matter relic abundance.  In
this paper, we numerically compute this relic abundance shift in the
context of explicit singlet extensions of the Standard Model and
explore the properties of the phase transition which would lead to the
observationally most favorable scenario.
%For the case of the Standard
%Model, the observable effect is at the level of a fraction of a
%percent.  However, for beyond the Standard Model scenarios, the
%observable effect may be as large as order unity.  
Through the numerical exploration, we explicitly identify a parameter
space in a singlet extension of the standard model which gives order
unity observable effects.  We also clarify the notion of a temperature
dependence in the vacuum energy.
\end{abstract}

\maketitle

\tableofcontents

% sec:Introduction
\section{Introduction}\label{sec:Introduction}

The hypothesis that the cosmological constant (CC) energy
density today is a result of a tuning between UV and IR contributions
\cite{Linde:1974at,Ellis:1975ap} 
%{[}MORE REFS{]}.  Although one often
%makes this assumption implicitly and without justification in order to
%perform routine early universe computations, it seems to be a favored
is favored according to some versions of the string landscape proposal
(see e.g.~\cite{Polchinski:2006gy}).  Furthermore, this hypothesis has
always been the default assumption in particle physics model building
(see e.g.~\cite{Nilles:1983ge,Weinberg:1988cp}).  Unfortunately, this
conjecture is notoriously difficult to test with lab experiments, such
as those at colliders.

One of the predictions of the tuning hypothesis is that there can be
an electroweak scale effective CC in the early
universe if there was a phase transition (PT) at that scale.  A
well-known reason to suspect that there was an electroweak scale PT in
the early universe is the thermally supported electroweak symmetry
restoration phenomenon in the context of the Standard Model (SM) of
particle physics \cite{Kirzhnits:1972iw,Kirzhnits:1972ut}.  Hence, if
lab experiments, such as particle colliders, can eventually measure
the field content and couplings of the scalar sector at the
electroweak scale with sufficient accuracy, then one may be able to
predict the CC energy density existing at around
the time of the PT.  Such an energy density would
interact with gravity to modify the expansion history of the universe.
Indeed, Kolb and Wolfram \cite{Kolb:1979bt} were one of the first to
state that this computable energy density arising from the Standard
Model Higgs condensate may have an observationally acceptable yet
significant effect in cosmology.

In a recent paper \cite{Chung:2011hv}, we proposed that dark matter
freeze out can be used to probe PTs, including the
properties of such a computable CC, through its effect on the
expansion rate of the universe during freeze out.  Such an idea is
abstractly very similar to the well known big bang nucleosynthesis
idea, as well as generic particle probes of cosmology (see e.g.~
\cite{Alpher:1953zz,Barrow:1982ei,Kolb:1990vq,Kamionkowski:1990ni,Salati:2002md}). In
particular, if a weakly interacting massive particle (WIMP) dark
matter candidate is discovered with a mass of the order of TeV, then
its freeze out dynamics would be sensitive to the value of the CC
during the electroweak scale PT at a temperature of the order of $100
\GeV$.  Therefore accurate measurements of the dark matter and scalar
sector properties will, in principle, make possible a lab test of the
tuning of the CC.  More accurately, what is being tested is the
absence of self-tuning mechanisms and/or modified gravitational
theories \cite{Aslanbeigi:2011si,Agarwal:2011mg,deRham:2010tw,Smolin:2009ti,Wetterich:2010kd,Kim:2010fb,Davidson:2009mp,Koroteev:2007yp,Klinkhamer:2007pe,Dvali:2007kt,Itzhaki:2006re,Lee:2003wg,Kachru:2000hf} that would eliminate or significantly change the effects of
vacuum energy during a PT.

For non-first order PTs, it was found that the shift in the relic
abundance due to the CC energy density effects is suppressed by
$\Delta n_X / n_X=\ord{g_{E}^{-1}}$ where $g_{E}$ is the number of of
relativistic degrees of freedom contributing to the energy
density. For first order PTs, it was found that this fractional shift
can be generically enhanced by supercooling such that the CC effects
can be $O(1)$ with a $1\%$ parameteric tuning.  In all cases, the
sought after CC ``signal'' is buried in the
dominant ``background'' coming from the adiabatic change in the number
of degrees of freedom and possibly the entropy release near the time
of the dark matter freeze out.  The adiabatic change in the number of
degrees of freedom and the vacuum energy effect tend to increase the
relic density today while the entropy production effect decreases the
relic density.

The purpose of this paper is to complement the previous short paper
\cite{Chung:2011hv} in several ways:
\begin{enumerate}
\item Present an explicit singlet extension of the Standard Model (SM)
  that gives a large supercooling with a first order PT at the
  electroweak scale.

\item Clarify the notion of how an effective vacuum energy (which is
  Lorentz invariant in the flat space limit) can depend on temperature
  (which manifestly breaks Lorentz invariance).

\item Compare numerical results with analytic results presented in
  \cite{Chung:2011hv}.

\item Provide relevant technical details that were left out in
  \cite{Chung:2011hv} to aid future research efforts in this
  direction.  

\end{enumerate}
In addition to giving a generic singlet scalar model coupled to a
Dirac fermion that gives a significant supercooling, we analyze xSM,
i.e. a real singlet coupled to the SM, and identify a parametric
region in which significant supercooling occurs.  As anticipated in
\cite{Chung:2011hv}, an $O(10^{-2})$ tuning is sufficient to induce an
$O(1)$ supercooling effect on the relic abundance.  
%analyze several thermal dark matter freeze out scenarios to understand
%the observable consequences of first order and smoother, non-first
%order phase transitions.  
%Hubble expansion rate 
%$\Delta H/H=O(g_{E}^{-1})$ 

The order of presentation is as follows.  In Section
\ref{sec:A-Brief-ReviewrofPT} we review the physics of PTs and focus
on the myriad ways in which a PT may impact dark matter freeze out.
We clarify the notion of a temperature dependence of vacuum energy
density in this section.  In Section \ref{sec:AnalyticEst} we
analytically compute the fractional shift of the relic abundance
$\delta n_X(t_0)$ due to an electroweak scale PT in the limit in which
the PT represents a small perturbation to the usual freeze out.  In
Section \ref{sec:IllustrativeModels} we compute the relic abundance
deviation in the SM and minimal singlet extensions (both supplemented
by a generic dark matter which is assumed to play a negligible role in
determining the properties of the PT).  In Section
\ref{sec:Conclusion} we conclude with a summary and suggestions for
future work.  An extensive set of appendices detail technicalities
useful for the material presented in the body of the paper.

Throughout the paper, we assume a flat Friedmann-Robertson-Walker
(FRW) metric, $ds^{2}=dt^{2}-a^{2}(t)|d {\bf x}|^{2}$, and use the
reduced Planck mass $M_{p}\approx2.4\times10^{18}$ GeV.

% sec:A-Brief-ReviewrofPT
\section{A Brief Review of the Physics of Phase Transitions\label{sec:A-Brief-ReviewrofPT}}

In this section, we review the physical features that accompany a
cosmological PT.  Each of these features modifies one of the
relationships, $\rho \sim T^4$, $T \sim a^{-1}$, or $\sv = \sv(T)$,
which are assumed in the usual freeze out calculation.  One of the
topics discussed in this section is how to understand
the thermal dependence of vacuum energy, which a priori is an
oxymoron.  Readers interested in mostly the phenomenology can skip to
the next section.

The standard cosmological model assumes an expanding FRW universe
which leads to the temperature of the relativistic species in the
universe decreasing as a function of time except during the time
periods when entropy is generated.  As the temperature decreases,
there may exist critical temperatures at which the thermodynamic
quantities are not analytic as a function of temperature and/or the
symmetries of the effective Lagrangian governing the dynamical degrees
of freedom changes.  Following the typical convention in the
literature, we refer to the passages through these critical
temperatures as PTs.

In order to calculate thermodynamic quantities in the system described
above, we will use the thermal effective potential (see
\cite{Quiros:1999jp} for a review).  The thermal effective potential
$V_{\rm eff} ( \phi_{c}, T )$, derived from Legendre transforming the
partition function coupled to external sources, represents the free
energy density of the plasma at temperature $T$ dynamically
interacting with a homogeneous scalar field background $\phi_{c}$ which
may affect the masses and interactions of particles in the plasma.
%The potential can be calculated perturbatively, and to lowest order,
%one evaluates $\phi_c$ at the vacua of $S[\phi]$ given by
%$v^{(s/b)}(0)$, where $s$ and $b$ represent ``symmetric'' and
%``broken'' vacua, which can be two different perturbative expansion point
%The interactions of $\phi$ with the thermal background generically
%generate tadpoles which tend to shift the expectation value of $\phi$.
%Resumming these tadpoles changes the expansion point of the theory to
%$\phi_c = v^{(s/b)}(T)$ which are defined to be the extrema of $V_{\rm
%  eff} ( \phi_{c}, T )$.  Note that although $v^{(s)}(T)$ is defined
%at high temperatures before the phase transition and $v^{(b)}(T)$ at
%low temperatures after the phase transition, we cannot always combine
%the two functions to form a single $v(T)$, because in the case of a
%first order phase transition the two phases coexist during the phase
%transition.
A local minimum $\phi_c=v(T)$ is called the thermal vacuum, and PTs
occur near critical temperatures $T_c$ which will be defined more
precisely below.\footnote{We will leave out the adjective ``thermal''
  in ``thermal vacuum'' whenever no confusion should arise.}

The critical temperature $T_c$ in the case of what is conventionally
referred to as a first order PT is defined by the existence of two or
more degenerate minima $\phi=v(T_c)$ existing for the thermal
effective potential $V_{\rm eff} ( \phi, T_c )$.  In such cases, we
refer to the vacuum of the universe just prior to the PT as
$v^{(s)}(T)$ (where the ``s'' superscript denotes ``symmetric''
vacuum) whether or not there is a symmetry in the thermal effective
potential prior to the PT.  The vacuum solution after the first order
PT is referred to as $v^{(b)}(T)$ where ``b'' denotes ``broken.''  A
non-first order PT (sometimes loosely referred to as a second order
PT) is characterized by a single continuous function $v(T)$ before and
after the PT: i.e.~$v^{(b)}(T_c)=v^{(s)}(T_c)$.  Even in
such situations, it is sometimes useful to define $v^{(s)}(T)$ to be
the vacuum before the PT whenever there is a restored
symmetry prior to the PT.  The quantity $v^{(s)}(T)$ can
then be taken as an order parameter associated with spontaneous
symmetry breaking.

The thermal vacua $v^{(s/b)}(T)$ can be obtained from summing up
thermal tadpole corrections obtained from expanding perturbatively
about the zero temperature vacua $v^{(s/b)}(0)$.  Despite the
suggestive notation of the thermally shifted vacuum $v^{(s/b)}(T)$,
the resummation of tadpoles is nothing more than a reorganization of
perturbation theory, and the vacuum energy represented by the
Lorentz-invariant part of the energy-momentum tensor, is not shifted
by the manifestly Lorentz-noninvariant thermal tadpoles.  Note that
%the thermal tadpole resummation also cannot by itself lead to symmetry
%restoration 
symmetry restoration cannot be inferred from the thermal tadpole 
resummation alone 
since the thermal perturbation theory breaks down when the
perturbations are expanded about the inflection points of the effective
potential.

Let us now establish some more notation for the quantities introduced
above.  The thermal effective potential and $v^{(s/b)}(T)$ can be used
to construct the thermodynamic quantities
\begin{subequations}\label{eq:therm_defs}
\begin{align}
%	\mathcal{F}_{s/b}(T) & = V_{\rm eff} (v_{s/b}(T), T) \label{eq:therm_defs_F} \\
%	s_{s/b}(T) & = - \frac{d \mathcal{F}_{s/b}}{dT} \label{eq:therm_defs_s}\\
%	\rho_{s/b}(T) & = \mathcal{F}_{s/b} + T \, s_{s/b} \label{eq:therm_defs_rho}
	\mathcal{F}^{(s/b)}(T) & = V_{\rm eff} (v^{(s/b)}(T), T) \label{eq:therm_defs_F} \\
	s^{(s/b)}(T) & = - \frac{d}{dT} \mathcal{F}^{(s/b)} \label{eq:therm_defs_s}\\
	\rho^{(s/b)}(T) & = \mathcal{F}^{(s/b)} + T \, s^{(s/b)} \label{eq:therm_defs_rho}
\end{align}
\end{subequations}
representing the free energy density $\mathcal{F}$, entropy density $s$, and energy density $\rho$ in the symmetric and broken phases.  
%Note that the thermodynamic quantities will depend on temperature both explicitly and implicitly through $v_{s/b}(T)$.  
A typical PT occurs as the universe cools, and the free energy of the
broken phase, in which the entropy and energy densities are high,
drops below the free energy of the symmetric phase, in which the
entropy and energy densities are low.   
It will be useful to define the critical temperature of the PT $T_c$ by 
\begin{align}\label{eq:Tc_def}
	\mathcal{F}^{(s)}(T_c) = \mathcal{F}^{(b)}(T_c) \, ,
\end{align}
but note that the PT may not actually occur until a much lower
temperature if the symmetric phase experiences supercooling.
% the broken phase, with a lower energy density than the symmetric phase, becomes energetically favored despite its lower entropy density.  
The PT is accompanied by a number of physical features, which we will
outline in the remainder of this section and which each have an impact
on dark matter freeze out.

The first feature that we would like to discuss is the vacuum energy
associated with the PT.  We assume that the energy density
$\rho^{(s/b)}(T)$ can be partitioned into the energy associated with
the plasma and the energy associated with the condensate (i.e. the
vacuum energy with an effective equation of state of $-1$), and we
define the latter as
\begin{equation}\label{eq:rhocc_def}
%	\rho_{cc} (v_T) \equiv V_{\rm eff} (v_T,0) \, ,
	\rho_{cc}^{(s/b)} (T) \equiv V_{\rm eff} (v^{(s/b)}(T),0) \, 
\end{equation}
%is smalle r in the broken phase after the phase transition than in the symmetric phase before (the use of the notation ``cc'' is clarified below).  
%In fact, each successive phase transition in the cosmological history causes $\rho_{cc}$ to decrease further.  
which has an observable consequence when coupled to gravity.  This
equation is artificial because the vacuum energy cannot be rigorously
separated from the particle energy with which it is associated for
most of the states populating the density matrix.  Nonetheless, it is
useful because it captures the CC type of contribution (i.e. negative
equation of state contribution) to the energy-momentum tensor.

Note that flat space thermal corrections to the zero temperature
effective potential cannot generate Lorentz invariant contributions to
the energy-momentum tensor because temperature $T$ dependent
quantities are not Lorentz invariant.  Since the CC contribution to
the energy-momentum tensor in the flat space limit is Lorentz
invariant, one may wonder whether Eq.~(\ref{eq:rhocc_def}) is valid
since it implies that thermal tadpoles are contributing to the vacuum
energy.  Furthermore, the fact that the effective vacuum energy takes
on a continuum of values while the only non-perturbatively stable
vacuum state is at $v^{(b)}(0)$ (which we will assume to be associated
with negligible vacuum energy) also leads one to be suspicious of
Eq.~(\ref{eq:rhocc_def}).

To semi-quantitatively resolve this puzzle, one notes that near the
time of the PT, there are $A \leftrightarrow B$ processes in
equilibrium where $A$ and $B$ schematically correspond to states of
the form $|\mbox{particles + vacuum energy}\rangle$ and
$|\mbox{particles}\rangle$, respectively.  These transitions are
mediated by non-perturbative processes since they are vacuum changing
processes.  Classically, the plasma (when these transitions are
efficient) is approximately described by inhomogeneous solutions in
Minkowski space.  This can easily be characterized by computing for
example the thermal two-point function.

The equation of state for such a plasma in the classical approximation
corresponds to neither that of quantum expectation values with respect
to states $A$ nor $B$, but is a mixture which from the quantum
perspective depends on the non-perturbative transition operators as
well as the relative statistical and/or coherent weighting of $A$ and
$B$ type of states.  The incoherent aspect of this mixture is what the
$T$ dependence of Eq.~(\ref{eq:rhocc_def}) reflects.\footnote{Note
  also unlike in flat spacetime, there are IR cutoffs associated with
  the expansion rate $H$ for a single causal domain during the PT and
  $H_0$ associated with the presently observable universe.  The former
  scale $H$ is also associated with one of the scales at wich
  quasi-equilibrium assumption breaks down.}  To corroborate this
picture, one can easily solve classical equations of motion in models
with spontaneous symmetry breaking to obtain {\em inhomogeneous
  background} field solutions which have an inhomogeneous equation of
state.  Since the Friedmann equation (which is the only gravitational
probe we will be concerned in this paper) approximately describes the
gravitational response to the spatial average of the energy-momentum
tensor, one can spatially average the energy density and the pressure.
This leads to an effectively homogeneous energy density and pressure
which is approximately the same as that due to particles plus a vacuum
condensate energy density.  The resulting effective vacuum condensate
energy density is somewhere between $V_{\rm eff} (v^{(s)}(0),0)$ and
$V_{\rm eff} (v^{(b)}(0),0)$, justifying the diagnostic quantity
defined by Eq.~(\ref{eq:rhocc_def}).

To renormalize the CC, we impose the tuning condition
\begin{equation}
%	\rho_{cc}(T = 2.7 \mathrm{\, K}) = \rho_{DE}
	\rho_{cc}^{(b)}(T = 0) = 0 \, ,
\label{eq:renormalizationcond}
\end{equation}
which states that the vacuum energy density today is on the order of
the $\mathrm{meV}^4$ dark energy density
\cite{Caldwell:2009ix,Komatsu:2010fb} and negligible as compared to
the PT scale.  Hence, we will refer to $\rho_{cc}^{(s/b)}(T)$ as the
``effective CC energy density.''  With this normalization, the CC
energy density before a PT at scale $M$ will typically be
\begin{equation}
\rho_{cc}^{(s)} (T \gtrsim M) \sim M^4,
\label{eq:CCfalsevac}
\end{equation} 
which can be measured, in principle, by gravitational probes such as
the Hubble expansion rate and its impact on dark matter freeze
out.\footnote{Although an in depth discussion of UV sensitivity of the
  CC is beyond the scope of this paper, one should keep in mind that
  using Eq.~(\ref{eq:renormalizationcond}) as a quantum
  renormalization condition leads to Eq.~(\ref{eq:CCfalsevac}) as
  a prediction only if assumptions about analyticity of the effective
  potential as well as Lorentz invariance structure of the UV cutoff
  is assumed.}  Any self-tuning/modified-gravity mechanism 
  %(for some
%attempts in this direction see
%e.g.~\cite{Aslanbeigi:2011si,Agarwal:2011mg,deRham:2010tw,Smolin:2009ti,Wetterich:2010kd,Kim:2010fb,Davidson:2009mp,Koroteev:2007yp,Klinkhamer:2007pe,Dvali:2007kt,Itzhaki:2006re,Lee:2003wg,Kachru:2000hf})
which decouples the vacuum energy or significantly modifies the vacuum
energy effect on gravity on a time scale shorter than that of the
expansion scale will have an effective $\rho_{cc}^{(s)}$ significantly
different from Eq.~(\ref{eq:CCfalsevac}).  It would be interesting in
future studies to compare various self-tuning/modified gravity models
which may have non-trivial time dependence in the effective vacuum
energy different from that in this paper.

The second PT feature is the decoupling of heavy degrees of freedom 
which become non-relativistic after the PT and cause the number 
of relativistic species, denoted here as $g$, to decrease.  
This has two 
consequences for the dark matter freeze out calculation.  First, the energy density 
of the plasma $\rho \sim g \, T^4$ 
and Hubble expansion rate $H \sim \sqrt{\rho}$ decrease more 
rapidly than usual after the PT.  Second, since temperature 
is related to the FRW scale factor by entropy conservation, which gives 
$T \sim g^{-1/3} a^{-1}$, the temperature decreases less rapidly 
than usual after the PT.  To estimate the magnitude of the effect on 
dark matter freeze out, 
consider the SM electroweak PT at $T \sim 100 \GeV$ and suppose 
that freeze out occurs at the same temperature.  Then during the 
residual annihilation stage 
of freezeout, which lasts until $T \sim 10\GeV$, $g$ will decrease by approximately 
$20 \%$ corresponding to the decoupling of the top, Higgs, and massive 
gauge bosons.  In the usual freeze out calculation, changes in $g$ are neglected, 
because freeze out occurs much later than the electroweak PT when $g$ is 
insensitive to $T$.  When we arrange for the two events to occur at the 
same scale, $g$ decreases significantly and can have a large effect on the relic 
abundance.

The third feature is related to the coupling between the PT sector and
the rest of the particle physics model.  As the phase changes at the
PT, in general the masses and interactions of particles in the plasma
can change as well.  In particular, it is possible for the scalar
field to couple to dark matter in such a way that the dark matter's
mass and/or annihilation cross section is different in the symmetric
and broken phases.  This scenario, studied by \cite{Cohen:2008nb,
  Feng:2009mn}, may allow dark matter to rethermalize and can have a
significant effect on the relic abundance.

If the PT is of the first order, then it possesses a number of additional 
features.  A first order PT can be divided into two stages.
The first stage, known as supercooling, occurs while the universe remains in the 
symmetric phase after it has become metastable at  $T \approx T_c$.  As the 
temperature decreases 
and the CC energy density remains approximately constant, the total energy 
density can deviate from the standard $\rho \sim T^4$ scaling (i.e., first feature
above).  Supercooling ends when it becomes energetically favorable for 
bubbles of the broken phase to nucleate.  Determining the temperature 
$T_{PT}^-$ at which bubble nucleation begins requires one to solve for the 
non-perturbative bounce solution and evaluate the decay rate of the 
metastable phase \cite{Linde:1981zj}.  
%This calculation cannot be performed analytically to the necessary accuracy and represents our largest numerical challenge and uncertainty when working with first order phase transitions.  
During the second stage, known as reheating, the expanding bubbles release 
%a latent heat
an energy density
\begin{equation}\label{eq:L_def}
%	L =\rho_{cc}\left( v_T^{sym}(T_{PT}^-) \right) -  \rho_{cc} \left( v_{T}^{brk}(T_{PT}^+) \right)
	\Delta \rho_{\rm ex} =\rho_{cc}^{(s)}(T_{PT}^-)  -  \rho_{cc}^{(b)}(T_{PT}^+) 
\end{equation}
which is converted into radiation and heats the gas from $T_{PT}^-$
before the PT to $T_{PT}^+ > T_{PT}^-$ after the PT.  We assume that
reheating occurs rapidly as compared with the expansion rate of the
universe\footnote{ A third stage, known as phase coexistence, can
  occur if a large latent heat is released by the expanding bubbles
  and the plasma is reheated to the point where the pressure gradient
  across the bubble wall vanishes \cite{Witten:1984rs}.  Subsequently,
  the bubbles expand only insofar as the universe expands, and the PT
  completes on a time scale $t \sim H^{-1}$.  Typically, this stage
  does not occur during an electroweak-scale PT because the number of
  relativistic species $\ord{100}$ is too many to allow for sufficient
  reheating.  }, which allows us to treat reheating as an abrupt
process at time $t_{PT}$ when $a = a_{PT}$.  Reheating is accompanied
by a non-adiabatic entropy increase.  This entropy growth modifies the
relationship between temperature and the FRW scale factor in such a
way that the universe is relatively larger for a given temperature.
As a result, the dark matter number density undergoes a longer period
of dilution and the relic abundance can be significantly smaller
\cite{Wainwright:2009mq}.  Finally, just as massive species can
adiabatically decouple after the electroweak PT occurs,
heavy particles can undergo a non-adiabatic decoupling at the time of
a first order PT if they abruptly acquire a mass $m \gtrsim T_{PT}$.

% sec:AnalyticEst
\section{\label{sec:AnalyticEst}An Analytic Estimate of the Change in the Dark Matter Abundance}

In this section, we estimate the change in the dark matter relic
abundance due to the presence of a PT, and the CC energy density in
particular, during freeze out.  Our final result is the fractional
deviation of the relic abundance, denoted $\delta n_X(t_0)$ and given
by \eref{eq:finalfractionalchange}, in which we have linearized in the
various effects of the PT on freeze out.  Although most of the results
in this section have already been presented in \cite{Chung:2011hv}, we
repeat some of the results for self-containedness as well as serving
as introduction for more complete results such as Eqs.~(\ref{eq:c1e1})
and (\ref{eq:c2e2}).  The main point of this section is to present a
formalism to understand analytically the effects outlined in Sec.~\ref{sec:AnalyticEst}.

Throughout the calculation, we will take $a$ as the independent
variable and rewrite functions of temperature using $T = T(a)$ given
by \eref{eq:Tofa}.  In particular, we will assume that freeze out
occurs at a temperature $T_f = T(a_f)$ before the PT at $a = a_{PT}$.
Since all of the thermodynamic quantities depend on the phase of the
system which changes at $a = a_{PT}$, the formulas in this section
would become unnecessarily obscure if we persisted in writing all the
$(s/b)$ superscripts and distinguishing the $a < a_{PT}$ and $a>
a_{PT}$ cases.  Therefore, we introduce the following shorthand.
Whenever a temperature-dependent function $F^{(s/b)}(T)$ appears
without the $(s/b)$ superscript, the intended meaning is
\begin{align}\label{eq:Fofa}
	F(a) = \begin{cases} F^{(s)}(T(a)) & a < a_{PT} \\ F^{(b)}(T(a)) & a > a_{PT} \end{cases}.  
\end{align}
In particular, one always has $F(a_f) = F^{(s)}(T_f)$ since $a_f < a_{PT}$ by assumption.

We calculate the thermal relic abundance of dark matter by integrating the thermally averaged Boltzmann equation, 
\begin{align}\label{eq:Bmaneqn}
	\frac{1}{a^{3}}\frac{d}{dt}\left(a^{3}n_{X}\right)=-\langle\sigma v\rangle\left(n_{X}^{2}-n_{X}^{\text{eq} \, 2}\right) ,
\end{align}
over the era of residual annihilations from freeze out at $a = a_f$ until today.  
Subject to general assumptions (see Appendix \ref{sub:simplederiv} for more details), we obtain 
\begin{equation}\label{eq:goodformula} 
	n_{X}(t_{0})=\left(\int_{0}^{\ln a_{0}/a_{f}}\frac{d\ln(a/a_{f})}{H} \sv \frac{a_{0}^{3}}{a^{3}}\right)^{-1}
\end{equation}
for the number density of dark matter today at $a = a_0$ and $t = t_0$.  
In this expression, the quantities that will be affected by the PT are the Hubble expansion rate $H(a)$, the thermally averaged cross section $\sv(a)$, and the dilution number since the time of the freeze out to today $a_{0}/a_{f}$, which is related to $T(a)$.  
As a fiducial reference value, we also calculate the ``usual'' relic abundance $n_X^{(U)}(t_0)$ by assuming that the PT does not occur, but instead that the universe remains radiation dominated 
%with $g_{E/S}^{(U)} = 106.75$ relativistic degrees of freedom 
and has the standard scaling relations 
\begin{align}\label{eq:usual_scaling}
	H^{(U)}  \sim a^{-2}, \qquad \qquad
	\sv^{(U)}  = \sv(T(a)), \qquad {\rm and} \qquad
	T^{(U)}  \sim a^{-1}
\end{align}
throughout freeze out.  
We define the relic abundance fractional deviation as
\begin{align}\label{eq:fractional_dev}
	\delta n_X(t_0) = \frac{n_X(t_0)}{n_X^{(U)}(t_0)} - 1
\end{align}
and expect this quantity to depend on the way in which $H$, $\sv$, and $a_0 / a_f$ deviate from the usual freeze out scenario.  
We will consider each effect in turn. 

Before addressing each of the factors in \eref{eq:goodformula}, let us discuss 
the partitioning of energy.  
The Hubble expansion rate, which appears in \eref{eq:goodformula}, is related to the total energy density $\rho^{(s/b)}(T)$.  
However, we are particularly interested in determining the impact of the effective CC on the calculation of dark matter freeze out.  
Therefore we will assume that the energy can be partitioned as 
\begin{equation}\label{eq:energy_partitioning}
	\rho\approx\mbox{(particle degrees of freedom + exotic energy component)} \, .
\end{equation}
In general, the exotic energy component can arise from physics other
than the effective CC, such as quintessence
(e.g.~\cite{Salati:2002md,Kamionkowski:1990ni,Rosati:2003yw,Profumo:2003hq,Pallis:2005hm,Chung:2007cn,Chung:2007vz})
or late-decaying massive particles
(e.g. \cite{Weinberg:1982zq,Coughlan:1983ci,McDonald:1989jd,Banks:1993en,Moroi:1999zb,Berkooz:2005sf,Dine:2006ii,Acharya:2008bk}).
To maintain a minimal degree of generality throughout our analytic
estimates (without accumulating distasteful notational complication),
we will parametrize the exotic energy component as $\rho_{\ex} \,
\kappa (a)$.
%where 
%\begin{equation}\label{eq:x_def}
%	x\equiv\frac{a}{a_{f}}  \, .
%\end{equation}
%However, since the exotic energy component is of greatest interest for us
%is the effective CC, we will write
However, since our primary interest is in the case that the exotic energy 
component represents an effective CC, we will write
\begin{align}
	\rho_{\rm ex} \, \kappa(a) = \rho_{cc}(a)
\end{align}
where $\rho_{cc}^{(s/b)}(T)$ is defined by \eref{eq:rhocc_def}, and we
have used the shorthand \eref{eq:Fofa}.  The remaining energy density
can be attributed to relativistic particles in the plasma, which we
will denote by\footnote{ Contributions from non-relativistic species
  are Boltzmann suppressed.  Defined in this way, $\rho_R^{(s/b)}$
  includes a term proportional to $d v^{(s/b)} / dT$ which arises from
  the derivative in \eref{eq:therm_defs_s}.  This term represents
  kinetic energy in the scalar field and, strictly speaking, should
  not be included in $\rho_R$.  Nevertheless, we do not separate out
  the kinetic term, because it is typically negligible.  }
\begin{equation}\label{eq:rhoR_def}
	\rho_{R}^{(s/b)}(T) = \rho^{(s/b)}(T)- \rho_{cc}^{(s/b)}(T) \, .
\end{equation}
To connect with a familiar and intuitive notation, we let the functions $g_E$ and $g_S$ be defined implicitly by 
\begin{align}
%	\rho_{R}(T,v_T) &=\frac{\pi^{2}}{30} \, g_{E}(T,v_T) \, T^{4} \label{eq:gE_def} \\
%	s(T,v_T) &=\frac{2\pi^{2}}{45} \, g_{S}(T,v_T) \,T^{3} \label{eq:gS_def}
	\rho_{R}^{(s/b)}(T) &=\frac{\pi^{2}}{30} \, g_{E}^{(s/b)}(T) \, T^{4} \label{eq:gE_def} \\
	s^{(s/b)}(T) &=\frac{2\pi^{2}}{45} \, g_{S}^{(s/b)}(T) \,T^{3} \label{eq:gS_def}
\end{align}
such that they represent the number of relativistic degrees of freedom at temperature $T$ in either the symmetric or broken phase.  As shown in Appendix \ref{sub:genotgs}, one must have $g_{S}(T)\neq g_{E}(T)$ if entropy and energy are to be conserved during the time when a species adiabatically decouples.  
%Hence, we will approximate $g_{S}$ as
%\begin{equation}
%	g_{S}(T,v_T)\approx g_{E}(T,v_T)(1+K)\label{eq:approxgsgerel}
%\end{equation}
%where $0<K\ll1$ is a constant.

Now, we will begin our investigation of the quantities in \eref{eq:goodformula}.
First, consider the effect on the Hubble expansion rate $H(a)$ which is obtained by solving the Friedmann equation.  To do so, we partition the energy as described above and assume that $\rho_{\ex} \ll \rho_R(a_f)$ such that we can treat the CC energy density as a perturbation.  With these assumptions, we obtain
\begin{eqnarray}
%	H (x) & = & \frac{1}{\sqrt{3} M_p} \sqrt{\rho(T, v_T)} \label{eq:FriedmannEqn} \\
%	 & \approx & \frac{T^{2}}{3M_{p}}\sqrt{\frac{\pi^{2}}{10} \, g_{E}(T,v_T)}\left[1+\frac{1}{2}\frac{\rho_{\ex} \, \kappa(x)}{\frac{\pi^{2}}{30} \, g_{E}(T, v_T) \, T^{4}}\right] \label{eq:FriedmannEqn_approx}
	H (a) & = & \frac{1}{\sqrt{3} M_p} \sqrt{\rho(a)} \label{eq:FriedmannEqn} \\
	 & \approx & \frac{T^{2}}{3M_{p}}\sqrt{\frac{\pi^{2}}{10} \, g_{E}(a)}\left[1+\frac{1}{2}\frac{\rho_{\ex} \, \kappa(a)}{\frac{\pi^{2}}{30} \, g_{E}(a) \, T(a)^{4}}\right] \, .\label{eq:FriedmannEqn_approx}
\end{eqnarray}
where we have used the shorthand \eref{eq:Fofa}.  
During the PT, we can approximate $\kappa(a)$ as 
\begin{equation}\label{eq:fitfunctionalform}
%	\kappa(x)\approx\Theta(1+\delta-x)+\Theta(x-1-\delta) \left( 1-\frac{\Delta\rho_{\ex}}{\rho_{\ex}} \right) \kappa_{2} ( x\frac{a_{f}}{a_{PT}} )
	\kappa(a)\approx\Theta(a_{PT} - a)+\Theta(a - a_{PT}) \left( 1-\frac{\Delta\rho_{\ex}}{\rho_{\ex}} \right) \kappa_{2} ( a )
\end{equation}
where $\Theta(z)$ is a step function,
%, $\delta\equiv\frac{a_{PT}}{a_{f}}-1\ll1$,
$\Delta\rho_{\ex} >0$ is given by \eref{eq:L_def}, and $\kappa_{2}(a)$ is a function which starts from $\kappa_{2}(a_{PT})=1$ and decreases as fast as 
\begin{equation}\label{eq:ndef}
%	\left(x\frac{a_{f}}{a_{PT}}\right)^{-n_d}
	\left(\frac{a}{a_{PT}}\right)^{-n_d}
\end{equation}
with $n_d \gtrsim4$.  
%Note that due to energy conservation, any change in the value of $\kappa(a)$ that occurs with a negligible change in $a$ must be accompanied by a change in the temperature (i.e. entropy release).  
%For the rest of this analysis, we will adopt the form of \eref{eq:fitfunctionalform}.  
If $\Delta\rho_{\ex}=0$, we have a continuous second order transition
or a crossover.  If $\Delta\rho_{\ex}=\rho_{\ex}$, then the
supercooling is sufficiently strong as to end up with no CC energy
just after the PT.
%If $\delta=0$, then freeze out occurs at the same time as the PT.  
The step functions represent the fact that the PT occurs with negligible change in the scale factor.  
With this assumption, $\Delta s$ and the corresponding change in the temperature become functions of $\Delta\rho_{\ex}$ according to \eref{eq:Deltas_intermsof_L} in Appendix \ref{sub:deriveTofa}.  
Finally, the 
%$\Theta(1+\delta-x)$ 
$\Theta(a_{PT} - a)$ 
term in \eref{eq:fitfunctionalform} should, in general, be multiplied by another smooth function unless there is some symmetry fixing $v^{(s)}(T)$, and consequently $\rho_{cc}^{(s)}(T)$, to a particular value in the high energy limit.  
However, we will neglect this detail in favor of cleaner notation, since the final result will be approximately unchanged.

As discussed in Section \ref{sec:A-Brief-ReviewrofPT}, particle species 
start becoming non-relativistic after the (electroweak) PT 
which causes $g_{E/S}(a)$ to decrease.  We will parametrize this decrease 
by focusing on the (non-)adiabatic decoupling of ($N_{PT}$) $N$ 
fermionic dynamical degrees of freedom and write
\begin{align}
%	g_{E/S} \left( T, v_T \right) & = g_{E/S}(T_f) - h(x) \label{eq:gES_param} \\
%	g_{E/S}(T_f) & \equiv g_{E/S} \left( T_f , v_T^{sym} (T_f) \right) \label{eq:gf_def} \\
%	h (x) & = \frac{7}{8}N_{PT} \, \Theta(x-(1+\delta))+\frac{7}{8}N \, f (x) \label{eq:h_def}
	g_{E/S} (a) & = g_{E/S}(a_f) - h(a) \label{eq:gES_param} \\
%	g_{E/S}(a_f) & \equiv g_{E/S}^{(s)}(T_f) \label{eq:gf_def} \\
	h (a) & = \frac{7}{8}N_{PT} \, \Theta(a-a_{PT})+\frac{7}{8}N \, f (a) \label{eq:h_def}
\end{align}
where $f(a)$, which rises from $0$ to $1$, is given by \eref{eq:fxeqapproxexpl}.  
Note that in reality, $h(a)$ is a smooth complicated function (particularly
$Nf(a)$), but here we are accounting for the change in the number
of degrees of freedom in a physically suggestive approximation.
As we will see below, this effect will be one of the dominant {}``backgrounds''
to the {}``signal'' of measuring the effects of the cosmological
constant.  We treat this effect as a perturbation to linear order, and we estimate 
the Hubble expansion rate to be 
\begin{equation}\label{eq:becomingnonrel}
%	H\approx\frac{T^{2}}{3M_{p}}\sqrt{\frac{\pi^{2}}{10}g_E(T_f)}\left[1-\frac{1}{2} \frac{h(x)}{g_E(T_f)}+\frac{1}{2}\frac{\rho_{\ex}\, \kappa(x)}{\frac{\pi^{2}}{30}g_E(T_f)T^{4}}\right] \, .
	H(a)\approx\frac{T(a)^{2}}{3M_{p}}\sqrt{\frac{\pi^{2}}{10}g_E(a_f)}\left[1-\frac{1}{2} \frac{h(a)}{g_E(a_f)}+\frac{1}{2}\frac{\rho_{\ex}\, \kappa(a)}{\frac{\pi^{2}}{30}g_E(a_f)T(a)^{4}}\right] \, .
\end{equation}
Writing $T(a)$ using \eref{eq:Tofa} and linearizing further with respect to 
small quantities, we have
\begin{equation}\label{eq:H_linear}
%	H\approx H^{(U)}(x)\left[1+\frac{\epsilon_{1}}{2}x^{4} \, \kappa(x)+\frac{2}{3}\epsilon_{2} \Theta(x-(1+\delta))+\frac{\epsilon_{31}\Theta(x-(1+\delta))+\epsilon_{32}f\left(x\right)}{6}\right]
	H(a)\approx H^{(U)}(a)\left[1+\frac{\epsilon_{1}}{2} \left( \frac{a}{a_f} \right)^{4} \, \kappa(a)+\frac{2}{3}\epsilon_{2} \Theta(a - a_{PT})+\frac{1}{6} \epsilon_{31}\Theta(a-a_{PT})+ \frac{1}{6} \epsilon_{32}f\left(a\right) \right]
\end{equation}
where
\begin{equation}
%	H^{(U)}(x)\equiv\frac{T_{f}^{2}}{3 \, M_{p} \, x^{2}}\sqrt{\frac{\pi^{2}}{10} \, g_{E}(T_{f})}
	H^{(U)}(a)\equiv\frac{T_{f}^{2}}{3 \, M_{p} \, \left( \frac{a}{a_f} \right)^{2}}\sqrt{\frac{\pi^{2}}{10} \, g_{E}(a_{f})}
\end{equation}
and
\begin{subequations}\label{eq:epsiloni_def}
\begin{align}
%	\epsilon_{1} & \equiv \frac{\rho_{\ex}}{\frac{\pi^2}{30} g_{E}(T_{f}) \, T_{f}^{4}}=\mbox{fractional energy of the exotic during freeze out} \label{eq:fractionalenergy} \\
	\epsilon_{1} & \equiv \frac{\rho_{\ex}}{\frac{\pi^2}{30} g_{E}(a_{f}) \, T_{f}^{4}}=\mbox{fractional energy of the exotic during freeze out} \label{eq:fractionalenergy} \\
%	\epsilon_{2} & \equiv \left(\frac{a_{PT}}{a_{f}}\right)^{3}\frac{\Delta s}{\frac{2 \pi^2}{45} g_{S}(T_{f}) \, T_{f}^{3}}=\mbox{fractional entropy increase during PT}\label{eq:fractionalentropyduringpt} \\
	\epsilon_{2} & \equiv \left(\frac{a_{PT}}{a_{f}}\right)^{3}\frac{\Delta s}{\frac{2 \pi^2}{45} g_{S}(a_{f}) \, T_{f}^{3}}=\mbox{fractional entropy increase during PT}\label{eq:fractionalentropyduringpt} \\
%	\epsilon_{31} & \equiv\frac{\frac{7}{8}N_{PT}}{g_{E}(T_{f})}=\mbox{ fractional decoupling degrees of freedom during PT} \label{eq:fractionaldecouplingduringPT} \\
	\epsilon_{31} & \equiv\frac{\frac{7}{8}N_{PT}}{g_{E}(a_{f})}=\mbox{ fractional decoupling degrees of freedom during PT} \label{eq:fractionaldecouplingduringPT} \\
%	\epsilon_{32}& \equiv\frac{\frac{7}{8}N}{g_{E}(T_{f})}=\mbox{ fractional decoupling degrees of freedom near freeze out} \label{eq:fractionaldecoupling}
	\epsilon_{32}& \equiv\frac{\frac{7}{8}N}{g_{E}(a_{f})}=\mbox{ fractional decoupling degrees of freedom near freeze out} \label{eq:fractionaldecoupling}
\end{align}
\end{subequations}
where $\Delta s$, denoting the entropy density change at the time of
the PT, is given by \eref{eq:Deltas_intermsof_L}.  Although $H(a)$
appears to vary discontinuously at $a = a_{PT}$, its continuity is
ensured by the conservation of energy.  At the PT,
%The fact that there is no discontinuity in $H$ even though a step function appears and $T$ can change discontinuously when entropy is changed (as seen in \eref{eq:Tofa}) is due to the conservation of energy: 
the CC energy converts into radiation, which generates an entropy but
leaves the total energy density fixed (i.e., $\epsilon_{2}$
compensates for the discontinuity of the $\epsilon_{1}$ term) because
the volume remains approximately constant through the duration of the
PT.  The fact that $H$ is boosted by $\epsilon_{31}$ and
$\epsilon_{32}$ is intuitive for the following reason.  When a
particle species becomes non-relativistic, the effective equation of
state becomes smaller, such that the energy dilutes less, which in
turn leads to a larger expansion rate for the same scale factor.  The
term $\epsilon_{31}$ accounts for the non-adiabatic change in the
number of degrees of freedom during the PT, while the term
$\epsilon_{32}$ accounts for the adiabatic change in the number of
degrees of freedom.

Next, consider the change in the cross section due to the PT.  
We parametrize this effect as 
\begin{equation}
	\sv =\sv^{(U)}\Bigl(1-\epsilon_{4}\, \Theta(a-a_{PT})\Bigr)
\label{eq:crosssecchange}\end{equation}
where
\begin{equation}
	\epsilon_{4}\equiv-\frac{\Delta_{\sigma}}{\langle\sigma v\rangle^{(U)}}
\label{eq:fractionalcrosssecchange}\end{equation}
and $\Delta_{\sigma}$ is the change in $\langle\sigma v\rangle$
due to the PT.  Since the derivation of Eq.~(\ref{eq:goodformula}) assumes 
that the dark matter is decoupled after $T_f$, we will assume that 
$\epsilon_{4}\gtrsim0$ in order to prevent re-thermalization due to an 
increase in the cross section.  Hence,
we can evaluate Eq.~(\ref{eq:goodformula}) by linearizing in the $\epsilon$'s 
to obtain
\begin{equation}\label{eq:almostthere}
%	n_{X}(t_{0})\approx\left(\frac{a_{f}}{a_{0}}\right)^{3}\left(\int_{0}^{\ln a_{0}/a_{f}}\frac{d\ln x}{H^{(U)}(x)}\frac{\langle\sigma v\rangle^{(U)}}{x^{3}}\left[1+\sum_{n}\theta_{n}(x)\epsilon_{n}\right]\right)^{-1}
	n_{X}(t_{0})\approx\left(\frac{a_{f}}{a_{0}}\right)^{3}\left(\int_{0}^{\ln a_{0}/a_{f}}\frac{d\ln a / a_f}{H^{(U)}(a)}\frac{\sv^{(U)}}{(a/a_f)^{3}}\left[1+\sum_{n}\theta_{n}(a)\epsilon_{n}\right]\right)^{-1}
\end{equation}
where
\begin{align}
%	\sum_{n}\theta_{n}(x)\, \epsilon_{n}= & -\frac{\epsilon_{1}}{2}x^{4}\kappa(x)-\Theta(a_{f}x-a_{PT})\frac{2}{3}\epsilon_{2} \nonumber \\
%	&-\frac{1}{6} \left[\epsilon_{31}\, \Theta(x-(1+\delta))+\epsilon_{32}\, f\left(x\right) \right]-\epsilon_{4}\, \Theta(a_{f}x-a_{PT})
	\sum_{n}\theta_{n}(a)\, \epsilon_{n}= & -\frac{\epsilon_{1}}{2} \left( \frac{a}{a_f} \right)^{4}\kappa(a)-\Theta(a-a_{PT})\frac{2}{3}\epsilon_{2} \nonumber \\
	&-\frac{1}{6} \left[\epsilon_{31}\, \Theta(a-a_{PT})+\epsilon_{32}\, f\left(a\right) \right]-\epsilon_{4}\, \Theta(a-a_{PT})
\end{align}
implicitly defines the $\theta_n$.  
Note that the integral is dominated by contributions around $\ln a/a_f=0$.
On the other hand, the $(a_{f}/a_{0})^{3}$ prefactor should be evaluated with all the $g_{S}$ changes accounted for, not just the effects around $\ln x=0$.

Next, let's consider the effects on the $a_{f}/a_{0}$ factor determined
by the freeze out condition itself. The freeze out temperature $T_{f}$
can be solved using \cite{Kolb:1979bt}
\begin{align}
	\sv  n_{X}^{\mathrm{eq}}(T_{f})&=C \frac{m_{X}}{T_{f}}H(T_{f}) \label{eq:Tf_def} \\
	n_X^{\mathrm{eq}} (T)& \equiv g_X \left(\frac{m_{X}T}{2\pi}\right)^{3/2}\exp\left(-\frac{m_{X}}{T}\right) \label{eq:nXeq_def}
\end{align}
where $C$ is an order unity number whose optimum value to reproduce
numerical integration is cross section dependent (e.g., $C\approx2$), 
$g_X$ counts the real dynamical degrees of freedom of the dark
matter, and $m_X$ is the dark matter mass. 
Evaluating $H(T_f)$ with \eref{eq:H_linear} and assuming freeze out 
occurs before the PT, 
%($\delta >0$), 
\eref{eq:Tf_def} becomes
\begin{equation}
%	\sv g_X\left(\frac{m_{X}T_{f}}{2\pi}\right)^{3/2}\exp\left(-\frac{m_{X}}{T_{f}}\right)\approx\frac{C\, m_{X}\, T_{f}}{3M_{p}}\sqrt{\frac{\pi^{2}}{10}g_{E}(T_{f})}\left[1+\frac{\epsilon_{1}}{2}\right] \, .
	\sv g_X\left(\frac{m_{X}T_{f}}{2\pi}\right)^{3/2}\exp\left(-\frac{m_{X}}{T_{f}}\right)\approx\frac{C\, m_{X}\, T_{f}}{3M_{p}}\sqrt{\frac{\pi^{2}}{10}g_{E}(a_{f})}\Bigl[1+\frac{\epsilon_{1}}{2}\Bigr] \, .
\label{eq:freezeoutcondexpl}\end{equation}
Although
not solvable in closed form, one can linearize in the perturbation
again to obtain
\begin{equation}
	T_{f}\approx\frac{m_{X}}{\ln A}\left[1+\frac{\epsilon_{1}}{2}\left(\frac{1}{\ln A}+\ord{(\ln A)^{-2}\right)}\right]
\label{eq:freezeouttemp}\end{equation}
where 
\begin{equation}\label{eq:approxexpressforA}
%	A\equiv\frac{g_X3\sqrt{5}M_{p}\sqrt{m_{X}T_{f}}\langle\sigma v\rangle}{2C\pi^{5/2}\sqrt{g_{E}(T_{f})}}\sim\exp[20]
	A\equiv\frac{g_X3\sqrt{5}M_{p}\sqrt{m_{X}T_{f}}\langle\sigma v\rangle}{2C\pi^{5/2}\sqrt{g_{E}(a_{f})}}\sim\exp[20]
\end{equation}
for electroweak mass scales.  If we assume that there is only one period
of entropy production between freeze out and today, and that it occurs at 
the PT temperature $T_{PT}$, we can use entropy conservation in the 
form of \eref{eq:EntropyConsv} to write 
\begin{equation}
%	\frac{a_{0}}{a_{f}}=\left(\frac{g_{S}(T_{f})}{g_{S}(T_{0})}\right)^{1/3}\frac{T_{f}}{T_{0}}\left[1+\frac{1}{3}\epsilon_{2}\right] 
	\frac{a_{0}}{a_{f}}=\left(\frac{g_{S}(a_{f})}{g_{S}(a_{0})}\right)^{1/3}\frac{T_{f}}{T_{0}}\left[1+\frac{1}{3}\epsilon_{2}\right] 
\end{equation}
where 
%$g_S(T_0) = g_S(T_0, v_T^{brk}(T_0))$ and 
$T_0$ is the temperature 
today.  
Combining this with Eq.~(\ref{eq:freezeouttemp}), we find
\begin{equation}
%	\frac{a_{0}}{a_{f}}= \left. \frac{a_0}{a_f} \right|_{\mathrm{usual}}  \times 
	\frac{a_{0}}{a_{f}}= \left( \left. \frac{a_0}{a_f} \right|^{(U)} \right)
	\left[1+\frac{\epsilon_{1}}{2}\frac{1}{\ln A}+\frac{1}{3}\epsilon_{2}\right]
\label{eq:a0overaffin}\end{equation}
where 
\begin{equation}
%	\left. \frac{a_{0}}{a_{f}} \right|_{\mathrm{usual}}\equiv\left(\frac{g_{S}(T_{f})}{g_{S}(T_{0})}\right)^{1/3}\frac{m_{X}}{T_{0}}\frac{1}{\ln A} \, .
%	\left. \frac{a_{0}}{a_{f}} \right|_{\mathrm{usual}}\equiv\left(\frac{g_{S}(a_{f})}{g_{S}(a_{0})}\right)^{1/3}\frac{m_{X}}{T_{0}}\frac{1}{\ln A} \, .
	\left( \left. \frac{a_0}{a_f} \right|^{(U)} \right) \equiv\left(\frac{g_{S}(a_{f})}{g_{S}(a_{0})}\right)^{1/3}\frac{m_{X}}{T_{0}}\frac{1}{\ln A} \, .
\end{equation}
Putting \eref{eq:a0overaffin} into Eq.~(\ref{eq:almostthere}) results in
\begin{equation}\label{eq:voldilprefactornotaccountedyet}
%	n_{X}(t_{0})\approx \left(\frac{a_{f}}{a_{0}}\right)_{\mathrm{usual}}^{3}\left[1-\frac{3\epsilon_{1}}{2}\frac{1}{\ln A}-\epsilon_{2}\right] \left(E_{1}+\int_{0}^{\ln a_{0}/a_{f}|_{\mathrm{usual}}}\frac{d\ln x}{H^{(U)}(x)}\frac{\langle\sigma v\rangle^{(U)}}{x^{3}}\left[1+\sum_{n}\theta_{n}(x)\epsilon_{n}\right]\right)^{-1}
%	n_{X}(t_{0})\approx \left(\frac{a_{f}}{a_{0}}\right)_{\mathrm{usual}}^{3}\left[1-\frac{3\epsilon_{1}}{2}\frac{1}{\ln A}-\epsilon_{2}\right] \left(E_{1}+\int_{0}^{\ln a_{0}/a_{f}|_{\mathrm{usual}}}\frac{d\ln a/a_f}{H^{(U)}(a)}\frac{\sv^{(U)}}{(a/a_f)^{3}}\left[1+\sum_{n}\theta_{n}(a)\epsilon_{n}\right]\right)^{-1}
	n_{X}(t_{0})\approx \left( \left. \frac{a_0}{a_f} \right|^{(U)} \right)^{-3} 
	\left[1-\frac{3\epsilon_{1}}{2}\frac{1}{\ln A}-\epsilon_{2}\right] 
	\left(E_{1}+
	\int_{0}^{\ln (a_{0}/a_{f} |^{(U)}) }
	\frac{d\ln a/a_f}{H^{(U)}(a)}\frac{\sv^{(U)}}{(a/a_f)^{3}}
	\left[1+\sum_{n}\theta_{n}(a)\epsilon_{n}\right]
	\right)^{-1}
\end{equation}
where the endpoint contribution to the integral has been written as
\begin{equation}
	E_{1}\equiv\frac{\frac{\epsilon_{1}}{2}\frac{1}{\ln A}+\frac{\epsilon_{2}}{3}}{H^{(U)}(a_{0})}
	\frac{\langle\sigma v\rangle^{(U)}}{(a_{0}/a_{f} |^{(U)})^{3}} \, .
\end{equation}
The term $E_{1}$ is negligible because of the volume dilution factor in its denominator.
Linearizing the small factors gives
\begin{eqnarray}
	n_{X}(t_{0}) & \approx & n_{X}^{(U)}(t_{0})\left[1-\frac{3\epsilon_{1}}{2}\frac{1}{\ln A}-\epsilon_{2}-F_{u}^{-1}\sum_{n}\tilde{\theta}_{n}\epsilon_{n}\right]
\label{eq:nxintermedsimpl}\end{eqnarray}
where
\begin{equation}
%	F_{u}\equiv\int_{0}^{\ln a_{0}/a_{f}|_{\mathrm{usual}}}\frac{d\ln x}{H^{(U)}(x)}\frac{\langle\sigma v\rangle^{(U)}}{x^{3}}
	F_{u}\equiv\int_{0}^{\ln (a_{0}/a_{f} |^{(U)}) }\frac{d\ln a/a_f}{H^{(U)}(a)}\frac{\sv^{(U)}}{(a/a_f)^{3}}
\end{equation}
\begin{equation}
	n_{X}^{(U)}(t_{0})\equiv\left(\left. \frac{a_{0}}{a_{f}} \right|^{(U)}\right)^{-3}F_{u}^{-1}=\text{usual computation of relic abundance}
\end{equation}
\begin{equation}
%	\tilde{\theta}_{n}\equiv\int_{0}^{\ln a_{0}/a_{f}|_{\mathrm{usual}}}\frac{d\ln x}{H^{(U)}(x)}\frac{\langle\sigma v\rangle^{(U)}}{x^{3}}\theta_{n}(x) \, .
	\tilde{\theta}_{n}\equiv\int_{0}^{\ln (a_{0}/a_{f} |^{(U)}) }\frac{d\ln a/a_f}{H^{(U)}(a)}\frac{\sv^{(U)}}{(a/a_f)^{3}}\theta_{n}(a) \, .
\end{equation}
%Note that $\left| F_{u}^{-1}\tilde{\theta}_{n} \right|= \mathcal{O}(1)$ except for $n=1$, the most important feature of our interest.  [IS THIS TRUE?]
In particular, if we assume an $s$-wave cross section (i.e., constant $\sv$), we can express $\tilde{\theta}_{1}$ explicitly as
\begin{equation}
%	\left\{ \tilde{\theta}_{1}\approx -\frac{\langle\sigma v\rangle^{(U)}}{2H^{(U)}(1)}\left[\delta+\frac{\left(1+3\delta\right)}{n-3}\left(1-\frac{\Delta\rho_{\ex}}{\rho_{\ex}}\right)\right],\,\,\,\,\,\, F_{u}^{-1}\tilde{\theta}_{1}\approx-\frac{1}{2}\left[\delta+\frac{\left(1+3\delta\right)}{n-3}\left(1-\frac{\Delta\rho_{\ex}}{\rho_{\ex}}\right)\right] \right\}
%	\left\{ \tilde{\theta}_{1}\approx -\frac{\langle\sigma v\rangle^{(U)}}{2H^{(U)}(a_f)}\left[\delta+\frac{\left(1+3\delta\right)}{n_d-3}\left(1-\frac{\Delta\rho_{\ex}}{\rho_{\ex}}\right)\right],\,\,\,\,\,\, F_{u}^{-1}\tilde{\theta}_{1}\approx-\frac{1}{2}\left[\delta+\frac{\left(1+3\delta\right)}{n_d-3}\left(1-\frac{\Delta\rho_{\ex}}{\rho_{\ex}}\right)\right] \right\}
	 F_{u}^{-1}\tilde{\theta}_{1}\approx-\frac{1}{2}\left[\delta+\frac{\left(1+3\delta\right)}{n_d-3}\left(1-\frac{\Delta\rho_{\ex}}{\rho_{\ex}}\right)\right]
\end{equation}
where we have expanded in $\delta \equiv a_{PT} / a_f - 1 \gtrsim 0$
which represents the delay between freeze out and the PT.  The first
term in square brackets comes from integrating the CC energy density
from $a_f$ to $a_{PT}$, and the second term comes from integrating the
decreasing CC energy density after the PT.  This equation shows that
if $\frac{\Delta\rho_{\ex}}{\rho_{\ex}}\approx1$ (large supercooling)
there is a suppression of the $\epsilon_{1}$ effect by a factor of
order $\delta$.
%Hence, we cannot always neglect the term suppressed by $\ln A$ in Eq.~(\ref{eq:nxintermedsimpl}) depending on the size of $\delta$.  
Although we have linearized in $\delta$ along with $\epsilon_i$, terms
of the form $\epsilon_i \, \delta$ are not higher order.  The
expansion in $\epsilon_i$ reflects the fact that we treat the PT as a
perturbation, whereas the expansion in $\delta$ is performed merely to
simplify the expressions.  With the same assumptions, we can evaluate
the other $F_{u}^{-1}\tilde{\theta}_{n}$ terms:
\begin{eqnarray}
	F_{u}^{-1}\tilde{\theta}_{2}& \approx & -\frac{2}{3}\left( 1- \delta \right)\\
	F_{u}^{-1}\tilde{\theta}_{31}& \approx & -\frac{1}{6}(1-\delta) \\ 
	F_{u}^{-1}\tilde{\theta}_{32}& \approx & -\frac{1}{6}\int_{0}^{\ln (a_{0}/a_{f}|^{(U)})}\frac{d \ln a/ a_f}{(a/a_f)^{2}}f (a)  \\
	F_{u}^{-1}\tilde{\theta}_{4}& \approx & -\left( 1- \delta \right) .
\end{eqnarray}
Hence, for $s$-wave cross sections,
the change in the relic abundance due to small changes made by the
PT can be expressed as
\begin{equation}\label{eq:finalfractionalchange}
	\delta n_X(t_0)  = c_1 \, \epsilon_1 + c_2 \, \epsilon_2 + c_{31} \, \epsilon_{31} + c_{32} \, \epsilon_{32} + c_4 \, \epsilon_4
\end{equation}
where
\begin{subequations}
\begin{align}
	c_1 & \equiv \frac{1}{2}\left(\delta+\frac{\left(1+3\delta\right)}{n-3}\left(1-\frac{\Delta\rho_{\ex}}{\rho_{\ex}}\right)\right)-\frac{3}{2}\frac{1}{\ln A} \label{eq:c1_def}  \\
	c_2 & \equiv -\frac{1}{3}(1+2\delta) \label{eq:c2_def} \\
	c_{31} & \equiv \frac{1}{6}(1-\delta) \label{eq:c31_def} \\
%	c_{32} & \equiv \frac{1}{6}\int_{1}^{a_{0}/a_{f}|_{\mathrm{usual}}}\frac{dx}{x^{2}}f\left(x\right) \label{eq:c32_def} \\
	c_{32} & \equiv \frac{1}{6}\int_{0}^{\ln (a_{0}/a_{f}|^{(U)})}\frac{d \ln a / a_f}{(a/a_f)^{2}}f\left(a\right) \label{eq:c32_def} \\
	c_4 & \equiv 1-\delta \label{eq:c4_def} \, .
\end{align}
\end{subequations}
%and where $\epsilon_{i}$ are given by Eqs.~(\ref{eq:epsiloni_def}) and (\ref{eq:fractionalcrosssecchange}); $\delta=a_{PT}/a_{f}-1\approx\frac{T_{f}}{T_{PT}}-1$ represents the delay between freeze out and the phase transition; $\Delta\rho_{\ex}$ is the energy released during the phase transition to produce entropy; $\ln A=\frac{m_{X}}{T_{f}}\sim20$ is approximately given by Eq.~(\ref{eq:approxexpressforA}); and $n$ parameterizes the power law with which the {\color{red} vacuum} energy decays after the phase transition. 
The key point of \eref{eq:finalfractionalchange} is that despite the ``background'' represented by $\epsilon_{n\neq1}$, the ``signal'' contained in $\epsilon_{1}$ can be {}``measured'' and represents a prediction of the hypothesis of a tuned CC. 
It is a tuned but striking statement, nonetheless.  
Since this term is central to the rest of our calculation, we have reproduced the so called ``CC effect'' term here as 
\begin{align}\label{eq:c1e1}
%	c_1 \, \epsilon_1 = \left( \frac{\rho_{\rm ex}}{\frac{\pi^2}{30} g_E(T_f) T_{PT}^4} \right)
	c_1 \, \epsilon_1 = \left( \frac{\rho_{\rm ex}}{\frac{\pi^2}{30} g_E(a_f) T_{PT}^4} \right)
	\frac{1}{(1+\delta)^4} \left\{
	\frac{1}{2} \left[ 
	\frac{(1+\delta)^3-1}{3} + \frac{(1+\delta)^3}{n_d-3} \left(1 - \frac{\Delta \rho_{\rm ex}}{\rho_{\rm ex}} \right) \right] - \frac{3}{2} \frac{1}{\ln A} \right\}
\end{align}
%by writing $T_{f}  = T_{PT} ( 1 + \delta)$ and 
without linearizing in $\delta$.  
We also write the so called ``entropy effect'' as
\begin{align}\label{eq:c2e2}
	c_2 \, \epsilon_2 = - \frac{\delta + \frac{1}{3}}{\delta + 1} \frac{\Delta s}{\frac{2 \pi^2}{45} g_{S}(a_{PT}) \, T_{PT}^{3}} \, .
\end{align}
Note finally that we can obtain a smooth non-first order PT by taking
the limit $\Delta\rho_{\ex}=\epsilon_{2}=\epsilon_{31}=\epsilon_{4} =
0$.

One should remember that all of the analysis has assumed that the entropy released from the PT (in the case of a first order PT) did not reheat the system to the point that the dark matter rethermalized after freeze out, i.e., $T_{PT}^+ \lesssim T_f$. 
This provides a lower bound on $\delta$ for a given $\Delta\rho_{\ex}$, which can be expressed as
\begin{equation}\label{eq:delta_bound}
%	\frac{1}{4} \epsilon_{31} + \frac{1}{4} \frac{\Delta \rho_{\rm ex}}{\frac{\pi^2}{30} \, g_E(T_f) \, (T_{PT}^-)^4} \lesssim \delta
	\frac{1}{4} \epsilon_{31} + \frac{1}{4} \frac{\Delta \rho_{\rm ex}}{\frac{\pi^2}{30} \, g_E(a_f) \, (T_{PT}^-)^4} \lesssim \delta
\end{equation}
by using \eref{eq:TPTplus} and assuming that 
%$f(\frac{a_{PT}}{a_{f}})$ 
$f(a_{PT})$ is negligible.

Note also that the range and independence of $\{\epsilon_{i},\delta\}$
that is achievable by choosing a beyond the SM Lagrangian is not easy
to compute nor to generalize.  For example, suppose we want to increase
$\delta$ while keeping $\epsilon_{1}$ fixed.  To increase $\delta$, we
increase $a_{PT}$ more than $a_{f}$.  Since $a_{f}$ is mostly
determined by the mass of the dark matter $m_{X}$ while $a_{PT}$ is
determined in part by the competition between the thermal mass support
and scalar field mass at the field origin, we can keep $a_{f}$ fixed
and increase $a_{PT}$ by decreasing the scalar field mass competing
with the thermal support.  This, however, typically changes the
fractional entropy increase $\epsilon_{2}$ during the PT.
Furthermore, this will change the index $n_d$ (defined in
\eref{eq:ndef}) which depends partly on the flatness of the
non-thermal part of the scalar potential.  Indeed, we see that if this
$n_d$ can be engineered to be as close to $3$ as possible (i.e. a flat
potential with no thermal particles decoupling), then the
$\epsilon_{1}$ signal can be enhanced.  One also sees that in the case
of a first order PT, the prediction for the effect of the cosmological
constant (i.e., the $\epsilon_{1}$ piece) depends on
$\Delta\rho_{\ex}$ and $\delta$, both of which depend on knowing
exactly when the PT occurs.  As described in
Sec.~\ref{sec:A-Brief-ReviewrofPT}, an accurate computation of this
will require a non-perturbative numerical treatment.  Hence, the first
order PT situation, which can give a larger CC dependent signal,
presents an interesting computational challenge of its own.

% sec:IllustrativeModels
\section{Illustrative Models}\label{sec:IllustrativeModels}

In this section, we present numerical calculations of $\delta
n_X(t_0)$ for various models.  This section represents one of the key
features of the paper that distinguish it from \cite{Chung:2011hv}, as
discussed in the introduction.  For each model we specify the
parameters of the scalar sector, which appear in the thermal effective
potential $V_{\rm eff} (\phi_{c},T)$, and the parameters of the dark
matter sector, $m_X$, $g_X$, and $\sv$.  We then calculate the relic
abundance shift using the methods of Section \ref{sec:AnalyticEst}.
Most of the numerical results have not been reported previously, and
the model dependent analysis of a real singlet extension of the
standard model is entirely new.

% sec:StandardModel
\subsection{Standard Model with Dark Matter}\label{sec:StandardModel}

\begin{figure}[t]
\begin{center}
\includegraphics[width=0.60\textwidth]{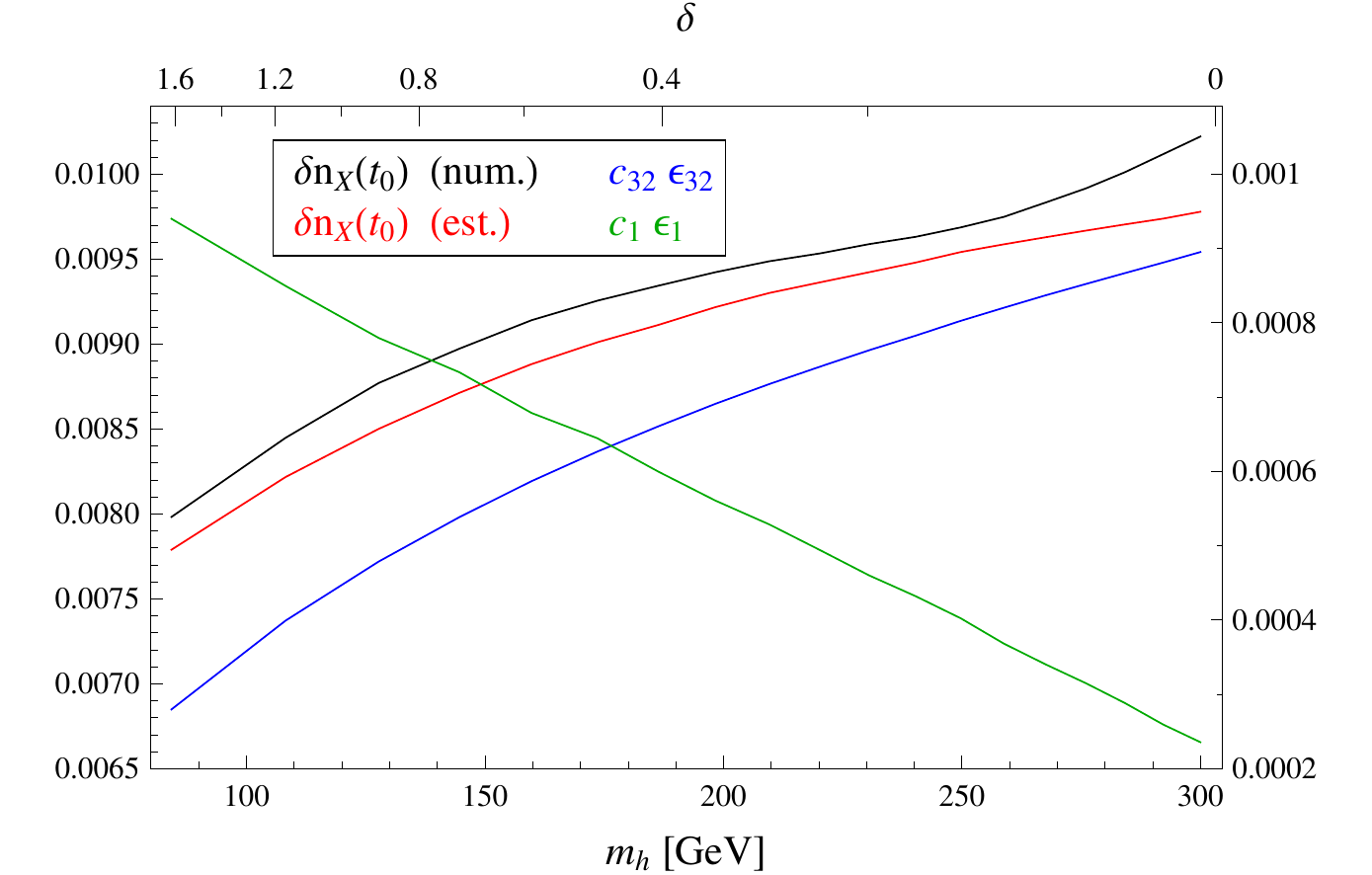}
\caption{\label{fig:versus_mh}  Fractional deviation of
  the relic abundance due to the SM electroweak PT.  The
  numerical calculation is represented by the black curve, the
  analytic estimate \eref{eq:finalfractionalchange} by the red
  curve, the CC effect ($c_1 \epsilon_1$ term) by the green
  curve, and the adiabatic decoupling effect ($c_{32}
  \epsilon_{32}$ term) by the blue curve.  The right axis
  shows the values of the $c_1 \epsilon_1$ curve only, and the left
  axis shows the values of the three other curves.  }
\end{center}
\end{figure}

\begin{figure}[t]
\begin{center}
\includegraphics[width=0.60\textwidth]{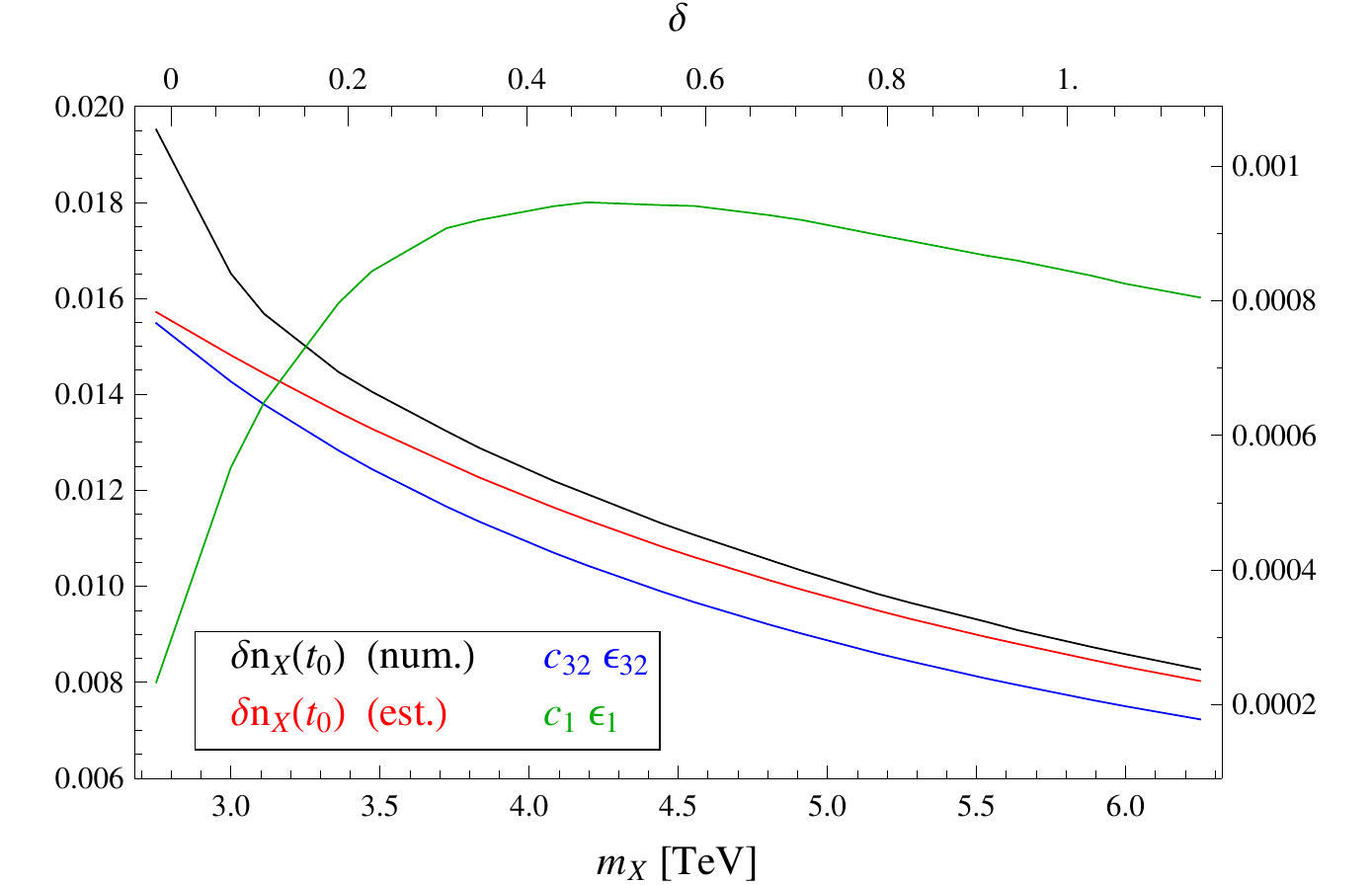}
\caption{\label{fig:versus_mX}  Fractional deviation of
  the relic abundance due to the SM electroweak phase transition,
  plotted against the WIMP mass and $\delta$.  The colors and axes are
  the same as in Figure \ref{fig:versus_mh} }
\end{center}
\end{figure}

We calculate here the relic abundance deviation due to the SM
electroweak PT.  The qualitative results were already given in
\cite{Chung:2011hv}.  The numerical details that we discuss in this
section can be summarized as $\delta n_X(t_0) = \ord{10^{-3}-10^{-2}}$
with the CC contributing $c_1 \epsilon_1 = \ord{10^{-4} - 10^{-3}}$.
With $m_h = 115 \GeV$, the largest CC effect occurs for $m_X \approx
4.2$ TeV where $c_1 \epsilon_1 \approx 9.5 \times 10^{-4}$.  Our
results are summarized in Figures \ref{fig:versus_mh} and
\ref{fig:versus_mX}.  In this section, we first discuss these figures
and then extend the analytic estimate of Section
\ref{sec:AnalyticEst}, now in the context of a concrete model, to
obtain \eref{eq:dnXt0_est2}, which lets us motivate extensions of the
SM that achieve larger $\delta n_X(t_0)$.  Some of the qualitative
discussion of \cite{Chung:2011hv} is reproduced for completeness.

In Appendix \ref{app:EffPot} we compute the SM thermal effective
potential $V_{\rm eff}(h_c,T)$ through one-loop order\footnote{It is
  well known that the one-loop approximation breaks down at the
  temperature of the SM electroweak PT \cite{Arnold:1992rz}, and that
  accurate results require lattice calculations \cite{Kajantie:1996mn,
    Kajantie:1996qd, Kajantie:1995kf}.  However, since the CC
  contribution already represents perturbative correction to dark
  matter freeze out, we will neglect higher-order corrections to the
  PT physics and simply apply the mean field approximation described
  in Section \ref{sec:A-Brief-ReviewrofPT}.  }, where $h(x) = \sqrt{2} \left|
H^{\dagger} H \right|^{1/2}$ is the radial component of the Higgs
field and $h_c = \left< h(x) \right>$.  It is important to point out
that the renormalization conditions, given by \eref{eq:SM_rencondit},
are chosen such that $V_{\rm eff}(h_c,0)$ has a minimum at $v = 246
\GeV$ where the curvature is $m_h^2$ and, most importantly the CC is
tuned by requiring $V_{\rm eff} (v,0) =0$.

% fig:SMrho
\begin{figure}[t]
\begin{center}
\includegraphics[width=0.6\textwidth]{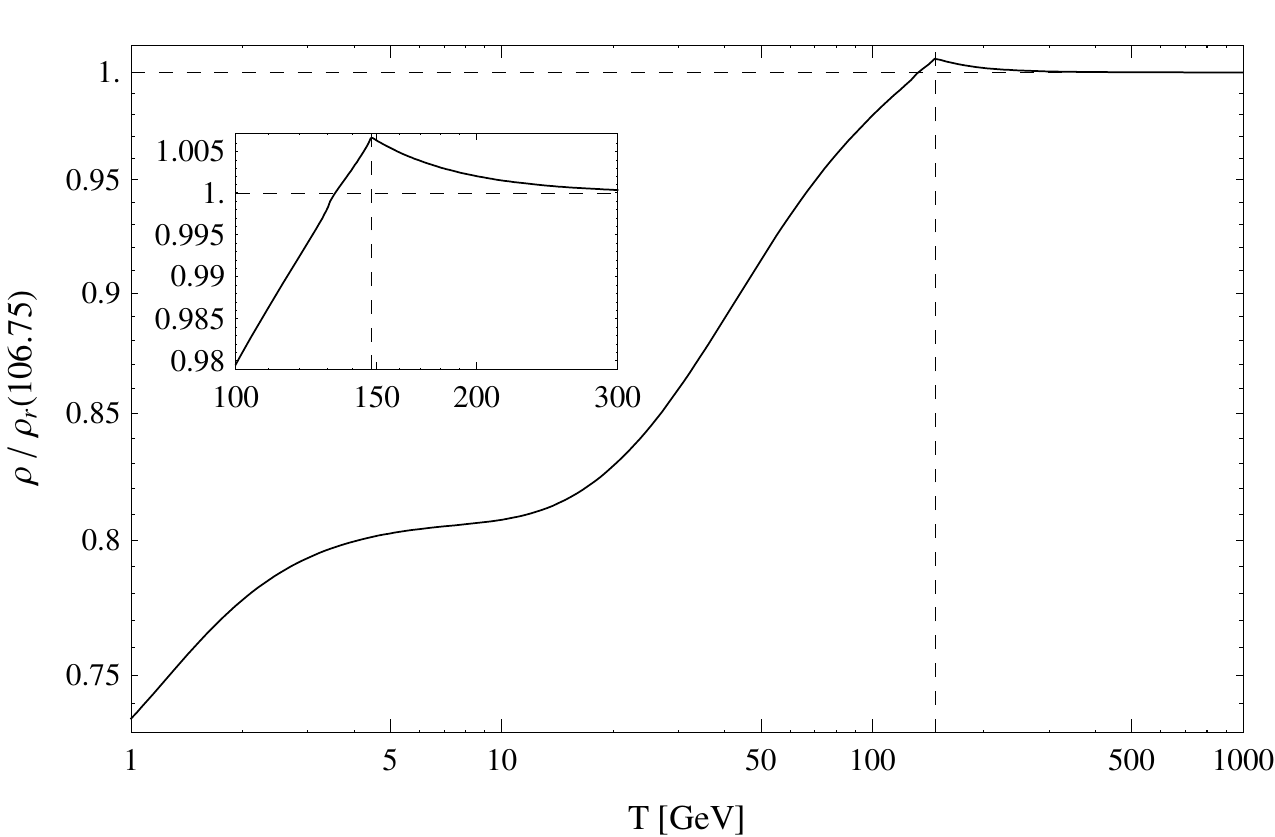}
\caption{\label{fig:SMrho} The energy density at the SM PT, $\rho(T) =
  \rho_{cc}(T) + \rho_R(T)$, relative to the energy density when the
  entire SM is relativistic, $\rho_r(106.75) \equiv \frac{\pi^2}{30}
  106.75 \, T^4$.  Just before the PT at $T \gtrsim 150 \GeV$, the
  energy density grows relative to $\rho_r$ due to the temperature
  independent CC contribution, $\rho_{cc}(T > T_{PT}) \approx $ const.
  After the PT, the top, bottom, Higgs, and massive gauge boson
  adiabatically decouple causing $\rho / \rho_r$ to drop below one.
  This adiabatic decoupling is the dominant feature of the SM PT that
  is relevant for freeze out.  }
\end{center}
\end{figure}

Before discussing the numerical results, it is useful to recall from
Section \ref{sec:AnalyticEst} that for a non-first order PT, freeze
out is only affected by modifications to the relations $H(T) \propto
\sqrt{\rho(T)} \propto T^{2}$ and $T^3 \propto g_S^{-1} \, a^{-3} \sim
a^{-3}$.  These modifications arise when the energy partitioning
deviates from radiation domination and the number of relativistic
degrees of freedom deviates from a constant value.  These deviations
can be visualized in Figure \ref{fig:SMrho}, where we plot $\rho(T)$
normalized by $\rho_r(106.75) \equiv (\pi^2/30) (106.75) \, T^4$, the
energy density of the SM as if all particles were relativistic.  We
have taken $m_h = 115 \GeV$ which gives a PT at $T_{PT} \approx 148
\GeV$.  As the temperature decreases toward $T_{PT}$ from above, $\rho
/ \rho_r$ grows to approximately $1.006$ due to the presence of the
additional CC energy density (i.e., $\lambda v^4 / (106.75 \,
T_{PT}^4) \approx 10^{-3}$).  Below $T_{PT}$ the massive species
decouple, the plasma loses about twenty relativistic degrees of
freedom, and $\rho / \rho_r$ decreases to approximately $0.8$.  This
figure illustrates that the adiabatic decoupling has an effect on
$\rho$ which is two orders of magnitude larger than that from the CC.
Therefore, we expect that the Standard Model electroweak effective
CC will have a subdominant effect on the relic
abundance as well.

% fig:versus_mh

% fig:versus_mX

The fractional shift $\delta n_X(t_0)$ is calculated using the perturbative, analytic expressions in Section \ref{sec:AnalyticEst} as well as by solving the Boltzmann equation numerically.  
In Figure \ref{fig:versus_mh} we have plotted $\delta n_X(t_0)$ by varying $m_h$ and fixing $m_X = 6 \mathrm{\ TeV}$, $g_X = 2$, and $\sv = 2.33 \times 10^{-39} \, \mathrm{cm}^2$.
%$\sv = 2.33$ fb.  
%$\sv = 6 \times 10^{-12} \GeV^{-2}$.  
As seen in the figure, the PT causes an $\ord{10^{-3}-10^{-2}}$
fractional increase in the relic abundance.  We have chosen the DM
mass to be $6 \mathrm{\ TeV}$ such that freeze out and the PT
coincide at $T \approx 303 \GeV$ for $m_h = 300 \GeV$.  For smaller
$m_h$, the PT is delayed with respect to freezeout.  The
analytic estimate, given by \eref{eq:finalfractionalchange}, only
receives contributions from the CC effect ($c_1 \epsilon_1$ term) and
the adiabatic decoupling effect ($c_{32} \epsilon_{32}$ term), because
the PT is not first order.  As we anticipated in the
discussion of the preceeding paragraph, the $\epsilon_{32}$ term
dominates.  The analytic formula consistently underestimates the
numerical calculation by $2 - 3 \%$, and moreover, in the large $m_h$
limit where $\delta \approx 0$, the deviation grows to approximately
$4.5 \%$.  Both of these features can be traced back to approximations
we have made in the analytic estimate.  The first is associated with
the approximation \eref{eq:neglect_1_in_denom}, which assumes the
number density per comoving volume decreases significantly due to
residual annihilations and introduces an $\ord{T_f / m_X} \lesssim 5
\%$ error at all $m_h$.  The second is associated with neglecting the
equlilibrium term $n_X^{\mathrm{eq}}$ in \eref{eq:Bmaneqn}, which is
not negligible at the start of the residual annihilation era.  The
scaling with $m_h$ also has a simple, intuitive explanation.  One can
understand why $\delta n_X(t_0)$ is small at small $m_h$, because in
this limit the PT occurs too late and becomes decoupled from freeze
out.  Considering the opposite limit, one may wonder if $\delta
n_X(t_0)$ continues to increase for $m_h \gtrsim 300 \GeV$ where
$\delta < 0$.  For $\delta < 0$ the PT occurs before
freeze out, as in the usual cosmology, and one would naively expect
$\delta n_X(t_0)=0$.  Nevertheless, $\delta n_X(t_0)$ does continue to
grow because of the way we have defined $n_X^{(U)}$.  To calculate the
usual relic abundance $n_X^{(U)}$ we assume that there are $106.75$
relativistic species at freeze out.  If the PT occurs much earlier,
the number of relativistic species at freeze out will be significantly
less than $106.75$ and $\delta n_X(t_0)$ will be non-zero.  The CC
contribution grows monotonically with decreasing $m_h$, since in this
limit the PT temperature decreases and $c_1 \epsilon_1
\sim \rho_{\ex} / T_{PT}^4$.

In Figure \ref{fig:versus_mX} we plot the relic abundance shift by fixing $m_h = 115 \GeV$ and varying $m_X$.  
At large $m_X$, freeze out occurs well before the PT, the two events decouple, and the relic abundance shift is small.  
At small $m_X \lesssim 2.8$ TeV, freeze out occurs after the PT, and the analytic estimate fails.  
The CC effect $c_1 \epsilon_1$ has a maximum of approximately $10^{-3}$ at $\delta_{max} \approx 0.5$.  
For $\delta > \delta_{max}$ the factor $\epsilon_1$, given by \eref{eq:fractionalenergy}, is small because $T_f$ in the denominator is large.  
For $\delta < \delta_{max}$ the factor $c_1$, given by \eref{eq:c1_def}, is small because the CC is only present over a short time during WIMP residual annihilations.  
The presence of this maximum suggests that $c_1 \epsilon_1$ will typically be more sensitive to variations in the parameters of the scalar sector (e.g., $m_h$) than in variations of the DM sector (e.g., $m_X$).  
With this in mind, we will focus the remainder of our discussion on determining the conditions that a scalar potential must satisfy to maximize $c_1 \epsilon_1$.

We will now extend the estimates of Section \ref{sec:AnalyticEst} in order to understand Figure \ref{fig:versus_mh} through a simple analytic approximation.  We focus on the CC contribution to $\delta n_X(t_0)$, given by \eref{eq:c1e1}, which is 
\begin{align}\label{eq:dnXt0_est}
	\delta n_X(t_0) \ni c_1 \epsilon_1 \sim \frac{1}{10} \frac{\rho_{\ex}}{ g_E \, T_{PT}^4} 
\end{align}
up to multiplication by an $\ord{1}$ function of $\delta$.  
The factor of $g_E \approx 106.75$ represents the SM relativistic degrees of freedom before the PT.  
If we assume that before the PT, the SM particles are light with respect to the temperature, then we can approximate $V_{\rm eff}$ using the so-called high-temperature approximation
\begin{align}\label{eq:Veff_1Dtoy}
	V_{\rm eff} (h_c,T) \approx \frac{\lambda_{eff}}{4} \left( h_c^2 - v^2 \right)^2 + c \, T^2 h_c^2 \, .
% + h\text{-independent / subdominant / log terms}
\end{align}
Here we have defined
\begin{align}
	\lambda_{eff} \equiv \frac{4}{v^4} \Bigl( V_{\rm eff}(0,0) - V_{\rm eff}(v,0) \Bigr)
\end{align} 
to be the one-loop effective self-coupling and $2\, c \, T^2$ is the
thermal mass acquired by Higgs particles passing through the plasma.
In the SM and subject to our renormalization scheme, these
dimensionless numbers are $\lambda_{eff}\approx \lambda_{SM}$ and
$c\approx c_{SM}$ where
\begin{subequations}\label{eq:cSM_and_lamSM}
\begin{align}
	c_{SM} & = \frac{1}{24 v^2} \left( 6 m_t^2 + 6m_b^2 + 6 m_w^2 + 3 m_z^2 + \frac{3}{2} m_h^2 \right) \approx 0.18 \\
	\lambda_{SM} & = \frac{m_h^2}{2 v^2} + \frac{1}{128 \pi^2 v^4} \left( 48 m_t^4 + 48 m_b^4 - 24 m_w^4 - 12 m_z^4 - (15+\log 4) m_h^4 \right) \approx 0.12
\end{align}
\end{subequations}
for $m_h \approx 115 \GeV$.  The PT occurs at a temperature $T_{PT}$ where $\partial^2_{h_c} V_{\rm eff}(0,T_{PT}) = 0$.  Solving for this temperature one obtains
\begin{align}\label{eq:SM_PTtemp}
	c \, T_{PT}^2 = \frac{\lambda_{eff}}{2} v^2 \, .
\end{align}
Before the PT, the CC energy density is 
\begin{align}
	\rho_{\ex} = V_{\rm eff}(0,0) = \frac{\lambda_{eff}}{4} v^4 
\end{align}
and we can estimate the deviation in the relic abundance using 
\eref{eq:dnXt0_est} to be 
\begin{align}\label{eq:dnXt0_est2}
	c_1 \epsilon_1 \sim \frac{1}{10} \frac{1}{g_E} \frac{c^2}{\lambda_{eff}} \, . 
\end{align}
For natural couplings one expects $c^2 / \lambda_{eff} \sim \ord{1}$ (e.g., $c_{SM}^2 / \lambda_{SM} \approx 0.28$) and finds $c_1 \epsilon_1 \sim 1/ (10 g_E) \sim 10^{-3}$.  
Recalling also that $\lambda_{eff} \sim m_h^2$, one sees that this estimate agrees well with both the magnitude and scaling shown in Figure \ref{fig:versus_mh}.  
Note that in the $\lambda_{eff} \to 0$ limit, we find that both $\rho_{\ex}$ and $T_{PT}$ approach zero, but the ratio $\rho_{\ex} / T_{PT}^4$ becomes large.  
This simple approximation suggests that the region of parameter space that maximizes the CC contribution to $\delta n_X(t_0)$ will have low temperature PTs.  
This is evident in Figure \ref{fig:versus_mh} because the CC effect grows at low $m_h$ where the PT temperature is low.  
Hence we will next consider a model in which a scalar singlet coupled to the Higgs is introduced to lower the PT temperature.

%sec:Z2xSM
\subsection{SM Singet Extension with $\mathbb{Z}_2$}\label{sec:Z2xSM}

In this section, we briefly discuss an extension of the Standard Model in which the presence of an additional scalar field modifies the electroweak PT dynamics.  
However, we show that the dark matter relic abundance is not significantly enhanced, and we argue that we should consider models with first order PTs.  
Consider an extension of the SM in which a real, singlet, scalar field $s(x)$ is coupled to the Higgs $h(x)$ through interactions which respect the $\mathbb{Z}_2$ symmetry $s \to -s$.  
The renormalized potential for this theory can be written as 
\begin{align}\label{eq:V0_Z2xSM}
	U(\hs) =& 
%	\rho_{DE} + \frac{m_h^2}{8 v^2} \left( h^2 - v^2 \right)^2 
	\frac{m_h^2}{8 v^2} \left( h^2 - v^2 \right)^2 
%	\rho_{\rm ex} + \frac{m_h^2}{8 v^2} h^4 - \frac{m_h^2}{4} h^2 
	+ \frac{b_4}{4} s^4 
	+ \frac{1}{2} m_s^2 s^2 
	+\frac{a_2}{2} s^2 \left( h^2 - v^2 \right)  
%	+\frac{a_2}{2} s^2 h^2 
\end{align}
such that $\partial_h U(\left\{ v, 0 \right\}) = 0$, $\partial_h^2 U( \left\{ v, 0 \right\} ) = m_h^2$ and $\partial_s^2 U( \left\{ v, 0 \right\} ) = m_s^2$.  
We require 
\begin{align}\label{eq:vacstab_Z2xSM}
	m_s^2-a_2 \, v^2 > 0 && \text{and} && 2 a_2 + b_4 + \frac{m_h^2}{2 v^2} > 0
\end{align}
to ensure $\left< s \right> = 0$.  This model, known as the
$\mathbb{Z}_2$xSM, has been previously studied in order to determine
the viability of $s$ as a dark matter candidate
\cite{Gonderinger:2009jp, Barger:2007im, McDonald:1993ex,
  Burgess:2000yq, He:2007tt, Davoudiasl:2004be}.  We will not restrict
ourselves to this scenario, but instead treat the dark matter as a
separate sector.
% as explained in the introduction to this section.  
The
role of $s$ is simply to modify the PT dynamics.

Since this model possesses a greater parametric freedom than the SM, we can attempt to verify the relationship \eref{eq:dnXt0_est2}, derived in the previous section, which relates $c_1 \epsilon_1 \sim c^2 / \lambda_{eff}$.  
This is accomplished by first mapping the parameters of the $\mathbb{Z}_2$xSM to $c$ and $\lambda_{eff}$, and second by performing a parameter scan while calculating $c_1 \epsilon_1$.  
We obtain $c$ and $\lambda_{eff}$ by calculating the thermal effective potential as described in the previous section (see also Appendix \ref{app:EffPot}).
If we assume that  the quanta of $s(x)$ are light with respect to the temperature, we can then extract $c$ and $\lambda_{eff}$ by matching the effective potential to \eref{eq:Veff_1Dtoy}.  
Doing so yields the expressions
\begin{align}
	c & = c_{SM} + \frac{a_2}{24} \label{eq:c_Z2xSM} \\
	\lambda_{eff} & = 
	\lambda_{SM} 	- \frac{\, a_2^2}{32 \pi^2} \, \psi \left( \frac{a_2 v^2}{m_S^2}  \right) \label{eq:lameff_Z2xSM}  \\
	\psi (x) & \equiv 3 - \frac{2}{x} - 2 \left( 1  - \frac{2}{x} + \frac{1}{x^2} \right) \log \left[ 1 - x \right] \label{eq:psi_def}
\end{align}
where the terms containing $a_2$ arise from 1-loop diagrams with an $s$-particle in the loop, and the function $\psi$ varies from $\psi(0) = 0$ to $\psi(1) = 1$.  
As a result of the minus sign in \eref{eq:lameff_Z2xSM}, there is an upper bound $a_2 \lesssim 5$ given by the constraint $\lambda_{eff} > 0$.  
Now we can see the impact of the singlet field on the PT.  
For $a_2 > 0$, the parameter $c$ is slightly larger and $\lambda_{eff}$ is slightly smaller than in the SM.  
Recall that the PT temperature, given by \eref{eq:SM_PTtemp}, scales like $T_{PT}^2 \sim \lambda_{eff} / c$.  
Hence, the singlet field lowers the PT temperature and makes the CC energy density relatively more significant, which causes the relic abundance shift to be greater.

\begin{figure}[t]
\begin{center}
\includegraphics[width=0.3\textwidth]{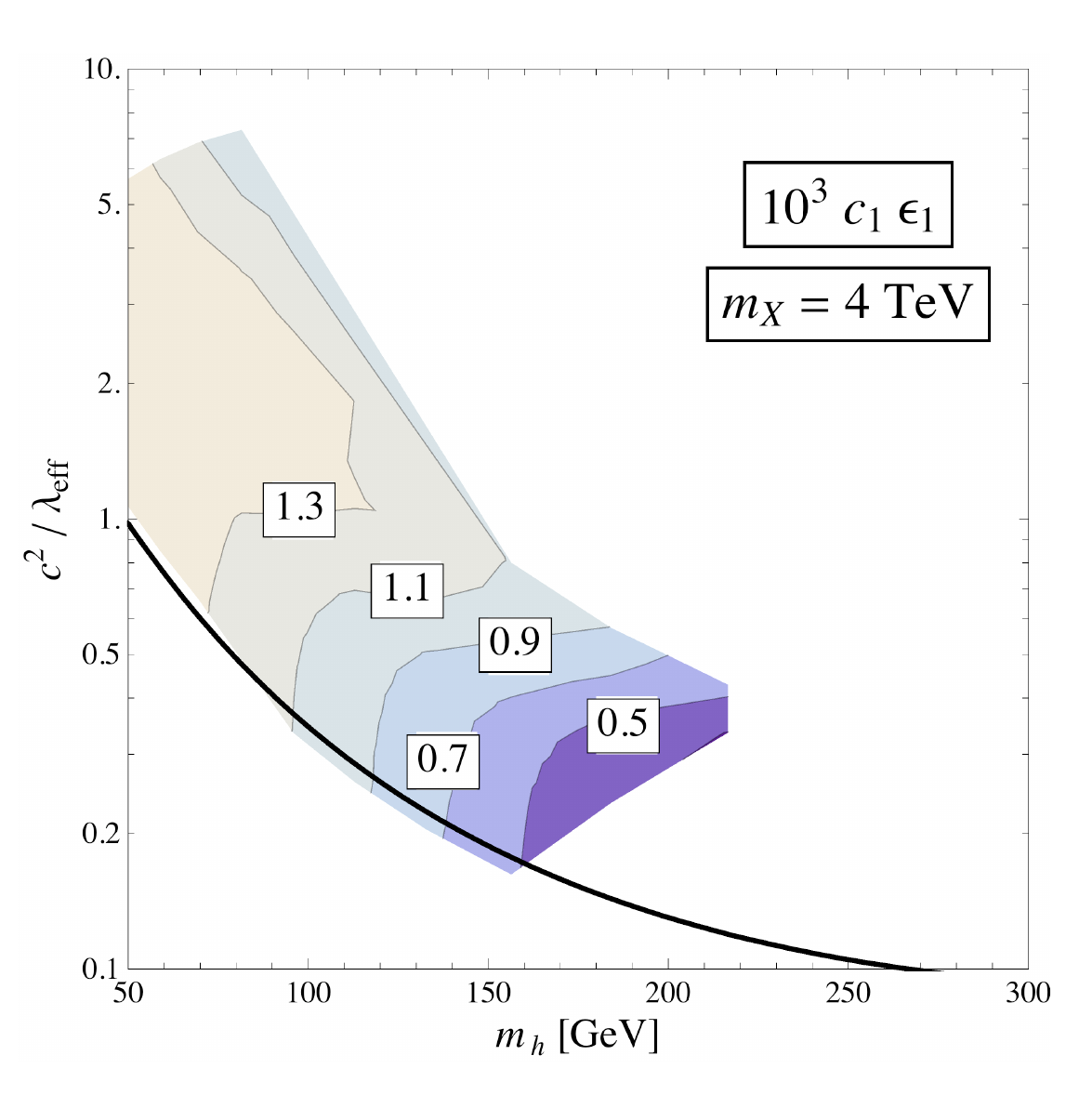} \hfill
\includegraphics[width=0.3\textwidth]{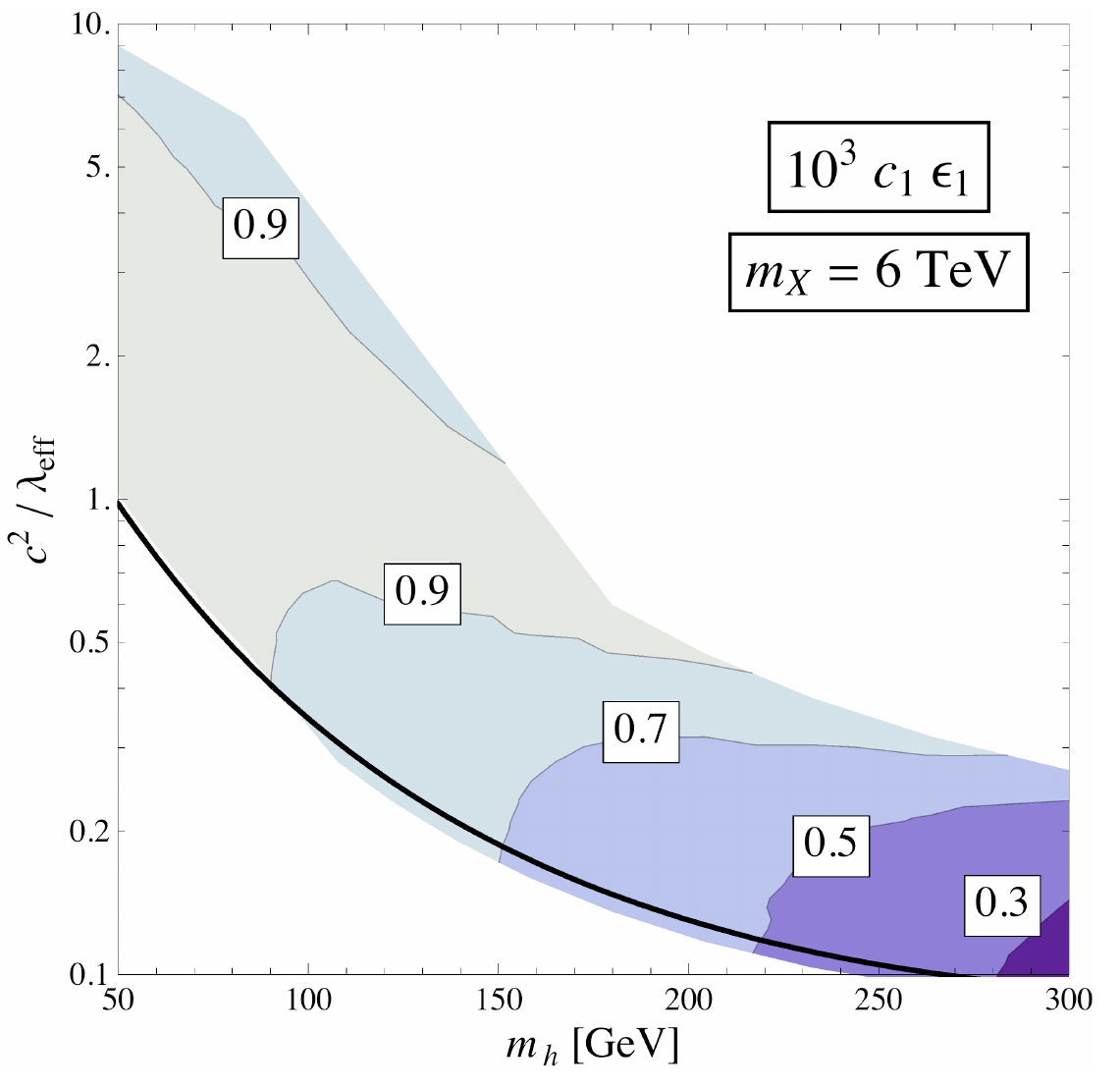} \hfill
\includegraphics[width=0.3\textwidth]{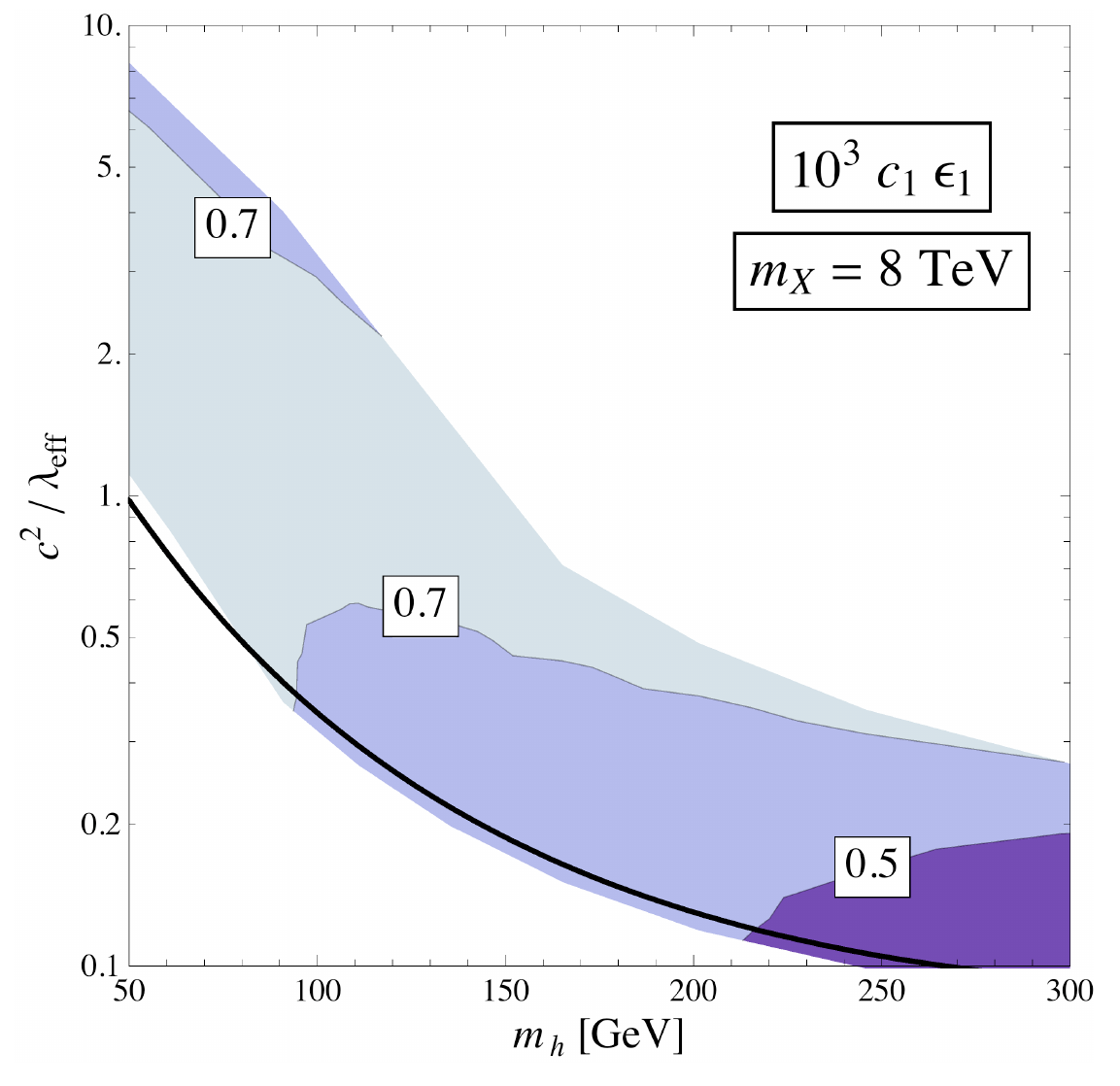}

\caption{\label{fig:Z2xSM}
The CC contribution to $\delta n_X(t_0)$, given by the $c_1 \epsilon_1$ term of \eref{eq:finalfractionalchange}, plotted over the $c^2 / \lambda_{eff}$--$m_h$ plane for three values of $m_X$.  The black line represents the SM ($a_2 = 0$).  
}
\end{center}
\end{figure}

To verify these analytic arguments, we calculate the PT temperature and $c_1 \epsilon_1$ numerically over a region of the theory space.  
We allow $m_h^2$ and $a_2$ to vary in the ranges $m_h^2 \in \left[ (50 \GeV)^2, (300 \GeV)^2 \right]$ and $a_2 \in \left[ -0.1, 4.0 \right]$, and we fix $b_4 = 0.25$ and $m_s^2 = \left( 500 \GeV \right)^2$.  
The range for $m_h$ is chosen to prevent the Higgs from becoming unacceptably light\footnote{Mixing with the singlet does not significantly reduce the LEP Higgs search bound \cite{Barger:2007im}.  Moreover, for small $m_h$ the electroweak breaking minimum may become metastable \cite{Gonderinger:2009jp, Profumo:2010kp}, and the PT becomes first order \cite{Csikor:1998eu}.  Nevertheless, we have allowed $m_h$ to be as small as $50 \GeV$ to illustrate the parametric dependence of the CC effect. 
},  while the range for $a_2$ is chosen to satisfy \eref{eq:vacstab_Z2xSM} and to avoid the unitarity bound.  
We map $m_h^2$ and $a_2$ to $c$ and $\lambda_{eff}$ using Eqs.~(\ref{eq:c_Z2xSM}) and (\ref{eq:lameff_Z2xSM}).  
In Figure \ref{fig:Z2xSM}, we have plotted the contribution to $\delta n_X(t_0)$ from the CC effect ($c_1 \epsilon_1$) over the $c^2 / \lambda_{eff}$--$m_h$ plane.   
This figure shows that the CC effect grows with increasing $c^2 / \lambda_{eff}$ and decreasing $m_h$, as we anticipated in \eref{eq:dnXt0_est2}.  
The largest value of $c_1 \epsilon_1$ is approximately $1.3 \times 10^{-3}$, which is only about $40 \%$ larger than in the SM.  
The insignificant enhancement can be understood by observing that although $a_2 > 0$ tends to decrease $c$, given by \eref{eq:c_Z2xSM}, its contribution is suppressed by a factor of $24$.  
Since $c_{SM} \approx 0.18$ we run into the unitarity bound on $a_2$ before it contributes significantly to $c$.  
If we were to add $N$ light singlet fields instead of one, the contribution to $c$ would be $N a_2 / 24$, which can be order one even for small $a_2$.  
We have not take this approach here because the $N$ additional relativistic degrees of freedom would have a larger effect on the relic abundance by increasing the energy density of radiation than through the CC.  
We have also plotted $c_1 \epsilon_1$ for three different values of the WIMP mass from $4$ to $8$ TeV.  
This narrow range of viable parameters illustrates the tuning that is required to ensure that the PT and freeze out occur at the same time.  
If the WIMP mass is too large, freeze out occurs too long before the PT when the CC energy density was subdominant to the energy density of the plasma.  
As the WIMP mass is lowered, the delay between freeze out and the PT decreases and $c_1 \epsilon_1$ grows.  
If the WIMP mass is too small, freeze out occurs after the PT when the CC energy density has been converted into radiation.  
This is the case in the $m_h \gtrsim 200$ region of the $m_X = 4$ TeV plot.

% fig:Z2xSM

The examples of the SM and the $\mathbb{Z}_2$xSM demonstrate that it is challenging to obtain $c_1 \epsilon_1$ larger than $\ord{10^{-3}}$.  
Our discussion at the end of Section \ref{sec:StandardModel} and simple dimensional analysis illustrate why this is the case.  
In that calculation we obtained \eref{eq:dnXt0_est2} which can be written schematically as $c_1 \epsilon_1 \sim \rho_{\rm ex} / T_{PT}^4 \sim c^2 / \lambda_{eff}$.  
Note that the mass scale $v$, which controls both $\rho_{\rm ex}$ and $T_{PT}$, cancels out in the ratio $\rho_{\rm ex} / T_{PT}^4$.  
In light of \eref{eq:dnXt0_est2} we propose that the CC effect can be enhanced by working in a model that has multiple mass scales if there exists a hierarchy between them.  
We will explore different applications in the remainder of this section.

% sec:GenericSinglet
\subsection{Generic Single Scalar Model}\label{sec:GenericSinglet}

In this section we calculate the CC contribution to the relic
abundance shift in a generic single scalar model.  Although extensions
of the Standard Model typically contain multiple scalar degrees of
freedom related by symmetries, the thermal dynamics (supercooling and
reheating) of a symmetry breaking PT can often be modeled by a single
scalar degree of freedom which does not display the symmetries of the
full theory \cite{Profumo:2007wc,Chung:2011hv}.  With this
motivation in mind, we consider the theory of a real scalar field
$\varphi(x)$ coupled to $N$ Dirac fields $\psi_i(x)$.  The scalar
field will experience a first order PT during which dark matter
freezes out, and the light fermions will compose the hot thermal bath.
Using this construction, we will be able to calculate the CC effect,
which is related to the non-thermal energy density and the amount of
supercooling, but we cannot estimate the entropy and decoupling
effects since these depends on how $\varphi$ is coupled to the full
theory.  Therefore, in this section we assume no decoupling occurs
near the time of the PT and that the number of relativistic species is
fixed to $g_{E/S} \approx 106.75$, the relativistic SM background.
Let the action be given by
\begin{align}\label{eq:GenSing_Sphi}
	S[\varphi] = \int d^4x \left\{ \frac{1}{2} (\partial \varphi)^2 - U(\varphi) - \sum_{i=1}^N \bar{\psi}_i \left( i \slashed{\partial} - m_i - h_i \varphi \right) \psi_i+ \mathcal{L}_{\rm ct}  \right\}
\end{align}
where
\begin{align}\label{eq:GenSing_Uphi}
	U(\varphi) = \rho_{\rm ex} + \frac{1}{2} M^2 \varphi^2 - \mathcal{E} \varphi^3 + \frac{\lambda}{4} \varphi^4 
\end{align}
is the renormalized potential and $\mathcal{L}_{\rm ct}$ is the counterterm Lagrangian.  
Note that we have eliminated the tadpole term in $U(\varphi)$ by defining the origin in field space appropriately, but there is still a counterterm for the tadpole in $\mathcal{L}_{\rm ct}$.  
As discussed in Section \ref{sec:Z2xSM}, we expect that there will be a greater impact on the dark matter relic abundance if freeze out occurs during a first order PT with large supercooling.  
Hence, we would like to understand what region of parameter space yields a PT of this kind.  
In particular, we expect that large supercooling can be obtained if the theory $S[\varphi]$ possesses two vacua, which will correspond to the low- and high-temperature phases, and that the vacua are separated by a barrier.

\begin{figure}[t]
\begin{center}
\includegraphics[width=0.6\textwidth]{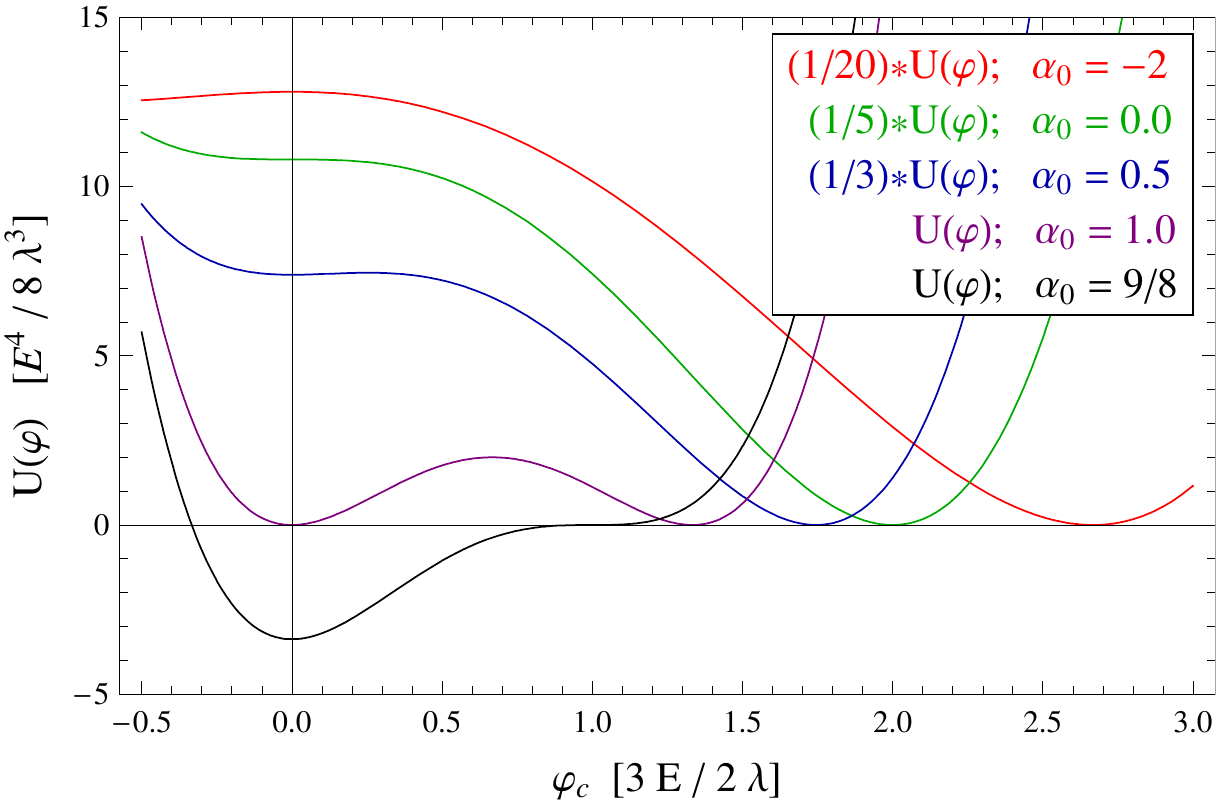}
\caption{\label{fig:Uphi}
An illustration of the $\alpha_0$ dependance of the potential given by \eref{eq:GenSing_Uphi}.  The curves represent $\alpha_0 = -2$ (red), $\alpha_0 = 0$ (green), $\alpha_0 = 0.5$ (blue), $\alpha_0 = 1$ (purple), and $\alpha_0 = 9/8$ (black).  
}
\end{center}
\end{figure}

We can determine the vacuum structure by identifying the minima of the effective potential, which is calculated in Appendix \ref{app:EffPot}.  
Provided that the non-thermal radiative corrections are negligible,
%[ELABORATE -- talk about tuning and renormalization conditions?]
the effective potential can be approximated as $V_{\rm eff}(\varphi_c, T=0) \approx U(\varphi_c)$.
It is convenient to eliminate $M^2$ for the dimensionless quantity $\alpha_0 \equiv \lambda M^2 / 2 \mathcal{E}^2$ while assuming $\lambda \mathcal{E} \neq 0$.  
We now see that the parameter $\alpha_0$ controls the shape of the potential $U(\varphi)$: for $\alpha_0$ = 1, the potential has two degenerate minima at $\varphi_c = 0$ and $\varphi_c = \left. v \right|_{\alpha_0 = 1}$ where
\begin{align}
	v = \frac{3 \mathcal{E}}{2 \lambda} \left( 1 + \sqrt{1 - \frac{8}{9} \alpha_0} \right) \, ;
\end{align}
for $\alpha_0 > 1$, $\varphi_c = 0$ is the global minimum; for $0 < \alpha_0 < 1$, $\varphi_c = v$ is the global minimum; and for $\alpha_0 < 0$, $\varphi_c = 0$ becomes a maximum (see also Figure \ref{fig:Uphi}).  
Therefore, provided that we take $0 \lesssim \alpha_0 \lesssim 1$, the theory possesses a metastable vacuum in which $\varphi_c \approx 0$ and a stable vacuum in which $\varphi_c \approx v$.  
In the stable vacuum, we impose the tuning condition $V_{\rm eff}(v,0) = 0$ to solve for 
\begin{align}\label{eq:GenSing_rhoex}
	\rho_{\rm ex} \approx \frac{\mathcal{E}^4}{8 \lambda^3} \left[ 27 - 36 \, \alpha_0 + 8 \, \alpha_0^2 + 27 \left(1 - \frac{8}{9} \alpha_0 \right)^{3/2} \right] + \ord{\hbar} \, ,
\end{align}
which represents the CC energy density prior to the PT.  
Finally, the barrier separating the two vacua has a ``height''
\begin{align}\label{eq:GenSing_Vbar}
	V_{\mathrm{barrier}} = U({\rm barrier}) - U(0) \approx  \frac{4 \mathcal{E}^4 \alpha_0^3}{27 \lambda^3} \Bigl[ 1 + \ord{\alpha_0} \Bigr]
\end{align}
relative to the metastable vacuum.  
Due to the factor of $\alpha_0^3$, the barrier vanishes rapidly as $\alpha_0$ approaches zero.  
This is illustrated by the $\alpha_0 = 0.5$ curve of Figure \ref{fig:Uphi} in which the barrier is already almost imperceptible to the eye.

% fig:Uphi --- alternate grayscale compatible file = BW_Uphi.pdf

Having established that this theory admits two vacua, we will study the PT using the thermal effective potential.  
Although the numerical calculations use the full effective potential, we can gain some intuition by making the high temperature approximation.  
We assume that the $\psi_i$-particles are light with respect to the temperature of the thermal bath, $m_i^2 \ll T^2$, and that the $\varphi$-particles are heavy.
In this limit, then the one-loop thermal effective potential may be approximated by the high temperature expansion 
\begin{align}\label{eq:GenSin_Veff}
	V_{\rm eff}(\varphi_c,T) \approx U(\varphi_c) + c \, T^2 \varphi_c^2 + \ord{m_i^2 / T^2} + \ord{\hbar}
\end{align}
where $c \approx \sum_{i=1}^N h_i^2 / 12$ is related to the couplings between $\varphi$ and $\psi_i$.  
Just as we introduced $\alpha_0$ to reparametrize $V_{\rm eff}(\varphi_c, 0)$, we can now introduce 
\begin{align}\label{eq:alphaT_def}
	\alpha(T) = \alpha_0 \left( 1 + \frac{\lambda c}{\mathcal{E}^2 \alpha_0} T^2 \right) \geq \alpha_0
\end{align}
to parameterize $V_{\rm eff}(\varphi_c, T)$.  
This definition is particularly convenient, because now Figure \ref{fig:Uphi} also illustrates the temperature dependence of $V_{\rm eff}(\varphi_c, T)$ (up to $\varphi_c$-independent terms) if one replaces $\alpha_0$ with $\alpha(T)$.  
We obtain the expectation values of $\varphi$ in the ``symmetric'' and ``broken'' phases, $v^{(s)}(T)$ and $v^{(b)}(T)$, by solving $(\partial / \partial \varphi_c) V_{\rm eff}(\varphi_c, T) = 0$ subject to the boundary conditions $v^{(b)}(0) = v$ and  $v^{(s)}(0) = 0$.  
We use the terms ``symmetric'' and ``broken,'' eventhough $S[\varphi]$ does not display a symmetry in order to connect with the notation of Section \ref{sec:A-Brief-ReviewrofPT}.

Provided that this model experiences a first order PT, the CC's effect on the relic abundance will depend sensitively on the amount of supercooling at the PT.  
This is seen by the factor of $(T_f)^4 \approx (T_{PT}^-)^4$ in \eref{eq:fractionalenergy}.  
Therefore, we will begin by investigating the parametric dependence of the amount of supercooling, and we will see that it has an interesting dependance on the parameter $\alpha_0$.  
The supercooling stage begins when the temperature drops below  
\begin{align}\label{eq:Tc}
	T_c \approx \mathcal{E} \sqrt{\frac{1 - \alpha_0}{\lambda \, c}} \, ,
\end{align}
defined by \eref{eq:Tc_def}, or equivalently when $\alpha(T_c) = 1$.  
During supercooling, the universe remains in the metastable, symmetric phase until bubbles of the broken phase begin to nucleate.  
Bubble nucleation is a non-perturbative process \cite{Miller:1974xx}, and it occurs at a rate per unit volume which carries the standard exponential suppression $\Gamma \sim T^4 {\rm exp} \left[ -S^{(3)} / T \right]$, where $S^{(3)}(T)$ is the action of the O(3) symmetric bounce \cite{Linde:1977mm, Coleman:1977py, Callan:1977pt}.  
Provided that $V_{\rm eff}(\varphi_c, T)$ can be expressed in the form of \eref{eq:GenSin_Veff}, then $S^{(3)}$ is well approximated by the empirical formula \cite{Dine:1992wr}
\begin{align}
%	\frac{S^{(3)}}{T} &= \frac{4.85 (M^2 + 2\, c\, T^2)^{3/2}}{\mathcal{E}^2 T} f(\alpha) = 9.7 \, \frac{M}{T} \frac{\alpha^{3/2}}{\lambda\,  \alpha_0^{1/2}} f(\alpha) \label{eq:S3overT}  \\
	\frac{S^{(3)}}{T} &\approx 13.7 \, \frac{\mathcal{E}}{T} \left(\frac{\alpha}{\lambda} \right)^{3/2} f(\alpha) \label{eq:S3overT}  \\
	f(\alpha) &\equiv 1 + \frac{\alpha}{4} \left( 1 + \frac{2.4}{1-\alpha} + \frac{0.26}{(1-\alpha)^2} \right)
\end{align}
with $\alpha = \alpha(T)$.  
Bubbles form rapidly once the bubble nucleation rate averaged over a Hubble volume $\Gamma \, H^{-3}$ is comparable to the Hubble expansion rate $H \sim T^2/M_p$.  
For an electroweak scale PT, this equality occurs when $S^{(3)} / T$ drops below approximately $140$ \cite{McLerran:1990zh, Anderson:1991zb}.  
Therefore, we can determine the amount of supercooling by solving $S^{(3)}/T  \approx 140$ for $T = T_{PT}^-$ and comparing this temperature with $T_c$.

Considerations of the equation $S^{(3)} / T \approx 140$ demonstrate that the nature of the PT is strongly dependent upon the vacuum structure of the theory, as parametrized by $\alpha_0$.  
We will discuss the two cases $\alpha_0 > 0$ and $\alpha_ 0 < 0$ separately.  
For $\alpha_0 > 0$, the vacuum with $\varphi_c = 0$ remains metastable as $T \to 0$.  
This implies that $T_{PT}^-$ can be arbitrarily low, and in this limit of large supercooling the CC effect may be arbitrarily large.  
However, in this case the barrier in $V_{\rm eff}(\varphi_c, T)$ persists as $T \to 0$, and it is possible that the PT does not occur at any temperature, but instead that the universe becomes trapped in the metastable vacuum.  
This follows from the observation that for $\alpha_0 > 0$, $S^{(3)} / T$ has a minimum at finite $T$: at low temperatures $S^{(3)} / T$ grows due to the explicit factor of $T$ in the denominator, and at high temperatures $f(\alpha)$ diverges as $T$ approaches $T_c$ and $\alpha \to 1$.  
For $\alpha_0 \lesssim 1$ the inequality $S^{(3)} / T \lesssim 140$ is not satisfied at any temperature, and the PT does not occur\footnote{At least, the PT does not occur as a thermal process, although it may still occur as a quantum tunneling process \cite{Coleman:1977py}.  However, since quantum tunneling typically proceeds on a longer time scale, the universe could enter an inflationary phase, which leads to a cosmological history that deviates significantly from the perturbations we consider in Section \ref{sec:AnalyticEst}.  
}.  
Therefore, if we require that the PT must occur via thermal bubble nucleation, we obtain an upper bound on $\alpha_0$.  
On the other hand, for the case $\alpha_0 < 0$, the PT necessarily occurs at a temperature $T_{PT}^- > 0$, since the symmetric phase becomes perturbatively unstable at low temperatures.  
This latter case has the drawback that supercooling cannot last an arbitrarily long time.

%fig:deltaSC
\begin{figure}[t]
\begin{center}
\includegraphics[width=0.6\textwidth]{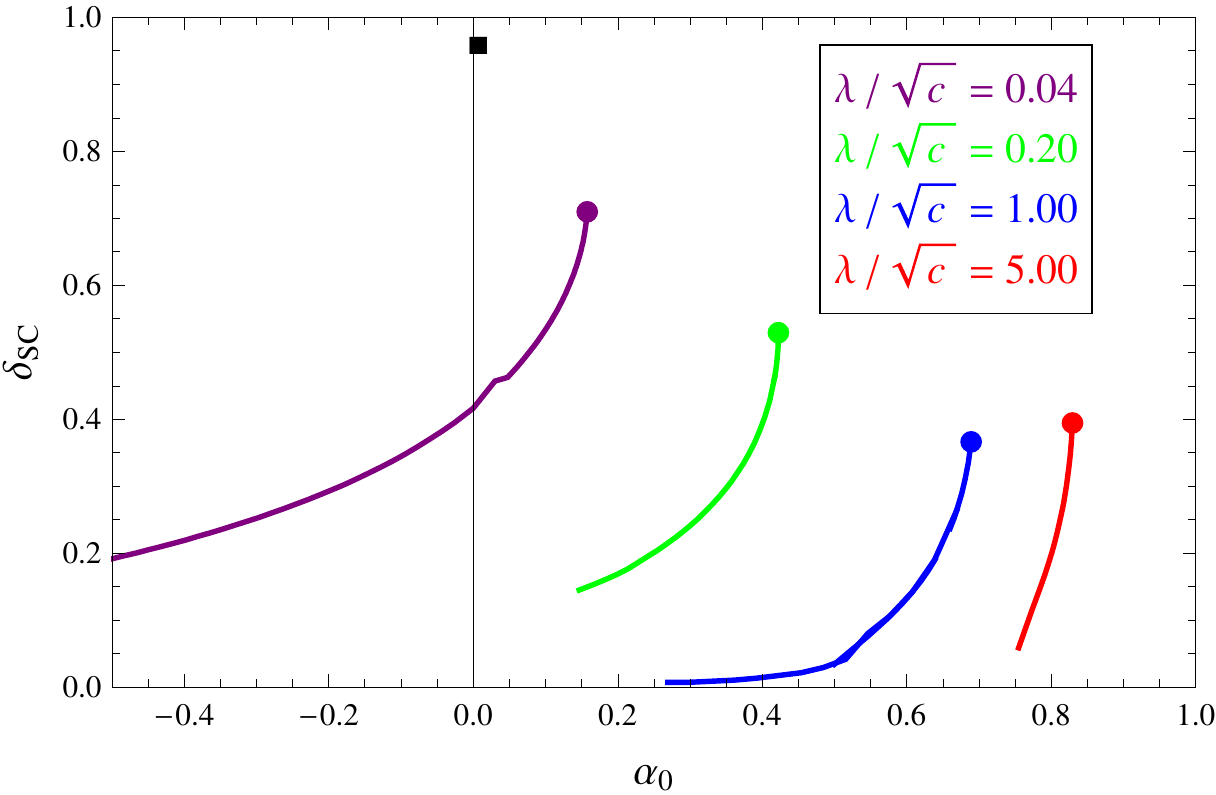} 
\caption{\label{fig:deltaSC}
The amount by which the PT temperature drops below the critical temperature, quantified by $\delta_{SC}$, is plotted against the parameter $\alpha_0$ which controls the height of the barrier.  The curves represent $\lambda / \sqrt{c} = 0.04$ (purple), $0.20$ (green), $1.00$ (blue), and $5.00$ (red).  
%These curves only depend on the parametric combination $\lambda / \sqrt{c}$.  The amount of supercooling grows as $\alpha_0$ is made larger, but reaches a finite maximum $ \delta_{SC}^{(max)} \lesssim \ord{1}$ at a value of $\alpha_0$ that depends on the ratio $\lambda / \sqrt{c}$.  
The square indicates the especially tuned parameter set given by \eref{eq:GenSing_tunedparams}.  
}
\end{center}
\end{figure}

Assuming that the PT does occur, we can quantify the amount of supercooling using  
\begin{align}\label{eq:GenSing_dSC}
	\delta_{SC} = 1- \frac{T_{PT}^-}{T_{c}} \, ,
\end{align}
which takes values between $0$ and $1$.  
Parametrizing the temperature dependance with $\delta_{SC}$, we can rewrite \eref{eq:S3overT} as
\begin{align}
	\left. \frac{S^{(3)}}{T} \right|_{T_{PT}^-} &\approx 13.7 \left( \frac{\lambda}{\sqrt{c}} \right)^{-1} \frac{\alpha^{3/2}}{\sqrt{1-\alpha_0}} \frac{f(\alpha)}{1 - \delta_{SC}} \\
	\alpha &= \alpha_0 + (1- \alpha_0)(1-\delta_{SC})^2 \, ,
\end{align}
which is now only a function of $\alpha_0, \lambda / \sqrt{c},$ and $\delta_{SC}$.  
Of course, this expression is approximate, since we assumed $V_{\rm eff}$ took the form of \eref{eq:GenSin_Veff}, but it suggests that the amount of supercooling will depend most sensitively on $\alpha_0$ and $\lambda / \sqrt{c}$.  
Now using the full thermal effective potential, we impose $\left. S^{(3)}/T \right|_{T_{PT}^-} = 140$ and solve for $\delta_{SC}$, which we have plotted in Figure \ref{fig:deltaSC} for various parameter sets:
\begin{align}\label{eq:GenSing_scanparams}
	\mathcal{E} = 5 \GeV \qquad  \lambda = \left\{ 0.004, \ 0.02, \ 0.10, \ 0.50 \right\} \nonumber \\
	N = 1 \qquad m = 10 \GeV \qquad h = 0.346 \qquad c \approx 0.01 \, .
\end{align}  
The supercooling grows with increasing $\alpha_0$ and decreasing $\lambda / \sqrt{c}$ as the barrier height and bounce action are made larger.  
%In the shaded region the PT is not first order.  
The amount of supercooling is typically $\delta_{SC} = \ord{0.5}$ which implies $T_{PT}^- = \ord{T_c / 2}$.  
Above a finite value of $\alpha_0$ (indicated by a dot) the barrier becomes insurmountably large, and the universe becomes trapped in the metastable vaccum.  
The largest amount of supercooling is achieved for $\lambda / \sqrt{c} \ll 1$ and $0 < \alpha_0 \ll 1$.  
In this parameter regime the CC is large (see \eref{eq:GenSing_rhoex}), and the metastable vacuum is separated from the true vacuum by small barrier (see \eref{eq:GenSing_Vbar}).  

%fig:c1e1
\begin{figure}[t]
\begin{center}
\includegraphics[width=0.60\textwidth]{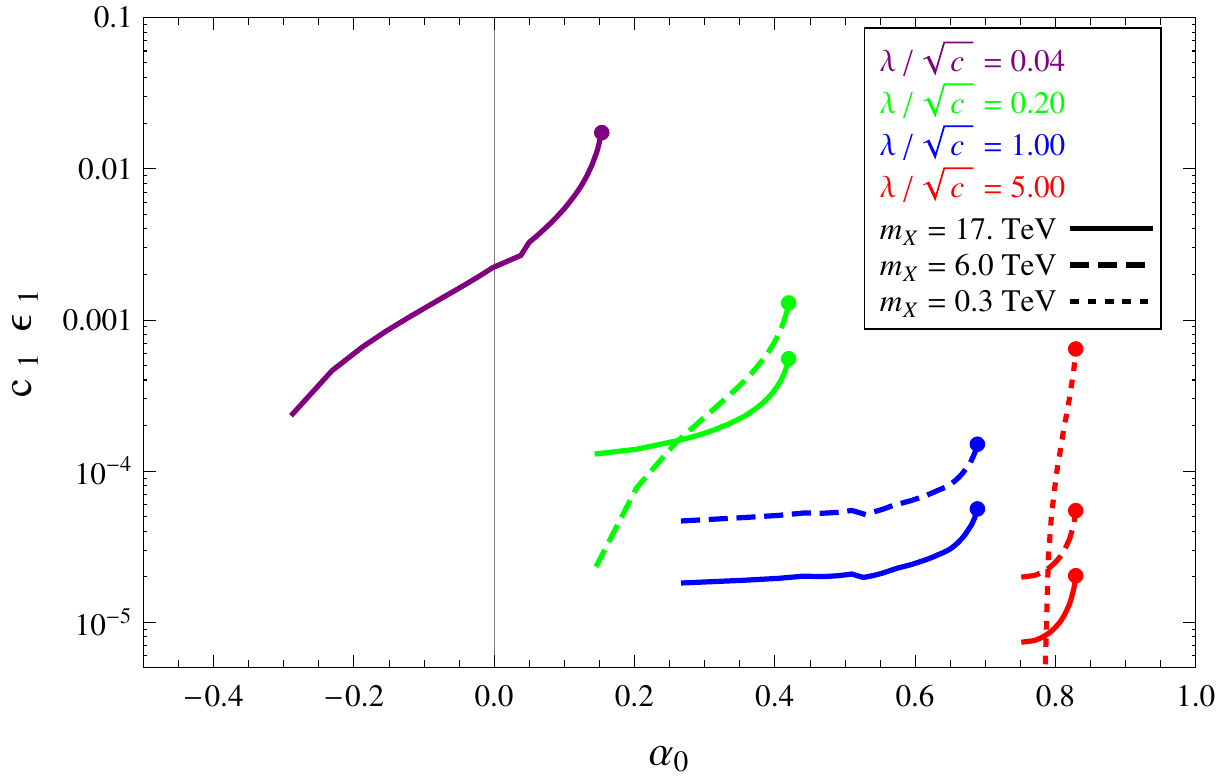} 
\caption{\label{fig:GenSing_ciei}
%The CC effect on the relic abundance $c_1 \epsilon_1$ plotted against $\alpha_0$ for $m_X =2$ TeV (solid) and 500 GeV (dashed).  The contours extend over a finite range of $\alpha_0$ because for larger $\alpha_0$ the PT does not occur and for smaller $\alpha_0$ the PT occurs before freeze out.  The $\lambda / \sqrt{c} = 0.04$ curve is absent for $m_X = 500 \GeV$ since freeze out always occurs at a lower temperature than the PT.  
%Contributions to the relic abundance shift coming from the CC effect $c_1 \epsilon_1$ (dashed) and entropy effect $c_2 \epsilon_2$ (solid).  We vary $\alpha_0$ and $\lambda / \sqrt{c}$ while setting $m_X =2$ TeV (left) and 500 GeV (right).  Certain contours extend over a finite range of $\alpha_0$ because for larger $\alpha_0$ the PT does not occur and for smaller $\alpha_0$ the PT occurs before freeze out.  
The CC effect on the relic abundance $c_1 \epsilon_1$ plotted against $\alpha_0$ for $m_X =17$ TeV (solid), 6 TeV (dashed), and 0.3 TeV (dotted).  For the contours which are absent, freeze out occurs after the PT when the CC is not EW-scale.  
}
\end{center}
\end{figure}

Having come to understand the parametric dependance of the amount of
supercooling as $\varphi$ experiences a first order PT, we turn our
attention back to calculating the impact of such a PT on dark matter
freeze out.  Using \eref{eq:c1e1} we calculate the effect of the CC on
the relic abundance shift and present the results in Figure
\ref{fig:GenSing_ciei}.  We have chosen the same parameters as
indicated in \eref{eq:GenSing_scanparams} and have taken
\begin{align}
	m_X = \left\{ 0.3, \ 6.0, \ 17 \right\} \ {\rm TeV} \qquad g_X = 2 \qquad \sv = 2.33 \times 10^{-39} {\rm cm}^{-2} \, 
\end{align}
as well.  
The figure illustrates that is possible to achieve $c_1 \epsilon_1 = \ord{0.01}$ in the tuned parametric regime where $\lambda / \sqrt{c}$ is small and $\alpha_0$ approaches its maximally allowed value.  
Some of the curves are absent for the smaller WIMP masses.  
This occurs because as $m_X$ is lowered, the temperature of freeze out decreases as well.
In the case that $\lambda / \sqrt{c}$ is small and the PT temperature is high (see \eref{eq:Tc}), freeze out will occur after the PT for small $m_X$.  
This statement about the relative times of freeze out and the PT also explains why $c_1 \epsilon_1$ is insensitive to $\alpha_0$ for certain parameter sets (e.g., $\lambda / \sqrt{c} = 1$, $m_X = 17$ TeV) and very sensitive for others (e.g., $\lambda / \sqrt{c} = 5$, $m_X = 0.3$ TeV).  
In the first case, freeze out occurs long before the PT while in the latter case, freeze out occurs just before and during the PT and there is a large impact on the relic abundance.

To conclude this section, we present a particular tuned parameter set which yields $c_1 \epsilon_1 = \ord{1}$.  
Suppose that we have only one fermion $\psi$ and the parameters of $S[\varphi]$ are given by 
\begin{align}\label{eq:GenSing_tunedparams}
	& \lambda = 5.4 \times 10^{-4} \qquad \qquad 
	h = 0.1 \nonumber \\
	\mathcal{E} = 0.27 \GeV & \qquad 
	M^2 = (1.89 \GeV)^2 \qquad 
	m = 10 \GeV \qquad
\end{align}
which leads to
\begin{align}
	v \approx 1497 \GeV \qquad 
	& \qquad \alpha_0 \approx 0.007 \nonumber \\
	c \approx 8.3 \times 10^{-4} \qquad & \qquad \frac{\lambda}{\sqrt{c}} \approx 0.018 
\end{align}
and PT temperatures 
\begin{align}
	T_c \approx 374 \GeV \qquad 
	T_{PT}^- \approx 16 \GeV \qquad
%	T_{PT}^+ \approx 50.2 \GeV \qquad 
	\delta_{SC} \approx 0.96 \, .
\end{align}
This parameter set is represented on Figure \ref{fig:deltaSC} by a square marker.  
In the dark matter sector we take 
\begin{align}
	m_X = 600 \GeV \qquad 
	g_X = 2 \qquad 
	\sv = 2.33 \times 10^{-39} {\rm cm}^{-2}
\end{align}
such that 
\begin{align}
	T_f \approx 34 \GeV \qquad {\rm and} \qquad \delta \approx 1.12 \, .
\end{align}
Using these values we can estimate the CC effect as
\begin{align}
	c_1 \epsilon_1 &\approx 6.1
%	c_2 \epsilon_2 &\approx -20.1 \, .
\end{align}
Note that the potential obtained with these parameters has a very shallow metastable vacuum at $\varphi_c \approx 0$, separated from the global vacuum at $\varphi_c \approx v$ by a very small barrier.

%[PUT DOT ON THE PLOTS TO REPRESENT THIS POINT] 
%[CALC ALL TERMS NOT JUST C1E1?]  
%[PARTICLE MASSES?] 
%[DURATION OF PT?]
%[RADIATIVE CORRECTIONS?]
%[SMALL BARRIER, PLOT?]
%[MAY BE CONCERNED THAT QUANTUM TUNNELLING IS UNSUPPRESSED -- CALC THIS TOO]

% sec:SingletExtension1PT
\subsection{Singlet Extension with First Order PT}\label{sec:SingletExtension1PT}

In this section, we consider a generalization of the SM extension studied in Section \ref{sec:Z2xSM}, in which we do not impose a $\mathbb{Z}_2$ symmetry on the singlet field $s(x)$.  
This leads to model known as the xSM \cite{Gonderinger:2009jp, Barger:2007im}.    
The xSM admits a first order electroweak PT \cite{Profumo:2007wc, Ahriche:2007jp}, and we seek to compute the effect on the relic abundance due to the effective CC at the PT.  
As discussed in Section \ref{sec:GenericSinglet}, the CC effect grows with the duration of supercooling.  
With this in mind, we will focus on a region of parameter space in which we expect to have first order PTs with large supercooling.  
Supercooling is an example of the hierarchy of mass scales which we argued in Section \ref{sec:Z2xSM} helps to obtain a larger CC effect.

We generalize the $\mathbb{Z}_2$xSM potential \eref{eq:V0_Z2xSM} by relaxing the $\mathbb{Z}_2$ symmetry. 
This allows us to write down the three additional operators $s h^2$, $s^3$, and $s$, but we eliminate the tadpole by an appropriate shift in the field space.   
We are left with the xSM renormalized potential 
\begin{align}\label{eq:V0_xSM}
	U(\hs) =& 
	\frac{m_h^2}{8 v^2} \left( h^2 - v^2 \right)^2 
	+ \frac{b_4}{4} s^4
	+ \frac{1}{2} m_s^2 s^2
	+ \frac{b_3}{3} s^3 
	+ \frac{1}{2} s \left( h^2 - v^2 \right) \left( a_1 + a_2 s \right) \, .
\end{align}
The thermal effective potential $V_{\rm eff}$ is calculated in Appendix \ref{app:EffPot}.  
With this parametrization, $V_{\rm eff}(\left\{ h_c, s_c \right\}, T=0)$ has a minimum at $\left\{ h_c, s_c \right\} = \vx$ where $V_{\rm eff}(\left\{ v,0 \right\}, T=0) = 0$ and the curvatures in the $h$ and $s$ directions are $m_h^2$ and $m_s^2$ respectively.  
The Higgs vacuum expectation value is fixed by electroweak constraints, but the six real numbers $\left\{ m_h^2, m_s^2, b_4, b_3, a_1, a_2\right\}$ are free parameters.

As in the previous section, we compute the bounce action $S^{(3)}$ in order to estimate the PT temperature $T_{PT}^-$ by solving $S^{(3)} / T \approx 140$.  
This calculation is made more challenging by the presence of the additional field direction.  
To obtain $S^{(3)}$ we make the approximation that the PT occurs along the trajectory $\bar{s}(h_c)$ satisfying
\begin{align}\label{eq:xSM_sbarofh}
	\left. \frac{d U(\left\{ h_c, s_c \right\}, 0)}{ds} \right|_{\bar{s}_c} = 0 && \mathrm{and} && \bar{s}_c(v) = 0  \, ,
\end{align}
which reduces the problem back to solving for the bounce in one dimension.  
In the region of parameter space on which we are focused, this approximation gives $T_{PT}^-$ to within a few percent (see Appendix \ref{sub:xSMBounce} for details).  
Note that the empirical formula \eref{eq:S3overT} cannot be applied here, because the effective potential is not well approximated by the form \eref{eq:GenSin_Veff}.  
%This leads to an over estimate of the bounce action (the bounce is a saddle point solution) at the order of a few percent.  Since we have focused on a region of parameter space with large supercooling, typically $T_{PT}^- \gtrsim T_0 > 0$.  In this limit, $S^{(3)}$ decreases very rapidly with $T$ and our error in $S^{(3)}$ translates into a much smaller error in $T_{PT}^-$.  

We have performed a parameter space scan and searched for a region with large corrections to the relic abundance from the CC.  
In the scan we fix the parameters $b_3 = -20 \GeV, b_4 = 0.2, a_1 = -25 \GeV,$ and $a_2 = 0.2$, and we vary $m_h^2 \in \left[(65 \GeV)^2, (170 \GeV)^2 \right]$ and $m_S^2 \in \left[ (40 \GeV)^2, (140 \GeV)^2 \right]$.  
In order to connect with the intuition garnered from the single field model of Section \ref{sec:GenericSinglet}, we have mapped the xSM parameter space to a single parameter $M^2$.  
This is accomplished by restricting to the trajectory given by \eref{eq:xSM_sbarofh} and defining 
\begin{align}
	M^2 \equiv \left. \frac{d}{dx^2} V_{\rm eff}\left( \left\{h_c(x), \bar{s}_c(h_c(x)) \right\}, T=0\right) \right|_{x=0}
\end{align}
where $x$ parametrizes the position along the curve $\bar{s}_c(h)$.  
The parameter $M^2$ controls the stability of the electroweak preserving vacuum:  if $M^2 > 0$ the symmetric phase remains metastable as $T \to 0$, whereas if $M^2 < 0$ the symmetric phase becomes perturbatively unstable at some finite temperature $T_0 > 0$.  
In this way, the potential depends on the parameter $M^2$ in the same way as the parameter  $\alpha_0$ from Section \ref{sec:GenericSinglet}.  
We cannot map the xSM parameter space to $\alpha_0$ directly because the effective potential along the trajectory \eref{eq:xSM_sbarofh} cannot be expressed in the form of \eref{eq:GenSin_Veff}.

%fig:xSM_pspace1
\begin{figure}[t]
\begin{center}
\includegraphics[width=0.60\textwidth]{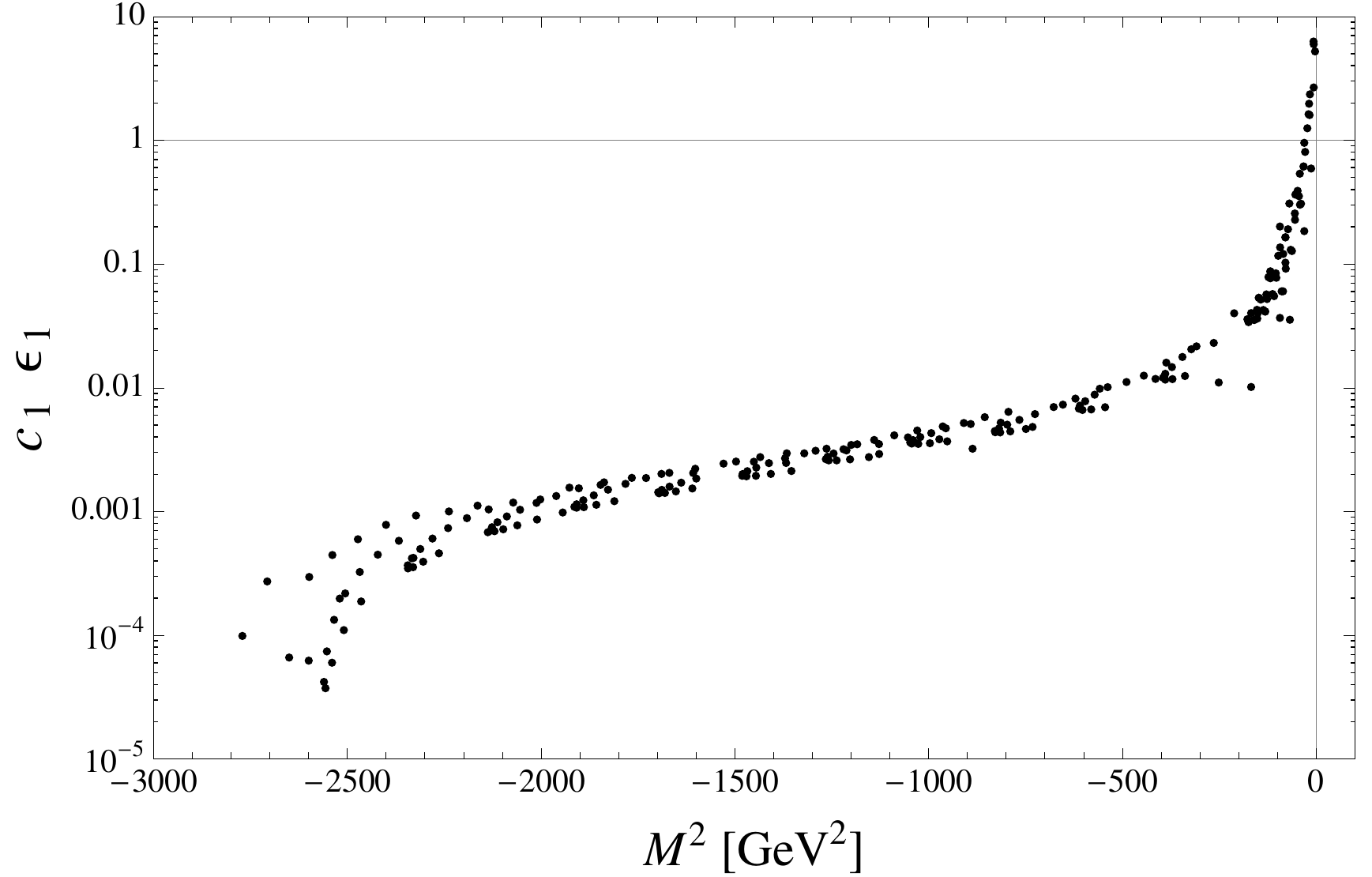}
\caption{\label{fig:xSM_pspace1}
The fractional deviation in the relic abundance of a 2 TeV WIMP due to the CC at the xSM electroweak PT.  The parameter $M^2$ controls the curvature of the zero temperature effective potential along the PT trajectory.  For $M^2 \lesssim 0$ large supercooling enhances the CC's effect.  For $M^2 \gtrsim 0$ the PT does not occur, and for $M^2 \lesssim -2500 \GeV^2$ the PT occurs before freeze out leading to a suppression of the relic abundance shift.  
}
\end{center}
\end{figure}

\begin{figure}[t]
\begin{center}
\includegraphics[width=0.60\textwidth]{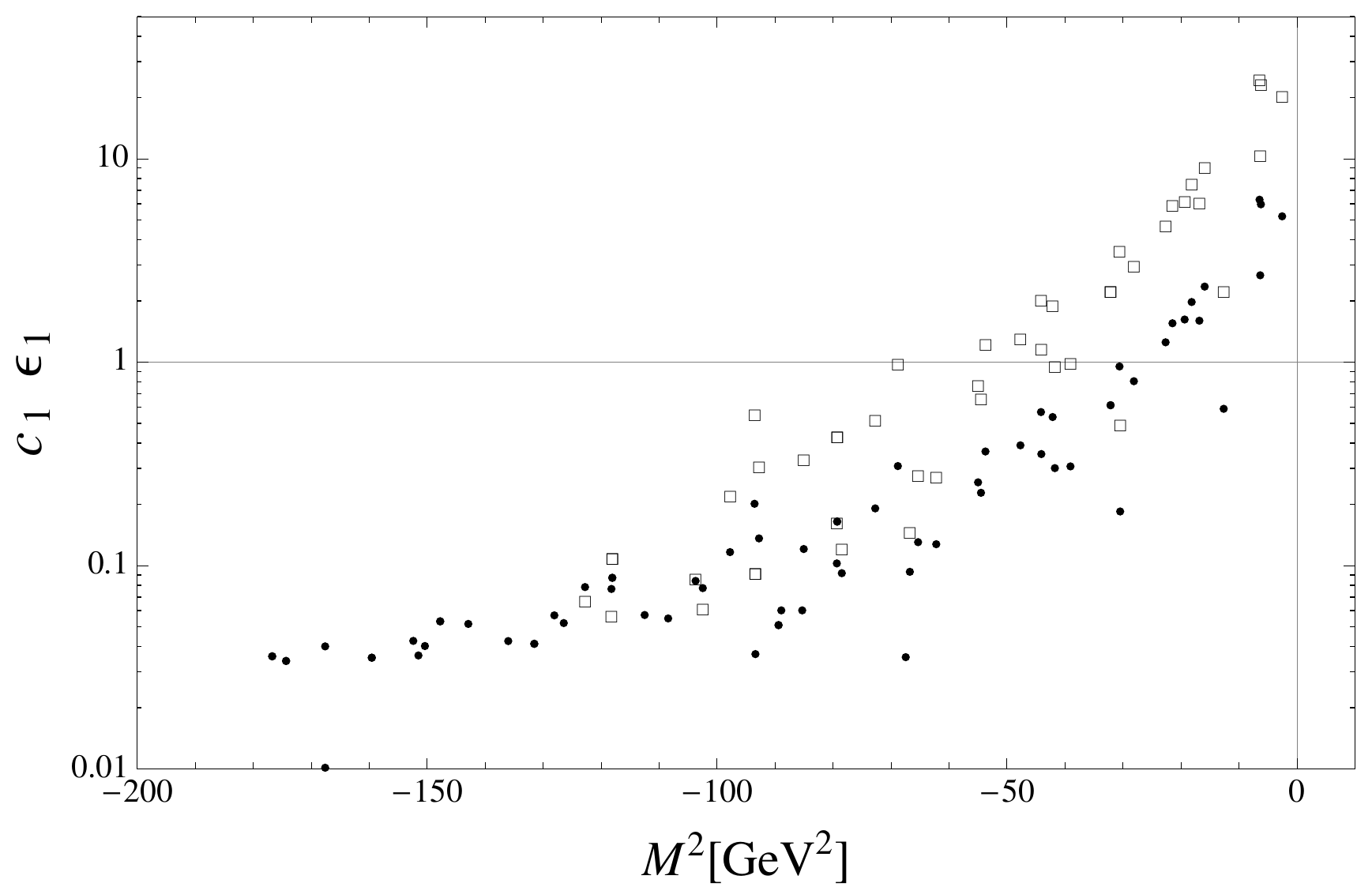}
\caption{\label{fig:xSM_pspace2}
This figure shows a subset of Figure \ref{fig:xSM_pspace1} as well as the CC effect for a 500 GeV WIMP represented by squares.  As the WIMP mass is reduced, freeze out occurs at a lower temperature.  This increases $c_1 \epsilon_1$ for $M^2 \lesssim 0$ where the PT temperature is low, but excludes points $M^2 \lesssim - 100 \GeV^2$ where freeze out occurs after the PT. 
}
\end{center}
\end{figure}

In Figure \ref{fig:xSM_pspace1}, we have plotted $c_1 \epsilon_1$, given by \eref{eq:c1e1}, by projecting onto the $M^2$ axis and choosing $m_X = 2$ TeV.  
For $M^2 \lesssim 0$ the CC has an $\ord{1}$ impact on the relic abundance.  
In this region, the supercooling is maximal and $T_{PT}^- \gtrsim T_0 = \ord{\mathrm{few} \GeV}$.  
For smaller values of $M^2$, the CC effect rapidly decreases and drops below $1 \%$ for $M^2 \lesssim 500 \GeV$.  
Therefore, in order for the CC to have a significant impact on the relic abundance, the parameters of the scalar sector must be tuned into a narrow band where supercooling is large.  
In Figure \ref{fig:xSM_pspace2}, we have allowed the WIMP mass to decrease to 500 GeV.  
This change lowers the freeze out temperature, reduces the delay $\delta$ between freeze out and the PT, and therefore increases the CC effect.  
However, this increase is small compared with the amount by which $c_1 \epsilon_1$ varies with $M^2$ in the $M^2 \lesssim 0$ region.  
For smaller values of $M^2$, the PT temperature is higher and for the 500 GeV WIMP, freeze out occurs after the PT causing the CC effect to be suppressed.  
These calculations lead us to the conclusion that the optimal region of parameter space is one in which the symmetric phase becomes perturbatively unstable at a low temperature and the effective potential is concave at zero temperature.  
We were unable to find any points with $M^2 > 0$ in which the PT completes.

%fig:xSM_pspace2

The following is a benchmark parameter point:  
\begin{align}\label{eq:xSM_benchmark}
	\left\{	a_1, b_3, m_h, m_s, m_X	\right\} = 
	\left\{	-25, -20, 128, 91.1, 2000	\right\} \mathrm{GeV}, \ \ \ \ 
	\left\{	a_2, b_4	\right\} = 
	\left\{	0.2, 0.2	\right\}, \nonumber \\
	M^2 = -47.7 \GeV^2 \ \ \ 
	\left\{	T_f, T_c, T_{PT}^+, T_{PT}^-, T_0	\right\} = 	
	\left\{107, 70.7, 30.0, 13.7, 12.7\right\} \GeV, \nonumber \\
	c_1 \epsilon_1 = 0.390,  \ \ \ \ 
	\rho_{\rm ex} = \left( 69.7 \GeV \right)^4 \, .
\end{align}
The scalar masses are given by the eigenvalues of \eref{eq:Mhs_xSM} which are
\begin{align}
	M_H = 141 \GeV \ \ \ \ \ \ \ &\left\{ 0.78, 0.22 \right\} \nonumber \\
	M_S = 70.7 \GeV \ \ \ \ \ \ \ &\left\{ 0.22, 0.78 \right\} 
\end{align}
with the respective squared eigenvectors indicated to the right.

% sec:Conclusion
\section{Conclusion \label{sec:Conclusion}}
We have considered a way to probe the hypothesis that the present-day,
minute CC energy density is the result of a tuning between UV
contributions of unspecified origin and IR contributions that arise
from cosmological PTs.  Prior to the electroweak scale PT, the UV
contribution would have been partially uncancelled leaving an
$\ord{M_W^4}$ energy density.  It is possible to probe this energy
density with the physics of dark matter freeze out provided that the
dark matter mass is greater than a few hundred GeV.  The dark matter
relic abundance is increased due to the effective CC's contribution to
the Hubble expansion rate during freeze out.  

The notion of how an effective vacuum energy (which is Lorentz
invariant in the flat space limit) can depend on temperature (which
manifestly breaks Lorentz invariance) has been clarified.  The
temperature is an approximation to the mixed vacua, inhomogeneous
states whose occupation is very probable near the time of the PT.
This leads to a spatially averaged equation of state that is expressed
in terms of an effective vacuum energy density that is somewhere
between the false and true vacuum energy densities.  The true
inhomogeneous field configurations may also lead to additional dark
matter freeze out effects that have not been investigated in this
paper.  This would be an interesting avenue for future investigations.

To provide a generic prediction associated with the established
physics and to provide the computational details missing in
\cite{Chung:2011hv}, we have analyzed the Standard Model with a $115
\GeV$ Higgs and a single WIMP dark matter degree of freedom, assuming
that the WIMP interaction effects on the dynamics of the PT is
negligible.  We have found that the CC causes an $\ord{10^{-3}}$
fractional increase in the relic abundance of a 4 TeV WIMPs.  This is
typical of non-first order PTs.

We have also investigated minimal singlet extensions of the SM and
searched for parametric regimes in which the CC effect on the relic
abundance is enhanced.  We find that a low temperature, first order PT
with large supercooling is the optimal scenario for maximizing the CC
effect.  In this limit, the effective CC energy density's contribution
to the Hubble expansion rate can be comparable to the radiation energy
density, and the CC effect can become order one.  In the context of a
generic single field model, we find that reaching this limit requires
a tuning of the scalar sector parameters and the WIMP mass.  Without
appropriate tuning, either 1) the PT will not occur at all by thermal
bubble nucleation, 2) the PT will occur before freeze out (when the
dark matter is still in equilibrium and the CC effect is suppressed),
or 3) the CC effect will not be large.

As a specific example, we have considered the xSM, an extension of the
SM that adds a real scalar singlet.  In that model, we find that the
CC may increase the relic abundance by as much as a factor of order
few.  To maximize the CC effect, the scalar parameters must be tuned
into a narrow band where fluctuations around the symmetric ``vacuum''
are slightly tachyonic, which allow for a long period of supercooling.
The magnitude of the CC effect is relatively insensitive to the WIMP
mass provided that the latter is sufficiently large such that freeze
out begins before the PT occurs.

The tests of CC fine tuning hypothesis are notoriously rare.  In the
context of a dark matter probe, it is encouraging that parametric
possibilities do exist within simple extensions of the SM.  It would
be interesting to further advance this exploration by computing the
dark matter implications of modified gravity/self-tuning models and
comparing the results with those of this paper.  Furthermore, it would
be interesting to cross correlate other astrophysical tests of those
modified gravity/self-tuning models with the dark matter predictions
made within those models.  Note also that there are other probes of
the cosmological constant during a PT such as gravity wave probes
\cite{Chung:2010cb} that will need more development as the gravity
wave spectrum calculational technology improves
\cite{Huber:2008hg,Caprini:2009yp,Kahniashvili:2009mf}.

\begin{appendix}

% sub:Renormalization-Scale
\section{Renormalization Scale\label{sub:Renormalization-Scale}}

Any measurable quantity is independent of the renormalization scale.
Hence, one should not expect that the running of the cosmological
constant parameter should affect any physical observable. Indeed,
the running of the other parameters in the Lagrangian will compensate
the running of the CC parameter to yield the same
$T_{00}$ governing the expansion rate $H$ which can be measured
for example by a test photon redshift. The renormalization scheme
and scale does however determine the manner in which radiative corrections
play a role. Furthermore, in any practical computations involving
finite order truncation in $\hbar$ expansion, there is a renormalization
scale dependence to next order in the perturbation power unless one
is able to explicitly keep exactly the terms of the relevant order
in $\hbar$.

Given that we are computing homogeneous quantities, one might also
naively worry that there is a coarse graining requirement down to
length scales of $H^{-1}$. To see why this is not the case and to see
what renormalization scales would minimize the radiative correction
dependence, consider the effective action generating the gravitational
equation of motion for the metric $g$:
\begin{equation}
e^{iS_{\rm eff}[g]}=e^{iS_{EH}[g]}\int D_{\Lambda}\phi \, e^{iS_{M}[g,\phi]}\label{eq:effectiveactiondef}
\end{equation}
where $S_{EH}$ is the Einstein-Hilbert action, the matter field schematically
written as $\phi$ satisfies the appropriate boundary conditions relevant
for the matter distribution, and we assume a renormalization scale
at $\Lambda$. Since we are going to resolve the one-particle thermal
states with masses of order the freeze out temperature $T_{f}$, we
should have $\Lambda\gtrsim T_{f}$. Semiclassically expanding about
the classical path $\phi_{0}$ on the right hand side of Eq.~(\ref{eq:effectiveactiondef}),
we have
\begin{equation}
	e^{iS_{\rm eff}[g]} = e^{i(S_{EH}[g]+S_{M}[g,\phi_{0}])}\mathcal{N}\int D_{\Lambda} \, \delta\phi \, e^{i\int d^4 x\frac{\delta\phi^{2}(x)}{2}\frac{\delta^{2}S_{M}[g,\phi]}{\delta\phi^{2}(x)}|_{\phi=\phi_{0}}+ \ldots }\end{equation}
where the path integral will have the usual perturbative renormalization.
Hence, one can consider the physical observables to be defined through
\begin{equation}
T_{\mu\nu}(y)=\frac{2}{\sqrt{g(y)}}\frac{\delta}{\delta g^{\mu\nu}(y)}\left(S_{M}[g,\phi_{0}]-i\ln\left\{ \mathcal{N}\int D_{\Lambda} \, \delta\phi\exp\left[i\int d^4 x\frac{\delta\phi^{2}(x)}{2}\frac{\delta^{2}S_{M}[g,\phi]}{\delta\phi^{2}(x)}|_{\phi=\phi_{0}}+\ldots \right]\right\} \right).\end{equation}
Note that in practice, we are expanding $g_{\mu\nu}$ perturbatively
about a homogeneous and isotropic FRW background before doing the
path integral. Hence, the inhomogeneities can be computed using classical
perturbation theory and the renormalization scale need not be at $\Lambda=H_{PT}$
even though it is at length scales longer than $H_{PT}^{-1}$ for
which homogeneity and isotropy are typically a good assumptions.

% sub:simplederiv
\section{Derivation of Eq.~(\ref{eq:goodformula})\label{sub:simplederiv}}

Start with the thermally averaged Boltzmann equation for $n_X(t)$ 
\begin{equation}
	\frac{1}{a^{3}}\frac{d}{dt}\left(n_{X} a^3\right)=-\langle\sigma v\rangle\left(n_{X}^{2}-n_{X}^{\text{eq} \, 2}\right)
\end{equation}
which says that $n_{X}$ tracks the equilibrium number density $n_{X}^{\text{eq}}$ until freeze out occurs at $t = t_f$.  
Long after freeze out, the equilibrium term can be neglected, and the equation asymptotically approaches
\begin{equation}
	\frac{d}{dt} \left( n_{X} a^3 \right) =-\langle\sigma v\rangle \left( n_{X} a^3 \right)^{2}\frac{1}{a^{3}} \, .
\end{equation}
One can solve for $n_X(t_0)$ by integrating
\begin{equation}\label{eq:relicabundanceeq}
	n_{X}(t_{0})=\frac{n_{X}(t_{f})\left(\frac{a_{f}}{a_{0}}\right)^{3}}{1+n_{X}(t_{f})\left(\frac{a_{f}}{a_{0}}\right)^{3}\int_{t_{f}}^{t_{0}} dt \sv
\frac{a_{0}^{3}}{a^{3}}} \, .
\end{equation}
The integral in the denominator accounts for residual annihilations of dark matter particles after freeze out.  
The freeze out time $t_{f}$ is not fundamental but instead an artifact of defining when the solution deviates ``significantly'' from the equilibrium distribution.   
For temperatures away from resonances and thresholds, one can typically parameterize $\sv$ as 
\begin{equation}
	\langle\sigma v\rangle=\tilde{a}+\tilde{b}\frac{T}{m_{X}},
\end{equation}
where $T$ is the temperature and $m_{X}$ is the mass of the dark matter.  
To further reduce \eref{eq:relicabundanceeq} we apply \eref{eq:Tf_def}, which implicitly defines $t_f$, and approximate $n_X(t_f) \approx n_X^{\mathrm{eq}}(t_f)$.  
Then, the denominator of \eref{eq:relicabundanceeq} satisfies
\begin{equation}\label{eq:neglect_1_in_denom}
	n_{X}(t_{f}) \left( \frac{a_{f}}{a_0} \right)^{3}\int_{t_{f}}^{t_{0}} dt \sv \frac{a_0^3}{a^{3}}\approx\frac{m_{X}}{T_{f}}\left(\frac{\tilde{a}+\frac{\tilde{b}}{2}\frac{T_{f}}{m_{X}}}{\tilde{a}+\tilde{b}\frac{T}{m_{X}}}\right)\gg1 
\end{equation}
for $T_f \approx m_X / 20$ the freeze out temperature.  
Using this approximation we can express the relic abundance as
\begin{equation}
	n_{X}(t_{0})=\left(\int_{0}^{\ln a_{0}/a_{f}}\frac{d\ln(a/a_{f})}{H}\langle\sigma v\rangle\frac{a_{0}^{3}}{a^{3}}\right)^{-1} 
\end{equation}
after also applying $dt = H^{-1} \, d \ln a$.

% sub:genotgs
\section{Difference Between Entropy and Energy Degrees of Freedom\label{sub:genotgs}}

In this appendix, we show that as the universe expands adiabatically during radiation domination, the relationship $g_{E}(T)=g_{S}(T)$ hold iff
\begin{equation}
	\frac{d\ln g_{E}}{d\ln T} = \frac{d\ln g_{S}}{d\ln T}=0
\end{equation}
where $g_{E}$ is the effective number of degrees of freedom for the thermal energy density and $g_{S}$ is the effective number of degrees of freedom for the entropy density. We also justify an ansatz that can be used to relate $g_{E}$ and $g_{S}$.

Assume that the CC energy density is negligible so that $\rho \approx \rho_R$, which is the case sufficiently far before or after the PT.  The entropy and energy densities of a gas are related by \eref{eq:therm_defs_rho}, which can be written as 
\begin{equation}\label{eq:entropydensity-1}
	\rho + P - T \, s = 0
\end{equation}
where the pressure $P$ of the gas is given by $P(T) = - \mathcal{F}(T)$.  The functions $g_E$ and $g_S$ representing the number of relativistic degrees of freedom were defined by \eref{eq:gE_def} and \eref{eq:gS_def} and are reproduced here for convenience: 
\begin{align}
	\rho =\frac{\pi^{2}}{30} \, g_{E}(T) \, T^{4}  \qquad {\rm and} \qquad
	s =\frac{2\pi^{2}}{45} \, g_{S}(T) \, T^{3} \, .\label{eq:modifiedentropydef} 
\end{align}
As the universe expands, energy conservation is enforced by 
\begin{align}
	d \left( \rho a^3 \right) + P \, d a^3 = 0 \, .
\end{align}
Using \eref{eq:entropydensity-1} and \eref{eq:modifiedentropydef} this becomes
%\begin{equation}\label{eq:generaleq}
%	a\frac{d}{da}\rho+3 \, T \, s=0  
%\end{equation}
%which using \eref{eq:modifiedentropydef} becomes 
\begin{equation}
	\frac{d\ln g_{E}(T)}{d\ln a}+4\frac{d\ln T}{d\ln a}+4 \frac{g_S}{g_E}=0 \, ,
\end{equation}
which can be resolved as
\begin{eqnarray}
	\frac{d\ln T}{d\ln a} & = & - \frac{g_{S}}{g_{E}} \, \left[1+\frac{1}{4}\frac{d\ln g_{E}(T)}{d\ln T}\right]^{-1}.
\end{eqnarray}
Next, impose adiabaticity $d (s a^3) / da = 0$ by first using Eq.~(\ref{eq:modifiedentropydef}) to write 
\begin{eqnarray}
	\frac{d\ln(s \, a^{3})}{d\ln a} & = & -\left[ \frac{d\ln g_{S}}{d\ln T}+3 \right] \frac{g_S}{g_{E} } \left[1+\frac{1}{4}\frac{d\ln g_{E}(T)}{d\ln T}\right]^{-1}+3 \, ,
\end{eqnarray}
and then setting this to zero and solving to find
\begin{align}\label{eq:gerelatedtogs}
%	\frac{d\ln g_{E}}{d\ln T}=-4 \left(1-\frac{g_{S}}{g_{E}} \right)+\frac{4}{3} \frac{g_S}{g_{E}} \frac{d\ln g_{S}}{d\ln T} \, . \\
	\frac{g_S}{g_E} = 1 + \frac{1}{4} \frac{d\ln g_{E}}{d\ln T} - \frac{1}{3} \frac{d\ln g_{S}}{d\ln T} \, .
\end{align}
This equation implies $g_{E}=g_{S}$ iff
\begin{equation}\label{eq:dlngdlnT}
	\frac{d\ln g_{E}}{d\ln T} = \frac{d\ln g_{S}}{d\ln T}=0 
\end{equation}
as claimed.

To obtain some intuition for this theorem consider the SM electroweak 
PT.  Before the PT, the entire spectrum is massless and 
\eref{eq:dlngdlnT} is satisfied exactly
so $g_E(T) = g_S(T) = \mathrm{const}$ for $T > T_{PT}$.  
After the PT, we can estimate how much difference between $g_{S}$
and $g_{E}$ is required for self-consistency and to justify an intuitive
parameterization, by considering a hypothetical situation in which one can
approximate
\begin{equation}
	g_{E/S}(T)=g_{E/S}(T_{i})\left[T/T_{i}\right]^{-12K}\label{eq:gescaling}
\end{equation}
where $K$ is a constant and $T_{i}$ is an initial condition temperature.
Then, one can solve Eq.~(\ref{eq:gerelatedtogs}) as
\begin{equation}
	\frac{g_{S}(T)}{g_{E}(T)}=\frac{1-3K}{1-4K}.
\end{equation}
Hence, if $0<K\ll1$, we have a situation in which $g_{E}(T)$ decreases
slowly as a function of time while satisfying both entropy conservation
and $g_{S}(T)\approx g_{E}(T)$. Presumably, $K$ can be viewed as
a leading term in a Taylor expansion regarding $g_{S}/g_{E}$. Hence,
we will approximate
\begin{equation}
	g_{S}(T)\approx(1+K) \, g_{E}(T)
\end{equation}
even though we are not necessarily making the assumption of Eq.~(\ref{eq:gescaling})
throughout the paper.

% sub:deriveTofa
\section{Derivation of $T_{PT}^+$, $\Delta s$, and $T(a)$}\label{sub:deriveTofa}

To find $T(a)$, we start with the temperature before the PT $T_{PT}^-$ and impose energy conservation to solve for the temperature after the PT $T_{PT}^+$.  
This allows us to calculate $\Delta s$ and $\epsilon_2$ in terms of $\Delta \rho_{\rm ex}$.  
Then, we require the entropy per comoving volume $S = s \, a^3$ to be conserved before and after the PT to find $T(a)$.

%Let $v_{ini} = v^{(s)}(T_{PT}^-)$ be the scalar field expectation value in the symmetric phase just before the PT and $v_{fin} = v^{(b)}(T_{PT}^+)$ be the broken phase expectation value just after the PT.  
Assuming that there is a negligible change in $a$ during reheating, we can impose energy conservation at $a_{PT}$.  
Using Eqs. (\ref{eq:L_def}), (\ref{eq:rhoR_def}), and (\ref{eq:gE_def}), energy conservation can be written as
\begin{align}\label{eq:EnergyConsv}
%	\frac{\pi^2}{30} \, g_E \left( T_{PT}^-, v_{ini} \right) \left( T_{PT}^- \right)^4 + L = \frac{\pi^2}{30} \, g_E \left( T_{PT}^+, v_{fin} \right) \left( T_{PT}^+ \right)^4  \, ,
	\frac{\pi^2}{30} \, g_E^{(s)} \left( T_{PT}^- \right) \left( T_{PT}^- \right)^4 + \Delta \rho_{\rm ex} = \frac{\pi^2}{30} \, g_E^{(b)} \left( T_{PT}^+ \right) \left( T_{PT}^+ \right)^4  \, ,
\end{align}
which implicitly defines $T_{PT}^+$.   
This equation can be solved analytically by expanding $T_{PT}^+ = T_{PT}^- \left( 1 + \Delta \tau \right)$ and linearizing in $\Delta \tau$ along with other small quantities.  
Using \eref{eq:gES_param} to expand $g_E^{(s)}(T)$ around $g_E^{(s)}(T_f)$, \eref{eq:EnergyConsv} becomes
\begin{align}
%	& \frac{\pi^2}{30} \left[ \lim_{\epsilon \to 0} h(a_{PT}/a_f+ \epsilon) - h(a_{PT}/a_f - \epsilon) \right] \left(T_{PT}^- \right)^4 + L \approx 4 \frac{\pi^2}{30} g_E(T_f) \left( T_{PT}^- \right)^4 \Delta \tau \, .
	& \frac{\pi^2}{30} \left[ \lim_{\epsilon \to 0} h(a_{PT}+ \epsilon) - h(a_{PT} - \epsilon) \right] \left(T_{PT}^- \right)^4 + \Delta \rho_{\rm ex} \approx 4 \frac{\pi^2}{30} g_E^{(s)}(T_f) \left( T_{PT}^- \right)^4 \Delta \tau \, .
\end{align}
where we have dropped higher order terms.  
Using \eref{eq:h_def}, the term in brackets is $(7/8) N_{PT}$.  Finally, 
the equation can be solved for $\Delta \tau = T_{PT}^+ / T_{PT}^- - 1$ to obtain 
\begin{align}\label{eq:TPTplus}
	T_{PT}^+ & \approx T_{PT}^- \left[1
	+ \frac{1}{4} \, \epsilon_{31} 
%	+ \frac{1}{4} \frac{L}{\frac{\pi^2}{30} \, g_E(T_f) \, (T_{PT}^-)^4}
	+ \frac{1}{4} \frac{\Delta \rho_{\rm ex}}{\frac{\pi^2}{30} \, g_E^{(s)}(T_f) \, (T_{PT}^-)^4}
	\right]
\end{align}
where $\epsilon_{31}$ is given by \eref{eq:fractionaldecouplingduringPT}.  
As expected, the energy released $\Delta \rho_{\rm ex} > 0$ controls the reheating from $T_{PT}^-$ to $T_{PT}^+$.  
Additionally, the reheating is larger when more particles non-adiabatically decouple (larger $\epsilon_{31}$), because the latent heat is distributed over fewer degrees of freedom after the PT.

Next we can calculate the entropy density increase at the PT given by
\begin{align}
%	\Delta s & \equiv s(T_{PT}^+, v_{fin}) - s(T_{PT}^-, v_{ini}) \label{eq:Deltas_def} \\
%	& =\frac{2\pi^{2}}{45}\left\{g_{S}(T_{PT}^{+}, v_{fin})(T_{PT}^{+})^{3}-g_{S}(T_{PT}^{-}, v_{ini})(T_{PT}^{-})^{3}\right\} \, .
	\Delta s & \equiv s^{(b)}(T_{PT}^+) - s^{(s)}(T_{PT}^-) \label{eq:Deltas_def} \\
	& =\frac{2\pi^{2}}{45}\left\{g_{S}^{(b)}(T_{PT}^{+})(T_{PT}^{+})^{3}-g_{S}^{(s)}(T_{PT}^{-})(T_{PT}^{-})^{3}\right\} \, .
\end{align}
Once again we will linearize in the perturbation by expanding $g_S$ using \eref{eq:gES_param} and writing $T_{PT}^+$ using \eref{eq:TPTplus}.  This gives 
%\begin{align}
%	\Delta s & \approx \frac{2 \pi^2}{45} \left\{
%	-\frac{1}{g_S(T_f)} \left[\lim_{\epsilon \to 0} h(a_{PT}/a_f+ \epsilon) - h(a_{PT}/a_f - \epsilon) \right]
%	+ 3  \Delta \tau
%	\right\} g_S(T_f) \left( T_{PT}^- \right)^3 \\
%	& \approx \frac{2 \pi^2}{45} \left\{
%	-\frac{g_E(T_f)}{g_S(T_f)} \epsilon_{31}
%	+ 3 \left[ 
%	\frac{1}{4} \, \epsilon_{31} + \frac{1}{4} \frac{L}{\frac{\pi^2}{30} g_E(T_f) (T_{PT}^-)^4}
%	\right]
%	\right\} g_S(T_f) \left( T_{PT}^- \right)^3
%\end{align}
\begin{align}
	\Delta s & \approx \frac{2 \pi^2}{45} \left\{
	-\frac{1}{g_S^{(s)}(T_f)} \left[\lim_{\epsilon \to 0} h(a_{PT}+ \epsilon) - h(a_{PT} - \epsilon) \right]
	+ 3  \Delta \tau
	\right\} g_S^{(s)}(T_f) \left( T_{PT}^- \right)^3 \\
	& \approx \frac{2 \pi^2}{45} \left\{
	-\frac{g_E^{(s)}(T_f)}{g_S^{(s)}(T_f)} \epsilon_{31}
	+ 3 \left[ 
	\frac{1}{4} \, \epsilon_{31} + \frac{1}{4} \frac{\Delta \rho_{\rm ex}}{\frac{\pi^2}{30} g_E^{(s)}(T_f) (T_{PT}^-)^4}
	\right]
	\right\} g_S^{(s)}(T_f) \left( T_{PT}^- \right)^3
\end{align}
As discussed in Section \ref{sub:genotgs}, we can approximate $g_S^{(s)}(T_f) \approx g_E^{(s)}(T_f)$.  
Then finally $\Delta s$ becomes
\begin{align}\label{eq:Deltas_intermsof_L}
%	\Delta s & \approx \frac{2 \pi^2}{45} g_S(T_f) \left( T_{PT}^- \right)^3 \left[ 
%	-\frac{1}{4} \, \epsilon_{31} + \frac{1}{4} \frac{L}{\frac{\pi^2}{30} g_E(T_f) (T_{PT}^-)^4} \right] \, .
	\Delta s & \approx \frac{2 \pi^2}{45} g_S^{(s)}(T_f) \left( T_{PT}^- \right)^3 \left[ 
	-\frac{1}{4} \, \epsilon_{31} + \frac{1}{4} \frac{\Delta \rho_{\rm ex}}{\frac{\pi^2}{30} g_E^{(s)}(T_f) (T_{PT}^-)^4} \right] \, .
\end{align}
Using \eref{eq:fractionalentropyduringpt} and noting $T_{PT}^- \, a_{PT} = T_f \, a_f$ up to higher order terms, we also obtain 
\begin{align}\label{eq:epsilon2_intermsof_L}
%	\epsilon_2 & \approx- \frac{1}{4} \, \epsilon_{31} + \frac{1}{4} \frac{L}{\frac{\pi^2}{30} g_E(T_f) (T_{PT}^-)^4} \, .
	\epsilon_2 & \approx- \frac{1}{4} \, \epsilon_{31} + \frac{1}{4} \frac{\Delta \rho_{\rm ex}}{\frac{\pi^2}{30} g_E^{(s)}(T_f) (T_{PT}^-)^4} \, .
\end{align}
Both of these equations illustrate that the entropy increase at the PT is  controlled by the amount of latent heat released and the number of particles that non-adiabatically decouple.

Finally we will solve the equation of entropy conservation for $T(a)$.  The entropy per comoving volume $S = s \, a^3$ is conserved excepting the entropy injection at reheating which is assumed to occur rapidly at $a_{PT}$.  Entropy conservation may be expressed as
\begin{align}\label{eq:EntropyConsv}
%	g_{S}(T, v_T) \, T^{3}a^{3}=g_{S}(T_{f}) \, T_{f}^{3}a_{f}^{3}+\Theta(a-a_{PT}) \, a_{PT}^{3} \, \left( \frac{2 \pi^2}{45} \right)^{-1} \Delta s \, ,
%	g_{S}^{(s)}(T) \, T^{3}a^{3} &=g_{S}^{(s)}(T_{f}) \, T_{f}^{3}a_{f}^{3} \\
%	g_{S}^{(b)}(T) \, T^{3}a^{3} &=g_{S}^{(s)}(T_{f}) \, T_{f}^{3}a_{f}^{3}+ a_{PT}^{3} \, \left( \frac{2 \pi^2}{45} \right)^{-1} \Delta s
	g_{S}(a) \, T(a)^{3}a^{3}=g_{S}^{(s)}(T_{f}) \, T_{f}^{3}a_{f}^{3}+\Theta(a-a_{PT}) \, a_{PT}^{3} \, \left( \frac{2 \pi^2}{45} \right)^{-1} \Delta s \, ,
\end{align}
which implicitly defines $T(a)$.  
To solve for $T$ we use \eref{eq:gES_param} to expand $g_S(a)$ then linearize in $h$ and $\Delta s$ to obtain 
\begin{align}
	T(a) \approx T_f \frac{a_f}{a} \left[ 1 
%	+ \frac{1}{3} \frac{h(a/a_f)}{g_S(T_f)} 
	+ \frac{1}{3} \frac{h(a)}{g_S^{(s)}(T_f)} 
%	+ \Theta \left( a - a_{PT} \right) \frac{1}{3} \left( \frac{a_{PT}}{a_f} \right)^3 \frac{\Delta s}{\frac{2 \pi^2}{45} g_S(T_f) \, T_f^3} \right]
	+ \Theta \left( a - a_{PT} \right) \frac{1}{3} \left( \frac{a_{PT}}{a_f} \right)^3 \frac{\Delta s}{\frac{2 \pi^2}{45} g_S^{(s)}(T_f) \, T_f^3} \right]
\end{align}
Further expanding $h$ using \eref{eq:h_def}, approximating $g_S(T_f) \approx g_E(T_f)$, and applying \eref{eq:fractionalentropyduringpt} we obtain the final expression,
\begin{align}\label{eq:Tofa}
	T(a) \approx T_f \frac{a_f}{a} \left[ 1 
%	+ \frac{1}{3} \epsilon_{32} \, f(a / a_f)
	+ \frac{1}{3} \epsilon_{32} \, f(a )
	+ \Theta \left( a - a_{PT} \right) \frac{1}{3} \left( 
	 \epsilon_{31} 
	+ \epsilon_2
	\right)
	\right]
\end{align}
After the PT, the exotic energy component behaves approximately adiabatically.

\section{Derivation of PT induced change in the degree of freedom\label{sub:Derivation-of-PT-induced-dof-change}}

We begin with the well-known formula for the energy density of a gas of fermions at temperature $T$ with $N$ dynamical degrees of freedom:
\begin{align}
	\rho(T) = N \int \frac{d^3 p}{(2 \pi)^3} \frac{E_p}{1 + e^{E_p/T}} \, .
\end{align}
The gas has an effective number of degrees of freedom $g_E$ given implicitly by $ \rho(T) =\frac{\pi^{2}}{30} \, g_{E}(T) \, T^{4}$.  
We can parameterize the decrease in $g_E$ due to the decoupling of the fermionic gas by writing 
\begin{align}
	g_E(T) = g_E(T_f) - \frac{7}{8} \, N f \left( a/a_f \right)
\end{align}
where 
\begin{equation}
	f\left(a\right)=\left(\frac{7}{8} \frac{\pi^2}{30}\right)^{-1}\int\frac{d^{3}p}{(2\pi)^{3}}E_{p}\left[\frac{1}{T_{f}^{4}}\frac{1}{e^{\frac{E_{p}}{T_{f}}}+1}-\frac{1}{T^4(a)}\frac{1}{e^{\frac{E_{p}}{T(a)}}+1}\right]\label{eq:fxeq} \, .
\end{equation}
The temperature $T = T(a)$ is given by \eref{eq:Tofa} to leading order in the perturbations $\epsilon_i$.  
Since $f$ already multiplies a small term in \eref{eq:h_def}, we need only keep the leading factor in \eref{eq:Tofa} which is $T = T_f \, a_f / a$.  
This lets us write \eref{eq:fxeq} as 
\begin{equation}
	f\left(a\right)=\frac{8}{7}\left(\frac{30}{\pi^{2}}\right)\int\frac{d^{3}p}{(2\pi)^{3}}\frac{E_{p}}{T_{f}^{4}}\left[\frac{1}{e^{\frac{E_{p}}{T_{f}}}+1}-\frac{(a/a_f)^{4}}{e^{\frac{aE_{p}}{a_f T_{f}}}+1}\right]\label{eq:fxeqapproxexpl} \, .
\end{equation}
Note that $f(a)$ increases from $f(a_f)=0$ to $f(\infty) \approx 1$.  
Due to the exponential temperature dependence, the transition to $f \approx 1$ occurs at $T \approx m_{N}$ and is smoothly steplike over a time scale $\Delta t \approx 1/H$. 
In this discussion we have assumed $E_p = \sqrt{ {\bf p}^2 + m_N^2}$ with $m_N$ constant, that is, we neglect any change in the mass of the particle as a function of 
time. 
This assumption is valid sufficiently far after the PT such that the scalar field expectation value and field-dependent masses have approximately stopped varying.

% sub:SMeffpot
%\section{Standard Model Thermal Effective Potential}\label{sub:SMeffpot}
\section{Thermal Effective Potential Details}\label{app:EffPot}

We have calculated the thermal effective potential through one-loop order for each of the models in Section \ref{sec:IllustrativeModels}.  
Our calculation employs the standard techniques \cite{Coleman:1973jx, Dolan:1973qd, Jackiw:1974cv}, and the case of the Standard Model is particularly well documented \cite{Carrington:1991hz, Ford:1992mv, Arnold:1992rz}.  
As such, we do not feel the need to reproduce the entire calculation here.  
However, we have chosen to use renormalization schemes which are convenient for our calculation, but not standardly employed.  
Hence, we will use this appendix to write down the thermal effective potentials for each of the models in Section \ref{sec:IllustrativeModels} and to spell out our renormalization conditions.  

%For all the thermal corrections to the scalar effective potential, we
%do not include the dark matter thermal corrections as such neglect
%constitutes an excellent approximation to the approximately factor of
%2 accuracy that we are trying to achieve in most of the computation
%below.
In calculating thermal corrections to the scalar effective potential, we
do not include contributions from the dark matter sector.  
This is an excellent approximation provided that freeze out occurs 
prior to the phase transition (as we have assumed), such that the dark 
matter is decoupled from the plasma during the phase transition.

%sec:TEP_SM
\subsection{Thermal Effective Potential: Standard Model}\label{sec:TEP_SM}

Let $h(x) = \sqrt{2} \left| H^{\dagger} H \right|^{1/2}$ be the radial component of the SM Higgs field and let $h_c = \left< h \right>$.  
In calculating the radiative corrections, we need not include the contributions from every field in the Standard Model.  
With regards to the non-thermal corrections, light particles which couple weakly to the Higgs can be neglected, and with regards to the thermal corrections, particles which are light and do not decouple during freeze out can be treated as massless.  
Since we expect that freeze out will coincide with the PT at a mass scale of about $100 \GeV$ and that residual annihilations will occur down to a mass scale of about $10 \GeV$, we can neglect particles with a mass below that of the bottom quark (i.e., 4.2 GeV).  
We retain the top quark, bottom quark, physical Higgs, and massive gauge bosons\footnote{We work in the Landau gauge ($\xi = 0$) for which the scalar polarization mode and ghost propagators are independent of $h_c$ \cite{Coleman:1973jx}.  } which have field dependent masses
\begin{subequations}\label{eq:SM_masses}
\begin{align}
	M^2_{t / b / Z / W }(h_c) & = \left( \frac{m_{t / b / Z / W}}{v} \right)^2 \, h_c^2 \\
	M^2_h(h_c) & = \frac{m_h^2}{2 v^2} \left( 3 h_c^2 - v^2 \right)  \label{eq:SM_masses_Mh}
%	M^2_{\chi}(h) & = -\frac{m_h^2}{2 v^2} \left( v^2 - h^2 \right)
\end{align}
\end{subequations}
where $m_t = 172.6 \GeV, m_b = 4.2 \GeV, m_Z = 91.2 \GeV$, and $m_W = 80.4 \GeV$ \cite{PDG}.  
The non-thermal corrections can be expressed as functions of the Coleman-Weinberg potential \cite{Coleman:1973jx}.  
Regulating in $(d = 4 - 2 \epsilon)$ spacetime dimensions, the unrenormalized potential is given by 
\begin{align}
	V_{\rm cw} (M^2)  & = \frac{M^4}{64 \pi^2} \left( \log \frac{M^2}{\mu^2} - \frac{3}{2} - C_{\rm uv} \right)
\end{align}
where $C_{\rm uv} = \epsilon^{-1} - \gamma_E + \ln 4 \pi$ and $\mu$ is the t'Hooft scale.  
The thermal corrections can be expressed in terms of the bosonic and fermionic thermal functions \cite{Dolan:1973qd, Kapusta:1989}
\begin{subequations}\label{eq:ThermalFuncs}
\begin{align}
	J_B \left( y \right) & \equiv \int_0^{\infty} dx \,  x^2 \log \left( 1 - e ^{- \sqrt{x^2 + y}} \right)  = -\sum_{n=1}^{\infty} \frac{1}{n^2} y \, K_2 \left( n \sqrt{y} \right) \label{eq:JB_def} \\
	J_F \left( y \right) & \equiv \int_0^{\infty} dx \,  x^2 \log \left( 1 + e ^{- \sqrt{x^2 + y}} \right) = -\sum_{n=1}^{\infty} \frac{\left( -1 \right)^n}{n^2} y \, K_2 \left( n \sqrt{y} \right)  \label{eq:JF_def} 
\end{align}
\end{subequations}
where $K_2(z)$ is the modified Bessel function of the second kind.  
Putting the pieces together, the Standard Model thermal effective potential (through one-loop order and before renormalization) is given by 
\begin{align}\label{eq:VeffT_SM}
	V_{\rm eff}^{\rm (SM)}(h_c,T) & \approx 
	\frac{m_h^2}{8 v^2} \left( h_c^2 - v^2 \right)^2 
	 + \Biggl\{ \delta \Omega + \frac{1}{2} \delta m^2 \, h_c^2 
	+ \frac{\delta \lambda}{4} \, h_c^4 \nonumber \\
	& - 12 \, V_{\rm cw} \left(M^2_t(h_c) \right) 
	- 12 \, V_{\rm cw} \left( M^2_b(h_c) \right)  %\nonumber \\ &
	+ 3 \, V_{\rm cw} \left( M^2_Z(h_c) \right)  
	+ 6 \, V_{\rm cw} \left( M^2_W(h_c) \right) 
	+ V_{\rm cw} \left( M^2_h(h_c) \right) \Biggr\}  \nonumber \\
	&+ \Biggl\{ - \frac{\pi^2}{90} 75.75 \, T^4
	+ \frac{T^4}{2 \pi^2} \Bigl[ 
	-12 \, J_F \left( M_t^2(h_c) \, T^{-2} \right)
	-12 \, J_F \left( M_b^2(h_c) \, T^{-2} \right) 
	\nonumber \\
	&   
	+ 3 \, J_B \left( M_Z^2(h_c) \, T^{-2} \right) 
	+ 6 \, J_B \left( M_W^2(h_c) \, T^{-2} \right) 
	+ J_B \left( M_h^2(h_c) \, T^{-2} \right) 
	\Bigr] \Biggr\}+ \ord{\hbar^2} 
\end{align}
where $\delta \Omega$, $\delta m^2$, and $\delta \lambda$ are counterterms.  
We have also included the term $(- \frac{\pi^2}{90} 75.75 \, T^4)$, which represents the thermal radiative contribution from light quarks, leptons, and massless gauge bosons which are relativistic at temperatures $T \gtrsim 10 \GeV$.  
The renormalization conditions, 
\begin{subequations}\label{eq:SM_rencondit}
\begin{align}
	\left. \frac{\partial}{\partial h_c} V_{\rm eff}^{\rm (SM)}(h_c, 0) \right|_{h_c = v} &= 0 \\
	\left. \frac{\partial^2}{\partial h_c^2} V_{\rm eff}^{\rm (SM)}(h_c, 0) \right|_{h_c = v} &= m_h^2 \\
	V_{\rm eff}^{\rm (SM)}(v, 0) &= 0 \, ,
\end{align}
\end{subequations}
are chosen such that tadpole graphs vanish and $V_{\rm eff}^{\rm (SM)}(h_c, 0)$ has a minimum at $h_c = v$, self-energy graphs vanish and the Higgs mass\footnote{Since the effective potential is computed from diagrams with zero external momentum, the mass $\partial_{h_c}^2 V_{\rm eff} (h_c=v, 0) = m_h^2 $ differs from the Higgs pole mass by logarithmic corrections \cite{Casas:1994us}, which we verify are $\ord{\mathrm{few}\  \%}$.  As such, we will neglect this distinction and continue to refer to $m_h$ as the ``Higgs mass.''  } is $m_h$, and the CC is tuned against the vacuum energy density to zero.  
%Solving \eref{eq:SM_rencondit} for the counterterms, one obtains
%\begin{subequations}\label{eq:ct_SM}
%\begin{align}
%	\delta \Omega & = -V_{\rm cw}(v) + \frac{5}{8} v \left. \frac{\partial V_{\rm cw}}{\partial h} \right|_{v} - \frac{1}{8} v^2 \left. \frac{\partial^2 V_{\rm cw}}{\partial h^2} \right|_{v} \\
%	\delta m^2 & = - \frac{3}{2 v} \left. \frac{\partial V_{\rm cw}}{\partial h} \right|_{v} + \frac{1}{2} \left. \frac{\partial^2 V_{\rm cw}}{\partial h^2} \right|_{v} \\
%	\delta \lambda & = \frac{1}{2 v^3} \left. \frac{\partial V_{\rm cw}}{\partial h} \right|_{v} - \frac{1}{2v^2} \left. \frac{\partial^2 V_{\rm cw}}{\partial h^2} \right|_{v}  .
%\end{align}
%\end{subequations}

%sec:TEP_Z2xSM
\subsection{Thermal Effective Potential: $\mathbb{Z}_2$xSM}\label{sec:TEP_Z2xSM}

The $\mathbb{Z}_2$xSM potential was specified by \eref{eq:V0_Z2xSM}.  
Since we focus on the case $\left< s \right> = 0$, we need only calculate the effective potential as a function of $h_c$ and not $s_c = \left< s \right>$.  
That is, the presence of the singlet in this model simply add an additional degree of freedom, with field dependent mass 
\begin{align}
	M_s^2(h_c) & = \left( m_s^2 - a_2 v^2 \right) + a_2 h_c^2  \, ,
\end{align}
to the radiative corrections.  
We can construct the effective potential from the SM effective potential \eref{eq:VeffT_SM} as
\begin{align}
	V_{\rm eff}^{(\mathbb{Z}_2{\rm xSM})} \left( h_c, T \right) =& 
	V_{\rm eff}^{\rm (SM)}(h_c, T) 
%	+ \frac{b_4}{4} s^4 + \frac{1}{2} m_s^2 s^2 +\frac{a_2}{2} s^2 \left( h^2 - v^2 \right) \nonumber \\
%	+ \left[ 
%	V_{\rm cw} \left( \overline{M}^2_h(h_c, s_c) \right)
%	- V_{\rm cw} \left( M^2_h(h_c) \right) 
%	+ V_{\rm cw} \left( M^2_s( h_c, s_c) \right)
%	\right] \nonumber \\
	+ V_{\rm cw} \left( M^2_s( h_c) \right)
	+ \frac{T^4}{2 \pi^2} J_B \left( M^2_s(h_c) T^{-2} \right) \, .
	% \nonumber \\
%	+ \frac{T^4}{2 \pi^2} \left[ 
%	J_B \left( \overline{M}^2_h(h_c, s_c) T^{-2} \right)
%	- J_B \left( M^2_h(h_c) T^{-2} \right) 
%	+ J_B \left( M^2_s(h_c, s_c) T^{-2} \right) 
%	\right]
%	 \nonumber \\
% 	&+ \frac{\delta b_4}{4} s^4 + \frac{1}{2} \delta m_s^2 s^2 +\frac{\delta a_2}{2} s^2 h^2 
\end{align}
%where $\delta b_4$, $\delta m_s^2$, and $\delta a_2$ are counterterms.  
An additional UV divergence arises from the term $V_{\rm cw}(M_s^2)$, and is cancelled by solving the renormalization conditions \eref{eq:SM_rencondit} once again for the counterterms.

%sec:TEP_GenSing
\subsection{Thermal Effective Potential: Generic Singlet}\label{sec:TEP_GenSing}

For the theory specified by the action \eref{eq:GenSing_Sphi}, we have the field dependent masses
\begin{align}
	M_{\varphi}^2(\varphi_c) & = M^2 - 6 \mathcal{E} \varphi_c + 3 \lambda \varphi_c \\
	M_{\psi_i}^2(\varphi_c) & = \left( m_i + h_i \varphi_c \right)^2 \, .
\end{align}
We construct the thermal effective potential as 
\begin{align}\label{eq:GenSing_Veff}
	V_{\rm eff}^{\rm (GS)}(\varphi_c, T) =&
	 \rho_{\rm ex} + \frac{1}{2} M^2 \varphi_c^2 - \mathcal{E} \varphi_c^3 + \frac{\lambda}{4} \varphi_c^4 + \frac{T^4}{2 \pi^2} \left[ 
	J_B \left( M_{\varphi}^2(\varphi_c) \right) - 4 \sum_{i=1}^N J_F\left( M_{\psi_i}^2(\varphi_c) \right)
	 \right]  \\
	 & + \Biggl\{ \delta \Omega + \delta t \, \varphi_c+ \frac{1}{2} \delta M^2 \varphi_c^2 - \delta \mathcal{E} \, \varphi_c^3 + \frac{\delta \lambda}{4} \varphi_c^4 % \nonumber \\ &
	 + V_{\rm cw}\left( M_{\varphi}^2(\varphi_c) \right) - 4 \sum_{i=1}^N V_{\rm cw}\left( M_{\psi_i}^2(\varphi_c) \right) \Biggr\}  \nonumber
%	 &  + \Biggl\{ - \frac{\pi^2}{90} 106.75 \, T^4  + \frac{T^4}{2 \pi^2} \left[ 
%	J_B \left( M_{\varphi}^2(\varphi_c) \right) - 4 \sum_{i=1}^N J_F\left( M_{\psi_i}^2(\varphi_c) \right)
%	 \right] \Biggr\}
\end{align}
where $\delta \Omega$, $\delta t$, $\delta M^2$, $\delta \mathcal{E}$, and $\delta \lambda$ are counterterms.  
We do not renormalize using the same renormalization conditions as we did for the SM.  
To simply the discussions of Section \ref{sec:IllustrativeModels}, we have attempted to choose the renormalization conditions such that the effective potential preserves certain features of the renormalized tree-level potential.  
For example, the renormalization conditions that we applied to the SM, \eref{eq:SM_rencondit}, ensured that the effective potential and the tree-level potential agreed to order $h_c^2$ as an expansion around $h_c = v$.    
In our analysis of Section \ref{sec:GenericSinglet}, we found it convenient to define the parameter $\alpha_0$ which controls the shape of the effective potential.  
This parameter is defined using the tree-level potential $U(\varphi)$, but we claim that it also describes the shape of the one-loop effective potential provided that the radiative corrections do not significantly distort the shape of the potential.  
For the tuned limit $0 \lesssim \alpha_0 \ll 1$, this parameter is particularly sensitive to the shape of the potential near the origin $\varphi_c = 0$ since the barrier is very small.  
The radiative corrections grow as $\varphi_c \to 0$, because the fermions $\psi_i$ become light, but these logarithmic corrections remain subdominant.  
However, with a renormalization scheme of the form of \eref{eq:SM_rencondit}, the counterterms pick up a finite piece, which depends on derivatives of logarithms at the renormalization point $\varphi_c \approx v$, and which contributes non-negligibly near $\varphi_c \approx 0$.  
If we were to use such a renormalization scheme in the limit where $U(\varphi_c)$ has a small barrier so $0 \lesssim \alpha_0 \ll 1$, then the radiative corrections may lift the minimum at $\varphi_c \approx 0$ and eliminate the barrier.  
Of course, there is nothing incorrect with using such a renormalization scheme except that it is inconvenient since we would not be able to characterize the shape of the potential using $\alpha_0$ derived from $U(\varphi_c)$.

In light of this discussion, we will use a renormalization scheme which preserves the location of the minimum at $\varphi_c  = v$ and also preserves the shape of the potential near $\varphi_c = 0$.  
This is accomplished by first writing \eref{eq:GenSing_Veff} for $T = 0$ as 
\begin{align}
	V_{\rm eff}^{\rm (GS)}(\varphi_c, 0) & = 
	\bar{\Omega}(\varphi_c) + \bar{t}(\varphi_c) \varphi_c + \frac{1}{2} \bar{M}^2(\varphi_c) \varphi_c^2 - \bar{\mathcal{E}}(\varphi_c) \varphi_c^3 + \frac{\bar{\lambda}(\varphi_c)}{4} \varphi_c^4
\end{align}
where
\begin{subequations}
\begin{align}
	\bar{\Omega}(\varphi_c) & = \rho_{\rm ex} + \delta \Omega + \frac{\hbar}{4 \pi^2} \Bigl[
	\frac{1}{16} M^4 f_{\varphi}(\varphi_c)  
	- \frac{1}{4} \sum_{i=1}^N m_i^4 f_{\psi_i}(\varphi_c)		
	\Bigr]\\
	\bar{t}(\varphi_c) & = \delta t + \frac{\hbar}{4 \pi^2} \Bigl[
	- \frac{3}{4} \mathcal{E} M^2 f_{\varphi}(\varphi_c) 
	- \sum_{i=1}^N m_i^3 h_i f_{\psi_i}(\varphi_c)		
	\Bigr] \\
	\bar{M}^2(\varphi_c) & = M^2 + \delta M^2 + \frac{\hbar}{4 \pi^2} \Bigl[
	\frac{3}{4} M^2 \lambda f_{\varphi}(\varphi_c) 
	+ \frac{9}{2} \mathcal{E}^2 f_{\varphi}(\varphi_c) 
	- 3 \sum_{i=1}^N  m_i^2 h_i^2 f_{\psi_i}(\varphi_c)		
	\Bigr]\\
	\bar{\mathcal{E}}(\varphi_c) & = \mathcal{E} + \delta \mathcal{E} + \frac{\hbar}{4 \pi^2} \Bigl[ 
	\frac{9}{4} \mathcal{E} \lambda f_{\varphi}(\varphi_c)
	+\sum_{i=1}^N  m_i h_i^3 f_{\psi_i}(\varphi_c)		
	\Bigr] \\
	\bar{\lambda}(\varphi_c) & = \lambda + \delta \lambda + \frac{\hbar}{4 \pi^2} \Bigl[ 
	\frac{9}{4} \lambda^2 f_{\varphi}(\varphi_c)
	- \sum_{i=1}^N h_i^4 f_{\psi_i}(\varphi_c)	 
	\Bigr]
\end{align}
\end{subequations}
and
\begin{align}
	f_{\varphi}(\varphi_c) & = \left( \ln \frac{M_{\varphi}^2(\varphi_c)}{\mu^2} - \frac{3}{2} - C_{\rm uv} \right)  \\
	f_{\psi_i}(\varphi_c) & = \left( \ln \frac{M_{\psi_i}^2(\varphi_c)}{\mu^2} - \frac{3}{2} - C_{\rm uv} \right)  \, .
\end{align}
Then the renormalization conditions can be expressed as
\begin{subequations}\label{eq:GenSing_RenCondit}
\begin{align}
	\bar{\Omega}(v) &= \rho_{\rm ex} \\
	\bar{t}(v) & = 0 \\
	\bar{M}^2(v) & = M^2 \\
	\bar{\mathcal{E}}(v) & = \mathcal{E} \\
	\bar{\lambda}(v) & = \lambda \, .
\end{align}
\end{subequations}
Near $\varphi_c \approx 0$, the radiative corrections are at most logarithmic.  

%sec:TEP_xSM
\subsection{Thermal Effective Potential: xSM}\label{sec:TEP_xSM}

%We calculate the thermal effective potential $V_{\rm eff}(\hs,T)$ using the approach of Section \ref{sub:SMeffpot}, but with the following modifications.  
%The classical potential \eref{eq:V0_SM} is replaced by 
%\begin{align}
%	U(\hs) = \rho_{\rm ex} + \frac{\mu^2}{2} h^2 + \frac{\lambda}{4} h^4 + b_1 s + \frac{b_2}{2} s^2 + \frac{b_3}{3} s^3 + \frac{b_4}{4} s^4 + \frac{a_1}{2} h^2 s + \frac{a_2}{2} h^2 s^2 \, .
%\end{align}
%One of the six parameters $b_i, a_j$ is redundant and can be fixed by shifting the origin along the $s$--axis.  In an alternate parametrization $U$ can be expressed as 
%\begin{align}
%	U(\hs) =& 
%	\rho_{DE} + \frac{m_h^2}{8 v^2} \left( h^2 - v^2 \right)^2 
%	+ \frac{b_4}{4} s^4
%	+ \frac{1}{2} m_s^2 s^2
%	+ \frac{b_3}{3} s^3 
%	+ \frac{1}{2} s \left( h^2 - v^2 \right) \left( a_1 + a_2 s \right)
%\end{align}

In Section \ref{sec:SingletExtension1PT} we wrote down the xSM renormalized potential in \eref{eq:V0_xSM}.  
For general $h_c$ and $s_c$, the Higgs and singlet fields mix.  
In order to calculate the radiative corrections, we must generalize the field-dependent Higgs mass $M_h^2$, given by \eref{eq:SM_masses_Mh}, to the Higgs-singlet mass matrix $M_{hs}^2$, which has components
%\footnote{In the remainder of this section, we use the terms ``Higgs mass'' and ``singlet mass'' to refer to the respective Higgs- and singlet-like eigenvalues of $M^2_{hs}(\vx)$.  The matrix $M_{hs}^2$ and its eigenvalues are unchanged by radiative corrections due to the renormalization conditions.  } 
\begin{subequations}\label{eq:Mhs_xSM}
\begin{align}
	\left[ M_{hs}^2(\left\{ h_c, s_c \right\}) \right]_{11}
	&= m_h^2 / (2 v^2) \left( 3 h_c^2 - v^2 \right) + s_c \left( a_1 + a_2\, s_c \right)  \\
	\left[ M_{hs}^2(\left\{ h_c, s_c \right\} ) \right]_{12}
	&= \left[ M_{hs}^2(\left\{ h_c, s_c \right\}) \right]_{21} 
	= h_c \left( a_1 + 2 \, a_2 \, s_c \right) \\
	\left[ M_{hs}^2(\left\{ h_c, s_c \right\} ) \right]_{22}
	&= m_s^2 + a_2 \left( h_c^2 - v^2 \right) + 2 b_3 \, s_c + 3 b_4 \,  s_c^2 \, .
\end{align}
\end{subequations}
Now we can write down the thermal effective potential in terms of $V_{\rm eff}^{\rm (SM)}$ by subtracting the contribution from the SM Higgs and adding the contribution from the mixed Higgs and singlet.  
Doing so we obtain
\begin{align}
	V_{\rm eff}^{\rm (xSM)}(\left\{ h_c, s_c \right\}, T) 
	=& V_{\rm eff}^{\rm (SM)}(h_c, T)
	+ \frac{b_4}{4} s_c^4 + \frac{1}{2} m_s^2 s_c^2 + \frac{b_3}{3} s_c^3 + \frac{1}{2} s_c \left( h_c^2 - v^2 \right) \left( a_1 + a_2 s_c \right)  \nonumber \\
	&+ \Biggl\{ \frac{\delta b_4}{4} s_c^4 + \frac{\delta b_3}{3} s_c^3 + \frac{1}{2} \delta b_2 s_c^2  + \delta b_1 s_c + \frac{1}{2} \delta a_2 s_c^2 h_c^2 + \frac{1}{2} \delta a_1 s_c h_c^2 + \delta \Omega \nonumber \\
	&- V_{\rm cw}\left( M_h^2(h_c) \right) + {\rm Tr} \, V_{\rm cw}\left( M_{hs}^2( \left\{ h_c, s_c \right\}) \right) \Biggr\}  \nonumber \\
	&+ \frac{T^4}{2\pi^2} \left[ - J_B\left( M_h^2(h_c) T^{-2} \right) + {\rm Tr} \, J_B\left( M_{hs}^2(\left\{ h_c, s_c \right\}) T^{-2} \right) \right]
\end{align}
where $\delta \Omega$, $\delta b_i$, and $\delta a_i$ are counterterms.
The trace is interpreted to mean evaluating $V_{\rm cw}$ or $J_B$ with the eigenvalues of $M_{hs}^2$.  
We generalize the SM renormalization conditions \eref{eq:SM_rencondit} to incorporate the additional fields, 
\begin{align}\label{eq:xSM_rencondit}
	& \left( \frac{\partial}{\partial h_c} \right)^{n_h}
	\left( \frac{\partial}{\partial s_c} \right)^{n_s}
	\left. V_{\rm eff}^{\rm (xSM)} (\left\{ h_c, s_c \right\}) \right|_{\vx} 
	=  \left( \frac{\partial}{\partial h_c} \right)^{n_h}
	\left( \frac{\partial}{\partial s_c} \right)^{n_s}
	\left. U(\left\{ h_c, s_c \right\}) \right|_{\vx} \nonumber \\
	\left\{ n_h, n_s \right\} &  = \left\{1, 0 \right\}, 
	\left\{2, 0 \right\}, \left\{0, 1 \right\}, \left\{0, 2 \right\}, \left\{1, 1 \right\}, 
	\left\{1, 2 \right\}, \left\{0, 3 \right\}, \left\{0, 4 \right\}, \left\{0, 0 \right\} 
\end{align}
where $U(\left\{ h_c, s_c \right\})$ is given by \eref{eq:V0_xSM}.
Once again, we require $V_{\rm eff}^{\rm (xSM)}(\vx,0) =0$ which tunes the CC.

\section{xSM Bounce Calculation}\label{sub:xSMBounce}

As discussed in Section \ref{sec:SingletExtension1PT}, the xSM electroweak PT is first order in the parametric regime of interest and proceeds through thermal bubble nucleation.  
In order to determine the bubble nucleation temperature $T_{PT}^-$ we estimate the action of the three dimensional bounce $S^{(3)}(T)$ and require $\left. S^{(3)} / T \right|_{T_{PT}^-} \sim 140$.  
The bounce field configuration $\phi_B(r)$ is a saddle point solution of the Euclidean equation of motion with an $\mathrm{O}(3)$ symmetry.  
Let $\vec{\phi} = \left\{ h, s \right\}$ be the field space coordinate and let $\vec{\phi}_{sym} = v^{(s)}(T)$ and $\vec{\phi}_{brk} = v^{(b)}(T)$ be the location of the symmetric and broken phases at temperature $T$.  
In this notation, the field equation and boundary conditions can be written as
\begin{align}
	\frac{d^2 \vec{\phi}}{dr^2} + \frac{2}{r} \frac{d \vec{\phi}}{dr} - \vec{\nabla}_{\vec{\phi}} V_{\rm eff}(\vec{\phi}, T) = 0& \label{eq:O3bounce_fieldeqn} \\
	\left. \frac{d \vec{\phi}}{dr} \right|_{r=0} = 0 \, , \ \ \ \ \ \  \lim_{r \to \infty} \vec{\phi}(r) = \vec{\phi}_{sym}&
\end{align}
where $r$ is the radial coordinate and $V_{\rm eff}$ is the thermal effective potential.  
The bounce solution is a curve $\vec{\phi}_B(r)$ which starts nearby to $\vec{\phi}_{brk}$ at $r = 0$ and approaches $\vec{\phi}_{sym}$ as $r \to \infty$.  
Once the solution $\vec{\phi}_{B}(r)$ is obtained, the bounce action is calculated as 
\begin{align}\label{eq:BounceAction}
	S^{(3)}(T) = 4 \pi \int_{0}^{\infty} r^2 dr \left[ \frac{1}{2} \left( \frac{d \vec{\phi}_{B}}{dr} \right)^2 + V_{\rm eff}(\vec{\phi}_{B}(r), T) \right] \, .
\end{align}
It is difficult to solve \eref{eq:O3bounce_fieldeqn} by brute force numerics, because the solution is unstable to perturbations about the initial point $\vec{\phi}_B(0)$, and the over shoot / under shoot method is non-trivial to apply in two dimensions.

Profumo et. al. \cite{Profumo:2010kp} have outlined a numerical procedure which reduces the calculation to iteratively solving the one-dimensional analog of \eref{eq:O3bounce_fieldeqn}.  
They suggest that one should decompose the field equation into a basis with unit vectors parallel and perpendicular to the solution curve $\vec{\phi}(r)$.  
Suppose that there exists a curve $\vec{\phi}(x)$ that interpolates between $\vec{\phi}(0) = \vec{\phi}_{sym}$ and $\vec{\phi}(L) = \vec{\phi}_{brk}$.  
Let
\begin{align}\label{eq:x_def}
	x = \int_{\vec{\phi}_{sym}}^{\vec{\phi}(x)} \left| d \vec{\phi} \right|
\end{align}
be the distance along the curve such that $L$ is the total length and 
\begin{align}\label{eq:basis}
	\hat{e}_{\parallel}  = \frac{d \vec{\phi}}{dx} && \mathrm{and} && 
	\hat{e}_{\perp}  = \begin{pmatrix} 0 & 1 \\ -1 & 0 \end{pmatrix} \,  \frac{d \vec{\phi}}{dx} \, 
\end{align}
are the unit vectors parallel and perpendicular to the curve at $x$.  
In this basis, \eref{eq:O3bounce_fieldeqn} becomes
\begin{align}
	\left\{ \frac{d^2 x}{dr^2} + \frac{2}{r} \frac{dx}{dr} - \frac{dV(\vec{\phi}(x))}{dx} \right\} \hat{e}_{\parallel}& = 0 \label{eq:fieldeqn1} \\
	\left\{ \left| \frac{d^2 \vec{\phi}}{dx^2} \right| \left( \frac{dx}{dr} \right)^2 - \left( \vec{\nabla}_{\vec{\phi}} V \right)_{\perp} \right\} \hat{e}_{\perp}& = 0 \label{eq:fieldeqn2} \, .
\end{align}
The authors of \cite{Profumo:2010kp} solve these equations numerically using an iterative procedure.  
%by 1) making an Ansatz for the path $\vec{\phi}(x)$, 2) solving \eref{eq:fieldeqn1} for $x(r)$, 3) calculating a ``normal force'' $\vec{N}(x)$ given by the left-hand side of \eref{eq:fieldeqn2} [which will be non-vanishing for $\vec{\phi}(x(r)) \neq \vec{\phi}_b(r)$], and 4) deforming the contour $\vec{\phi}(x)$ by an amount proportional to $\vec{N}(x)$.  Iterating this procedure converges on the solution $\vec{\phi}_b(r) = \vec{\phi}(x(r))$.  

Since we compute $T_{PT}^-$ by calculating the bounce at various temperatures in order to solve $S^{(3)} / T \approx 140$, the iterative procedure is too computationally intensive for our purposes.  
Fortunately, in the parametric regime of interest the bounce solution $\vec{\phi}_B(r)$ can be approximated by $\vec{\phi}_{\rm app}(x) = \left\{ h(x), \bar{s}(h(x)) \right\}$ where $\bar{s}(h)$ satisfies\footnote{
In the parametric region described in Section \ref{sec:SingletExtension1PT}, the solution of $d U / d s = 0$ is not generally a single-valued function of $h$.  However, the boundary condition $\bar{s}(v) = 0$ selects out a unique trajectory which tends to stay in the ``valley'' connecting the two minima and passes through the saddle point. }
\begin{align}\label{eq:dvds_approx}
	\left. \frac{d U(\hs, 0)}{ds} \right|_{\bar{s}} = 0 && \mathrm{and} && \bar{s}(v) = 0  \, ,
\end{align}
$U$ is the classical potential, and $\vec{\phi}_{\rm app}(x)$ is parametrized by its length $x$ given by \eref{eq:x_def}.  
Using $\vec{\phi}_{\rm app}(x)$, we solve \eref{eq:fieldeqn1} for $x(r)$ and calculate $S^{(3)}$ using \eref{eq:BounceAction}.

%fig:compare_profumo
\begin{figure}[t]
\begin{center}
\includegraphics[width=0.50\textwidth]{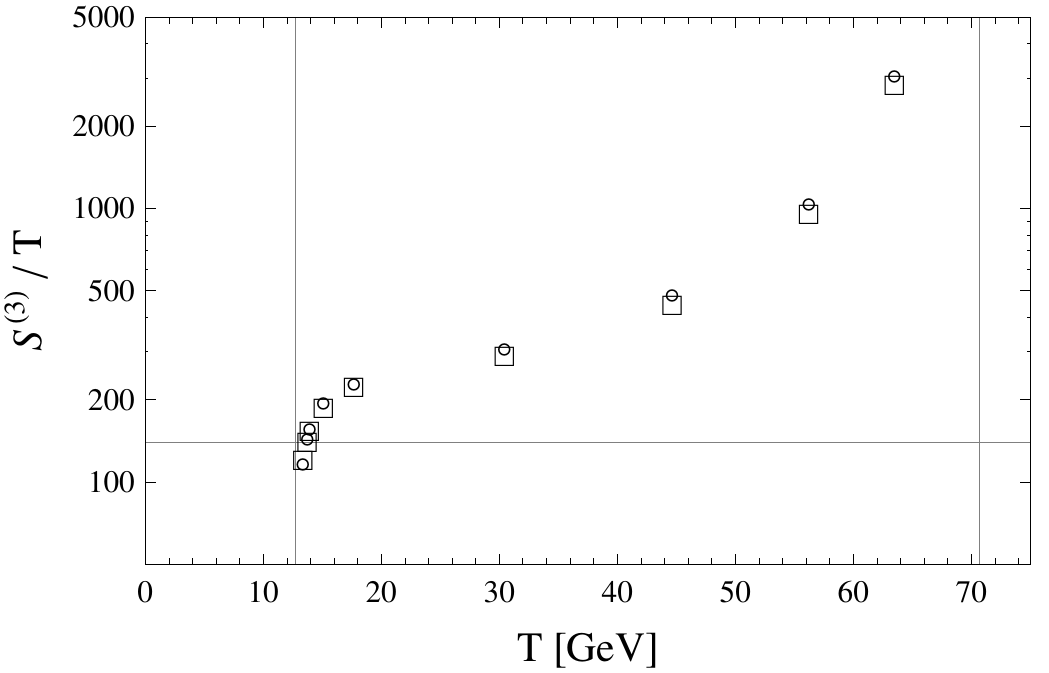} \hfill
\includegraphics[width=0.40\textwidth]{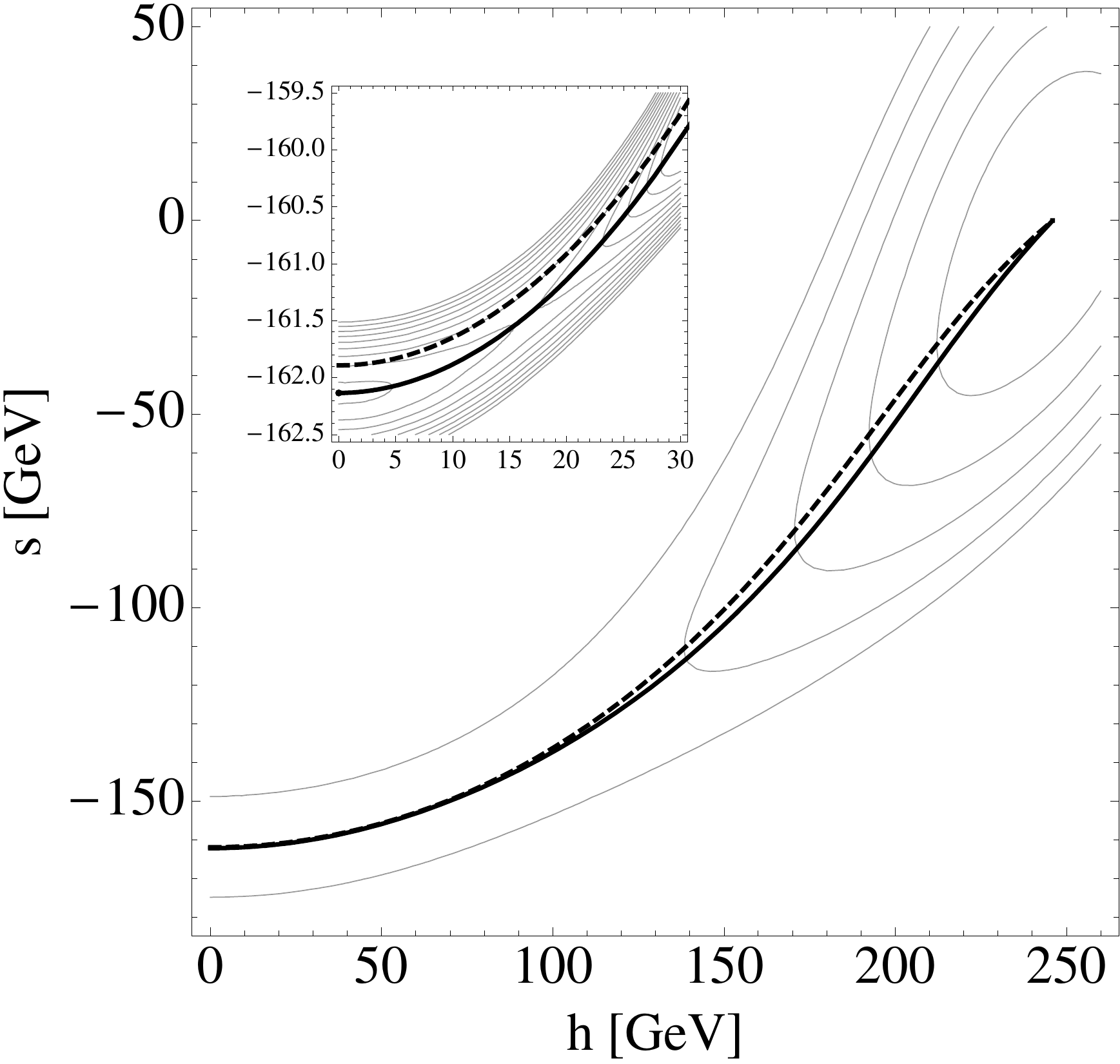} 
\caption{\label{fig:compare_profumo}
Comparisons of bounce calculations for the xSM benchmark point \eref{eq:xSM_benchmark}.  On the left, the bounce action computed at various temperatures between $T_0 = 12.7 \GeV$ and $T_c = 70.7 \GeV$ using the method of \cite{Profumo:2010kp} (squares) and our approximation (circles).  
On the right, the xSM thermal effective potential at $T_{PT}^- = 13.7 \GeV$.  The solid curve shows the trajectory $\vec{\phi}_B(x)$ obtained using the method of \cite{Profumo:2010kp}, and the dashed curve shows the approximation $\vec{\phi}_{\rm app}(x)$ given by \eref{eq:dvds_approx}.  The curves do not coincide at small $h$ because the minimum along the $h = 0$ axis shifts as the temperature is raised.  Nevertheless, the action along the two paths still agrees remarkably well.  
}
\end{center}
\end{figure}

To check our approximation, we also compute the PT temperature using the method of \cite{Profumo:2010kp} for a few parameter sets.  In Figure \ref{fig:compare_profumo} we contrast our approximation with the procedure of \cite{Profumo:2010kp} for the xSM benchmark point \eref{eq:xSM_benchmark}.   We find that our approximation tends to overestimate $S^{(3)}$ by a few percent generically.  However, $S^{(3)}$ is a rapidly increasing function of temperature, and  even an $\ord{5 \%}$ deviation in $S^{(3)}$ does not causes $T_{PT}^-$ to deviate appreciably.

\end{appendix}

\begin{acknowledgments}

We thank Lisa Everett, Lian-Tao Wang, and Sean Tulin for useful correspondence.  DJHC thanks the
hospitality of KIAS where part of this work was completed.  DJHC and
AJL were supported in part by the DOE through grant DE-FG02-95ER40896.

\end{acknowledgments}

\bibliographystyle{JHEP}
\bibliography{references}

\begin{comment}

\providecommand{\href}[2]{#2}\begingroup\raggedright\endgroup

\end{comment}

\end{document}